\documentclass[aps,twocolumn,prd,showpacs,nofootinbib]{revtex4}
\usepackage{amsmath}
\usepackage{graphicx}
\usepackage{dcolumn}
\usepackage{bm}
\usepackage{amssymb}
\usepackage{latexsym}

\def\setC{\mathbb{C}}

\def\setR{\mathbb{R}}
\def\setZ{\mathbb{Z}}
\def\setK{\mathbb{K}}
\def\setT{\mathbb{T}}

\newcommand{\ie}{\textsl{i.e.~}}

\newcommand{\eg}{\textsl{e.g.~}}
\newcommand{\etal}{\textsl{et al.~}}

\newcommand{\mP}{m_{_{\mathrm Pl}}}
\newcommand{\GReCO}{${\cal G}\setR\varepsilon\setC{\cal O}$}

\begin{document}

\title{Shift Symmetry and Inflation in Supergravity}

\author{Philippe Brax \footnote{Electronic mail: pbrax@cea.fr \\ Also
at Institut d'Astrophysique de Paris, \GReCO, 98bis boulevard Arago,
75014 Paris, France}} \affiliation{Service de Physique Th\'eorique,
CEA-Saclay, Gif/Yvette cedex, France F-91191}

\author{J\'er\^ome Martin \footnote{Electronic mail: jmartin@iap.fr}}
\affiliation{Institut d'Astrophysique de Paris, \GReCO, 98bis
boulevard Arago, 75014 Paris, France}

\date{\today}

\begin{abstract}
We consider models of inflation in supergravity with a shift
symmetry. We focus on models with one moduli and one inflaton field.
The presence of this symmetry guarantees the existence of a flat
direction for the inflaton field. Mildly breaking the shift symmetry
using a superpotential which depends not only on the moduli but also
on the inflaton field allows one to lift the inflaton flat
direction. Along the inflaton direction, the $\eta$-problem is
alleviated. Combining the KKLT mechanism for moduli stabilization and
a shift symmetry breaking superpotential of the chaotic inflation
type, we find models reminiscent of ``mutated hybrid inflation'' where
the inflationary trajectory is curved in the moduli--inflaton
plane. We analyze the phenomenology of these models and stress their
differences with both chaotic and hybrid inflation.

\vspace{0.3cm}

\end{abstract}

\maketitle

\section{Introduction}

The recent observations of the Cosmic Microwave Background (CMB)
fluctuations~\cite{wmap} give a strong hint in favor of an early
period of inflation~\cite{inflation} in the history of the
Universe. Inflation is an attractive scenario as it allows us to avoid
the difficulties which plague the standard hot big bang theory, \eg
the flatness and the horizon problems. Moreover, when combined with
quantum mechanics, inflation can also give rise to a satisfactory
model for structure formation, see Refs.~\cite{pert}. As a matter of
fact, inflation implies that the power spectrum of the cosmological
perturbations should be (almost) scale invariant~\cite{pert}, a
prediction which has been known for a long time to be in good
agreement with the astrophysical data. In fact, the data are now so
accurate that one can start probing the details of the inflationary
scenario. For instance, the small deviation from scale invariance in
the power spectrum predicted by inflation (for density perturbations
and for gravitational waves) directly encodes the underlying high
energy physics responsible for the phase of accelerated
expansion~\cite{slowroll} implying that the model building issue for
inflation is meaningful~\cite{LR}.

\par

The above-mentioned observations and the endeavor to better understand
space--like singularities like the big-bang have sparked a renewed
interest in cosmological models based on string theory and/or
supergravity~\cite{cosmological}. In this framework, alternatives to
inflation have also been studied like, for instance, the pre big-bang
scenario~\cite{PBB} or the ekpyrotic model~\cite{ekp}, but so far no
scenario has been as successful as inflation. Hence, finding a
satisfactory inflationary scenario from the most recent ideas in
string theory is an important challenge~\cite{stringinflation}. In all
these attempts, our universe is pictured as a moving brane embedded in
a compactification space. The acceleration of the universe is caused
by the motion of the brane~\cite{Dvali}. At the level of the four
dimensional effective description obtained after compactification, the
theory results in particular supergravity models. Then, in general,
two main issues must be addressed in order to obtain a satisfactory
scenario.

\par

The first problem is to obtain a sufficiently flat potential. One of
the crucial stumbling blocks of F--term inflation in supergravity is
the presence of large ${\cal O}\left(H \right)$ corrections (where $H$
is the Hubble parameter during inflation) to the inflaton mass, which
spoil the flatness of the inflaton potential. In order to lead to
realistic models, the string--inspired scenarios for inflation must
overcome this problem. Another highly conspicuous problem is the
stabilization issue. In string inflation, the 10--dimensional type IIB
theory is compactified on a Calabi-Yau manifold. Calabi-Yau
compactifications lead to two types of moduli fields describing the
K\"ahler and complex structure deformations. There is also another
scalar degree of freedom originating from the 10--dimensional
dilaton. Now all these moduli fields need to be stabilized in order to
guarantee the flatness of the inflaton potential, i.e. no runaway
behavior in the moduli directions.

\par

Different solutions to the above-mentioned questions have been
proposed. In particular, it has been noted recently that one can
obtain flat enough potentials by requiring that a shift symmetry
$\phi \to \phi + c$, where $c$ is a real constant, is a symmetry of
the K\"ahler potential, later broken mildly. This is particularly
natural within the low energy description of brane dynamics emerging
from string theory~\cite{Tye,shift,Hsu,D3,koyama}. On the other
hand, for the stabilization question, the complex structure moduli
and the dilaton can be generically stabilized when fluxes are turned
on.  This leaves only the K\"ahler moduli as flat directions. The
K\"ahler moduli can be also stabilized once strong coupling effects,
such as gaugino condensation, take place on a stack of $D7$ branes
wrapped around a four-cycle in the Calabi-Yau variety. This leads to
an ${\rm AdS}_4$ supergravity background. Now de Sitter space can be
achieved by incorporating an anti $D3$ brane whose energy density
makes the total energy density positive, this is the KKLT
stabilization mechanism~\cite{KKLT}. It was then realized that the
lifting of the vacuum energy can be performed in a supersymmetric
way using Fayet--Iliopoulos terms~\cite{Burgess2}. In supergravity,
it has been recently argued that this mechanism is problematic as
the Fayet--Iliopoulos terms of supergravity vanish in a supergravity
vacuum~\cite{nilles}.

\par

A first attempt to combine inflation with the KKLT mechanism was
carried out in the KKLMMT paper, see Ref.~\cite{KKLMMT}. In this
model, string inflation is obtained when a pair of $D3$-anti $D3$
branes is added to the configuration needed for the KKLT mechanism and
described above. The anti $D3$ is naturally sitting on top of the
other anti $D3$ branes, the distance between the $D3$ and anti $D3$
branes playing the role of the inflaton. The shift symmetry is
explicitly broken by the non-perturbative superpotential leading to
the stabilization of the K\"ahler moduli. The flat directions
corresponding the free motion of a $D3$ brane in the Calabi-Yau space
are then lifted in a strong manner. As a consequence, since the shift
symmetry is absent, the corresponding inflationary model suffers from
the $\eta$ problem. More precisely, it has been shown that the squared
mass of the inflaton is $2H^2$ spoiling the flatness of the
potential. This example is typical of the general flatness problem of
supergravity potentials when no shift symmetry is present.

\par

Subsequent papers have tried to overcome this problem by taking into
account the shift symmetry while still relying on the KKLT mechanism
for stabilizing the moduli fields. The first model~\cite{Hsu} to
combine both aspects was based on an interesting type of configuration
(different from the one envisaged in the KKLMMT paper) corresponding
to $D7$ branes evolving in the background of a very heavy stack of
$D3$ branes~\cite{Hsu,koyama,D3}. In this case, one may consider the
free motion of the $D7$ brane and the associated shift symmetry.  The
model realizes hybrid inflation in string inflation. The waterfall
fields are represented by the charged open strings between the $D3$
and the $D7$ branes. When the distance between the branes is large
enough, the configuration admits a flat direction which is lifted at
one loop as in supersymmetric hybrid inflation hence giving a
logarithmic slope to the potential in the inflaton direction. The
presence of a flat direction is directly linked to the fact that the
superpotential takes the form required to both preserve the shift
symmetry and stabilize the K\"ahler moduli. For small distances
between the branes, the waterfall fields condense and inflation ends.

\par

Another possibility was studied in Refs.~\cite{Tye}. It consists in
implementing the shift symmetry in the context of the original KKLMMT
scenario, where inflation is obtained when a pair of $D3$-anti $D3$
branes is present. In fact, the shift symmetry is present initially in
the K\"ahler potential. Imposing the shift symmetry invariance of the
superpotential leads to a flat potential for the inflation. Inflation
is then due to the small interaction potential between the $D3$ and
the anti $D3$ brane. As in usual brane inflation this requires an
adjustment of the brane--antibrane potential. In this context, it has
been noticed that threshold corrections in string theory give rise to
an explicitly shift symmetry breaking superpotential. As a
consequence, the model suffers from the $\eta $-problem unless the
stabilization of the complex structure moduli is fine-tuned.  We will
discuss this model (and compare it to what is achieved in the present
article) where the non-perturbative superpotential responsible for the
stabilization of the K\"ahler moduli is multiplicatively corrected by
loop effects~\cite{mac,berg}.  Finally the shift symmetry is also
instrumental in the race--track inflation model~\cite{race}.

\par

Notice that combining the KKLT approach to stabilization and the shift
symmetry in string theory requires the existence of isometries on
Calabi--Yau threefolds as originally argued in Ref.~\cite{shift} for
compactifications on $\setK_3\times \setT^2$. In the following, we
will concentrate on supergravity issues only and use string
motivations as a guideline only.

\par

Combining the flatness of the F--term inflation potential in
supergravity using the shift symmetry and the stabilization of
moduli (in particular using the KKLT mechanism) is the aim of this
paper. We will extract the main ingredients from the D3/D7 system
in string theory and deal with its supergravity description
exclusively. The two main new aspects of the model presented here
are the following. Firstly, we will consider a case where the
shift symmetry is initially present in both the K\"ahler and the
superpotential before being mildly broken by an explicit inflaton
dependence of the superpotential. In our case, the fact that the
superpotential can depend on the inflaton field allows us to give
a (small) slope to the potential already at the tree level without
having to compute the quantum corrections. Notice that the shift
symmetry breaking superpotential has coefficients constrained by
the COsmic Background Explorer (COBE) normalization and must
therefore be small. We also insist on obtaining the inflationary
potential from a strict $N=1$ supergravity context. Secondly, we
find that the end of inflation is due to the presence of the
moduli field playing the role of a waterfall field. Contrary to
the hybrid inflation case, the end of inflation is not triggered
by extra fields representing the charged open strings between the
branes. In fact, the name waterfall field is not very appropriate
in our case since we will see that the effective inflationary
model is reminiscent of mutated inflation and that inflation stops
due to the violation of the slow-roll conditions and not by
instability.

\par

The outline of the paper is the following. In section~II, we present
the flatness problem in F--term supergravity inflation and discuss
the realization of the shift symmetry as a K\"ahler transformation.
The shift symmetry alleviates the flatness problem and leads to
restrictions on the type of possible superpotentials. In
section~III, we give examples of chaotic inflation models with a
shift symmetry broken by the superpotential. The model has problems
such as runaway potentials. In Section~IV, we then introduce what we
call ``mutated chaotic inflation'' based on a chaotic inflation
superpotential and for which the inflationary trajectories become
curved. This model does not suffer from the $\eta $-problem and the
moduli is stabilized. We give a thorough analysis of the
inflationary parameter space. In particular, we find that the model
is different from hybrid inflation. In section~V, we combine the
KKLT stabilization mechanism and a chaotic inflation superpotential.
This gives another realization of mutated chaotic inflation. Then,
we discuss various aspects of our results. Finally, in section~VI we
present our conclusions.

\section{Shift Symmetry in Supergravity}

\subsection{The $\eta$-problem}

One of the stumbling blocks of F-term inflation in supergravity is the
natural presence of ${\cal O}\left(H\right)$ corrections to the
inflaton mass which would spoil the flatness of the potential (the
so--called $\eta$-problem). Let us consider the K\"ahler potential for
the inflaton of the form $K\left(\phi,\phi^\dagger \right) = \phi
\phi^\dagger$ and its role in the scalar potential of supergravity
\begin{equation}
\exp \left[\kappa K\left(\phi,\phi^\dagger\right)\right]V_{\rm inf}\,
,
\end{equation}
where $V_{\rm inf}$ is the inflationary potential when neglecting the
supergravity corrections and where $\kappa $ is defined by $\kappa
\equiv 8\pi/\mP^2$. Expanding the exponential leads to
\begin{equation}
\left[1+\kappa K\left(\phi,\phi^\dagger\right) +\cdots \right]V_{\rm
inf}\, .
\end{equation}
The first term leads to the inflation potential, while the second one
leads to a term in $H^2\phi\phi^\dagger$ which spoils the flatness of
the potential, \ie the quantity
\begin{equation}
\frac{m_{\rm Pl}^2}{8\pi}\frac{1}{V}\frac{\partial ^2 V}{{\rm \partial
  }\phi ^2}\, ,
\end{equation}
becomes of order one. However, one should also remark the
following. The above parameter (the so-called $\eta$-parameter) is not
the parameter which controls whether inflation is taking place or
not. Indeed, strictly speaking, the condition $\ddot{a}>0$, where
$a(t)$ is the Friedmann-Lema\^{\i}tre-Robertson-Walker scale factor,
is equivalent to $\epsilon\equiv -\dot{H}/H^2 <1$ or $m_{\rm
Pl}^2/(16\pi V^2)(\partial V/\partial \phi )^2 <1$. When the $\eta
$-parameter becomes of order one, we just have violation of the
slow-roll conditions although, in principle, inflation could still
proceed. Of course, one could argue, based on the well-known formula,
$n_{_{\rm S}}-1\simeq 2\eta -6\epsilon$, that a parameter $\eta $ of
order one would imply a scalar spectral index far from scale
invariance, \ie $\vert n_{_{\rm S}}-1\vert \gg 1$. Again, this
conclusion is not rigorous because the previous formula is derived
under the assumption that the slow-roll conditions are valid and,
hence, not applicable when $\eta $ is large. In principle, one could
imagine a situation where $\eta $ is large but where inflation
proceeds and leads to an almost scale-invariant spectrum. Admittedly,
this is probably not the most generic situation but this is possible
in principle; for interesting recent remarks on the $\eta $ problem,
see also Ref.~\cite{kinney}.

\subsection{Shift symmetry}

Let us now consider a typical ansatz, motivated by string inspired
theories~\cite{Hsu} given by
\begin{eqnarray}
\label{deltanf}
K &=& -\frac{3}{\kappa }\ln \left[\kappa ^{1/2}\left(\rho
+\rho ^{\dagger}\right)-\sigma \kappa {\cal K}\left(\phi, \phi
^{\dagger}\right)\right]
\nonumber \\
& & +s{\cal G}\left(\phi ,\phi
^{\dagger}\right)\, , \\
W &=& W\left(\rho ,\phi \right)\, ,
\end{eqnarray}
where $\sigma =0,1$ and/or $s=0,1$ according to the situation we want
to describe. In the above expression, the field $\rho $ is a moduli
while $\phi $ represents the inflaton. When $\sigma=1$ and $s=0$, the
K\"ahler potential describes the motion of a $D3$ brane within a
Calabi-Yau manifold (we have only retained one of the possible six
directions). In that case ${\cal K}\left(\phi,\phi^\dagger\right)$ is
the K\"ahler potential on the Calabi-Yau manifold. For small
arguments, one can expand ${\cal
K}\left(\phi,\phi^\dagger\right)=\phi\phi^\dagger+\cdots$. Similarly
when $\sigma=0$, ${\cal G}$ is also identified with the K\"ahler
potential of the Calabi-Yau manifold and $\phi$ corresponds to the
position of a $D7$ brane. These two cases will be exemplified later.

\par

Let us come back to the issue of the shift symmetry $\phi \to \phi+c$
where $c$ is real which guarantees that the real part of the inflaton
superfield is a flat direction. Following the above discussion, we
focus on the K\"ahler potential
\begin{equation}
\label{typicK}
K= -\frac{3}{\kappa }\ln \left[\kappa ^{1/2}(\rho+\rho^\dagger )
-\sigma \kappa \phi\phi^\dagger \right] +s\phi\phi^\dagger\, .
\end{equation}
which follows from Eq.~(\ref{deltanf}) in the small argument
limit. Let us also remind that a K\"ahler transformation is a
transformation which leaves the Lagrangian of a supergravity
theory invariant. It is of the form
\begin{eqnarray}
K(\rho ,\phi ) &\to & K(\rho ,\phi ) +\xi\left (\rho , \phi \right)+\xi
^\dagger \left(\rho ^\dagger, \phi ^\dagger \right)\, ,\\
W(\rho ,\phi ) &\to &{\rm e}^{-\kappa \xi(\rho ,\phi )}W(\rho ,\phi )\, ,
\end{eqnarray}
where $\xi $ is an arbitrary function. The shift symmetry, $\phi \to
\phi +c$, can be viewed as a K\"ahler transformation provided the
moduli field $\rho $ transforms in a specific way and the
superpotential possesses a given form, namely
\begin{eqnarray}
\label{shiftsym}
\rho &\to & \rho + \sigma c \kappa ^{1/2}\phi +\sigma \kappa
^{1/2}\frac{c^2}{2}\, ,\\
\label{shiftsym2}
W(\rho,\phi) &=& {\rm e}^{-s\kappa
\phi^2/2}W\left(\rho -\sigma \kappa ^{1/2}\frac{\phi^2}{2}\right)\, .
\end{eqnarray}
First of all, one can check that the quantities $\kappa
^{1/2}\left(\rho +\rho ^{\dagger }\right)-\sigma \kappa \phi \phi
^\dagger $ and $\rho -\sigma \kappa ^{1/2}\phi ^2/2$ are invariant
under the above transformation of the fields $\phi $ and $\rho
$. Secondly, if the superpotential has the form given in
Eq.~(\ref{shiftsym2}), then the corresponding K\"ahler transformation
is described by the function
\begin{equation}
\xi (\phi )=sc\phi +s\frac{c^2}{2}\, .
\end{equation}
Therefore, if we want to implement the shift symmetry, one must
restrict our considerations to models described by the K\"ahler
potential given in Eq.~(\ref{typicK}) and superpotential given by
Eq.~(\ref{shiftsym2}).

\par

This class of models can be simplified further (or transform to
another form). Two ingredients are necessary. The first one is a
K\"ahler transformation described by the function $\xi =-s\phi ^2/2$
(this K\"ahler transformation has of course nothing to do with the
other K\"ahler transformation considered before). In this case the
K\"ahler potential and superpotential of the shift symmetry invariant
model is equivalent to the one given by
\begin{eqnarray}
K &=& -\frac{3}{\kappa }\ln \left[\kappa ^{1/2}(\rho+\rho^\dagger )
-\sigma \kappa \phi\phi^\dagger \right] \nonumber \\
& & -\frac{s}{2}\left(\phi-\phi^\dagger\right)^2\, , \\
W(\rho,\phi) &=& W\left(\rho -\sigma \kappa ^{1/2}
\frac{\phi^2}{2}\right)\, .
\end{eqnarray}
The second ingredient consists in changing variables in a holomorphic
way (as required by supersymmetry) $\rho \to \rho -\sigma \kappa
^{1/2}\phi^2/2$. In terms of the new variables, the K\"ahler potential
becomes
\begin{eqnarray}
\label{sigma}
K &=& -\frac{3}{\kappa }\ln \left[\kappa ^{1/2}\left(\rho+\rho ^\dagger
\right)+\frac{\sigma }{2}\kappa
\left(\phi-\phi^\dagger\right)^2\right]
\nonumber \\
& & -\frac{s}{2}\left(\phi-\phi^\dagger \right)^2 \, ,\\
\label{sigma2}
W(\rho,\phi) &=& W(\rho)\, .
\end{eqnarray}
This representation guarantees that the inflaton field has a flat
direction. This is the representation with ${\cal K}=-\sigma
\left(\phi-\phi^\dagger\right)^2/2$ and ${\cal
G}=-s\left(\phi-\phi^\dagger\right )^2/2$ in Eq.~(\ref{deltanf}). The
previous model does not immediately lead to inflation in the $\phi$
direction as the scalar potential is exactly flat along the real $\Re
(\phi)$ direction due to the shift symmetry (but the potential could
be lifted by quantum corrections). A large class of supergravity
models with no $H^2$ corrections to the inflaton mass can be
constructed by modifying the previous superpotential and including an
explicit $\phi$ dependence which breaks the shift symmetry.  Notice
that the $H^2$ corrections will be absent as the K\"ahler potential is
still shift symmetric. We now turn to the construction of such models.

\vspace{0.5cm}

\section{Inflation and Shift Symmetry}

\subsection{The scalar potential}

Following the discussion of the previous section, we now generalize
the class of models invariant under the shift symmetry and consider
the K\"ahler potential given by (we remind that, in the present
context, the field $\phi $ will be viewed as the inflaton while the
field $\rho $ will represent a moduli)
\begin{equation}
K= -\frac{3}{\kappa }\ln \left[\kappa ^{1/2}\left(\rho+ \rho ^{\dagger
}\right)-\kappa {\cal K}\left(\phi-\phi^\dagger \right)\right]
+{\cal G}\left(\phi-\phi^\dagger\right)\,
\end{equation}
in such a way that the shift symmetry is explicitly present. In the
above expression, ${\cal K}$ and ${\cal G}$ are arbitrary functions,
the form of which is not specified at this stage. In order to
calculate the corresponding potential, one must first evaluate the
matrix $G_{A\bar{B}}$ defined by
\begin{equation}
G_{A\bar{B}}=\frac{\partial ^2}{\partial \varphi ^A
\partial \left(\varphi ^B\right)^{\dagger}}
\left[\kappa K+\ln \left(\kappa ^3\left\vert W
\right\vert^2\right)\right]\, ,
\end{equation}
where $\varphi ^A=\{\rho ,\phi \}$ and where $W=W\left(\rho ,\phi
\right)$ is the superpotential which, as announced, explicitly depends
on $\phi $. Explicitly, straightforward calculations lead to
\begin{widetext}
\begin{equation}
\displaystyle
G_{A\bar{B}}=\left\{
\begin{matrix}
\displaystyle 3\frac{\kappa }{\Delta ^2} & \, \, \, \, \, &
\displaystyle 3\frac{\kappa
^{3/2}}{\Delta ^2} \frac{\partial {\cal K}}{\partial \left(\phi -\phi
^\dagger \right)} \cr \cr
\displaystyle -3\frac{\kappa ^{3/2}}{\Delta ^2}
\frac{\partial {\cal K}}{\partial \left(\phi -\phi ^\dagger \right)}
& \, \, \, \, \, &
\displaystyle -3\frac{\kappa }{\Delta } \frac{\partial ^2{\cal
K}}{\partial \left(\phi -\phi ^\dagger \right)^2} -3\frac{\kappa
^2}{\Delta ^2}\left[\frac{\partial {\cal K}}{\partial \left(\phi
-\phi ^\dagger \right)}\right]^2 -\kappa \frac{\partial ^2{\cal
G}}{\partial \left(\phi -\phi ^\dagger \right)^2}
\end{matrix}\right\}\, ,
\end{equation}
where the quantity $\Delta $ is defined by $\Delta \equiv \kappa
^{1/2}\left(\rho +\rho ^{\dagger }\right)-\kappa {\cal K}\left(\phi
-\phi ^\dagger \right)$. These models are of the no-scale type with a
cancellation of the $-3\vert W\vert^2$ term in the scalar potential
$V={\rm e}^G\left(G^AG_A-3\right)/\kappa ^2$ [we remind here that the
function $G$ is given by $G\equiv \kappa K+\ln \left(\kappa ^3\vert
W\vert ^2\right)$]. The potential reads
\begin{eqnarray}
{\rm e}^{-\kappa {\cal G}}V (\rho ,\phi) &=& \frac{1}{3\Delta}\biggl
\vert \frac{\partial W}{\partial \rho }\biggr \vert^2 \times \left\{
1+\frac{\kappa }{\Delta {\cal D}} \left[\frac{\partial {\cal
K}}{\partial \left(\phi -\phi ^\dagger \right)}\right]^2\right\}
-\frac{1}{\Delta^2 {\cal D}} \left \vert\frac{\partial W}{\partial
\phi }\right \vert^2 \nonumber \\ & & -\frac{\kappa ^{1/2}}{\Delta^2}
\left(W\frac{\partial W^\dagger }{\partial \rho ^\dagger } +W^\dagger
\frac{\partial W }{\partial \rho }\right) \times \left[ 1-\frac{\kappa
}{{\cal D}} \frac{\partial {\cal K}}{\partial \left(\phi -\phi
^\dagger \right)} \frac{\partial {\cal G}}{\partial \left(\phi -\phi
^\dagger \right)}\right] \nonumber \\ & & +\frac{\kappa ^2}{\Delta
^2{\cal D}} \left[\frac{\partial {\cal G}}{\partial \left(\phi -\phi
^\dagger \right)}\right]^2\vert W\vert ^2 -\frac{\kappa ^{1/2}}{\Delta
^2{\cal D}} \frac{\partial {\cal K}}{\partial \left(\phi -\phi
^\dagger \right)} \left(\frac{\partial W}{\partial \rho }
\frac{\partial W^\dagger }{\partial \phi ^\dagger } -\frac{\partial
W^\dagger }{\partial \rho ^\dagger } \frac{\partial W }{\partial \phi
}\right) \nonumber \\ & & +\frac{\kappa }{\Delta^2{\cal D}}
\frac{\partial {\cal G}}{\partial \left(\phi -\phi ^\dagger \right)}
\left(W^\dagger \frac{\partial W }{\partial \phi } -W \frac{\partial W
^\dagger}{\partial \phi ^\dagger }\right)\, ,
\end{eqnarray}
where the coefficient ${\cal D}$ is defined by the following expression
\begin{equation}
{\cal D}\equiv
3\frac{\partial ^2 {\cal K}}{\partial
\left(\phi -\phi ^\dagger \right)^2}+\Delta
\frac{\partial ^2 {\cal G}}
{\partial \left(\phi -\phi ^\dagger \right)^2}\, .
\end{equation}
Let us notice that, if the fields $\rho $ and $\phi $ are real and if
the superpotential $W$ is also real when $\rho $ and $\phi $ are real,
then the above expression can be simplified further since the last two
terms cancel out. Moreover, if ${\cal G}=0$ and if the superpotential
does not depend on the field $\phi $ but only on the moduli $\rho$ ,
then one recovers the expression (5.12) found in Ref.~\cite{KKLMMT}
(which, therefore, appears to be a particular case of the most general
formula established above), namely
\begin{eqnarray}
V (\rho ) &=& \frac{1}{3\Delta}\biggl
\vert \frac{\partial W}{\partial \rho }\biggr \vert^2 \times \left\{
1+\frac{\kappa }{3\Delta }
\left[\frac{\partial ^2{\cal
K}}{\partial \left(\phi -\phi ^\dagger \right)^2}\right]^{-1}
\times \left[\frac{\partial {\cal
K}}{\partial \left(\phi -\phi ^\dagger \right)}\right]^2\right\}
-\frac{\kappa ^{1/2}}{\Delta^2}
\left(W\frac{\partial W^\dagger }{\partial \rho ^\dagger }
+W^\dagger \frac{\partial W }{\partial \rho }\right)\, .
\end{eqnarray}
In this paper, we will consider a different situation. The functions
${\cal K}$ and ${\cal G}$ can always be Taylor expanded according to
\begin{equation}
{\cal K}=\sum _{n=0}^{\infty }\frac{a_n}{n!}\left(\phi -\phi
^\dagger \right)^n\, , \quad {\cal G}=\sum _{n=0}^{\infty
}\frac{b_n}{n!}\left(\phi -\phi ^\dagger \right)^n \, .
\end{equation}
We will assume that the functions ${\cal K}$ and ${\cal G}$ satisfy
the properties
\begin{equation}
{\cal K}\vert_{\phi=\phi^\dagger}={\cal G}\vert_{\phi=\phi^\dagger}=0
\, , \quad
\frac{\partial {\cal K}}{\partial\left(\phi -\phi ^\dagger \right)
}\biggl \vert_{\phi=\phi^\dagger}=
\frac{\partial {\cal G}}{\partial\left(\phi -\phi ^\dagger \right)
}\biggl \vert_{\phi=\phi^\dagger}=0\, ,
\end{equation}
\ie that $a_0=a_1=b_0=b_1=0$, the remaining coefficients being
arbitrary.  These simple assumptions are sufficient to render the
matrix $G_{A\bar{B}}$ diagonal. Explicitly, the potential takes the form
\begin{equation}
\label{potgeneral}
V (\rho ,\phi)= \frac{1}{3\Delta}\biggl \vert \frac{\partial
W}{\partial \rho }\biggr \vert^2 -\frac{1}{\Delta^2}
\left[3\frac{\partial ^2 {\cal K}}{\partial \left(\phi -\phi ^\dagger
\right)^2}+\Delta \frac{\partial ^2 {\cal G}}{\partial \left(\phi
-\phi ^\dagger \right)^2} \right]^{-1}\left \vert\frac{\partial
W}{\partial \phi }\right \vert^2 -\frac{\kappa ^{1/2}}{\Delta^2}
\left(W\frac{\partial W^\dagger }{\partial \rho ^\dagger } +W^\dagger
\frac{\partial W }{\partial \rho }\right)\, .
\end{equation}
In the following, we will mainly focus on the ansatz (\ref{sigma}) for
which the potential becomes
\begin{equation}
\label{potgeneral2} V (\rho ,\phi)= \frac{1}{3\Delta}\biggl \vert
\frac{\partial W}{\partial \rho }\biggr \vert^2 +\frac{1}{\Delta^2}
\left(3\sigma +s \Delta \right)^{-1}\left \vert\frac{\partial
W}{\partial \phi }\right \vert^2 -\frac{\kappa ^{1/2}}{\Delta^2}
\left(W\frac{\partial W^\dagger }{\partial \rho ^\dagger } +W^\dagger
\frac{\partial W }{\partial \rho }\right)\, .
\end{equation}
\end{widetext}
This last equation is the main equation of this section. It gives the
general form of the scalar potential in a theory where the shift
symmetry is implemented in the K\"ahler potential and for a general
superpotential which can depend both on the inflaton field $\phi $ but
also on the moduli $\rho $. From the general form of the K\"ahler
potential, we obtain
\begin{equation}
\label{kinetic}
K_{\phi \phi ^\dagger }=-\frac{3}{\Delta } \frac{\partial ^2 {\cal
K}}{\partial \left(\phi -\phi ^\dagger \right)^2} -\frac{\partial ^2
{\cal G}}{\partial \left(\phi -\phi ^\dagger \right)^2}\, ,\quad
K_{\rho \rho ^\dagger }= \frac{3}{\Delta ^2}\, .
\end{equation}
which will be used to normalize the fields.  In the case of
(\ref{sigma}), this gives
\begin{equation}
\label{kinetic2} K_{\phi \phi ^\dagger }=\frac{3\sigma }{\Delta } +s
\, ,\quad K_{\rho \rho ^\dagger }= \frac{3}{\Delta ^2}\, .
\end{equation}

Let us stress once more why the shift symmetry is so crucial to
alleviate the $\eta$-problem. To do so, we will compare with the
results obtained in Ref.~\cite{KKLMMT}. In this specific model the
K\"ahler potential springs from ${\cal
K}(\phi,\phi^\dagger)=\phi\phi^\dagger$ and $\sigma=1$.  As argued in
section~II, this combination is in fact explicitly shift symmetric
(after using the transformations studied in that section). Now it is
assumed that the non-perturbative superpotential depends only on the
moduli $\rho$. Notice that this step is exactly where the shift
symmetry is explicitly broken.  Indeed the shift symmetry invariant
combination is $\rho -\kappa^{1/2}\phi^2/2$. Now the scalar potential
follows as before
\begin{widetext}
\begin{eqnarray}
V (\rho ) &=& \frac{1}{3\Delta}\biggl \vert \frac{\partial W}{\partial
\rho }\biggr \vert^2 \left( 1+\frac{\kappa }{3\Delta }
\phi\phi^\dagger\right) -\frac{\kappa ^{1/2}}{\Delta^2}
\left(W\frac{\partial W^\dagger }{\partial \rho ^\dagger } +W^\dagger
\frac{\partial W }{\partial \rho }\right)\, ,
\end{eqnarray}
where $\Delta\equiv \kappa^{1/2} (\rho+\rho^\dagger) -\kappa
\phi\phi^\dagger$. The potential is corrected by the presence of
an anti D3-brane leading to a total potential
\begin{eqnarray}
V (\rho ) &=& \frac{1}{3\Delta}\biggl \vert \frac{\partial
W}{\partial \rho }\biggr \vert^2 \left(1+\frac{\kappa }{3\Delta }
\phi\phi^\dagger\right) -\frac{\kappa ^{1/2}}{\Delta^2}
\left(W\frac{\partial W^\dagger }{\partial \rho ^\dagger }
+W^\dagger \frac{\partial W }{\partial \rho }\right)\
+\frac{E}{\Delta^2}\, ,
\end{eqnarray}
\end{widetext}
for a given constant $E$. The extra potential is explicitly shift
symmetric [again using the transformation on the fields discussed at
the beginning of this article in section~II, the denominator of the
new term becomes $\kappa^{1/2}(\rho +\rho ^{\dagger }) +\kappa (\phi
-\phi ^{\dagger })^2/2$].  Now assume that $W(\rho)$ has been chosen
in such a way that $\phi=0$ and $\rho=\rho_0$ is a minimum for a real
$\rho_0$, the potential reads close to the minimum
\begin{equation}
V=V(\rho_0)\left(1+\frac{2}{3}\kappa \Phi\Phi^\dagger\right)\, ,
\end{equation}
for the canonically normalized inflaton $\Phi=\phi\sqrt{3/2\rho}$.  We
can see that the inflaton potential is not flat and runs into the
$\eta$-problem.

\par

Our goal is now to find inflationary models where the moduli is
stabilized. We will consider a class of models where the
superpotential breaks the shift symmetry mildly and does not
jeopardize the flatness of the potential. Some examples are given in
the next section and some stringy motivations for such a form are also
presented. We find that the potentials lead to inflation along the
$\phi$ direction. Notice that there is no cosmological constant and
that the potential is a function of the inflaton field which is
polynomial when the superpotential has a polynomial dependence on the
inflaton field.

\subsection{Chaotic Inflation}

As a warm up, let us now discuss a simple example of chaotic inflation
as can be found in Ref.~\cite{linde} where a similar case is
treated. Explicitly, one assumes
\begin{eqnarray}
{\cal K} &=& -\frac{1}{2}\left(\phi -\phi ^{\dagger }\right)^2\, , \quad
{\cal G}=+\frac{1}{2}\left(\phi -\phi ^{\dagger }\right)^2\, ,\\
W(\rho, \phi ) &=& \frac{1}{\sqrt{2}}m\phi^2 \, .
\end{eqnarray}
The factor $1/\sqrt{2}$ in the inflationary part of the superpotential
is chosen for future convenience. Notice that the shift symmetry is
preserved by the K\"ahler potential while the superpotential breaks
the shift symmetry explicitly. Then, using Eqs.~(\ref{potgeneral}) and
(\ref{potgeneral2}), straightforward calculations lead to
\begin{equation}
V(\rho ,\phi)=\frac{1}{\Delta ^2} \frac{1}{3-\Delta}\left\vert
\frac{\partial W}{\partial \phi } \right\vert ^2\, .
\end{equation}
The moduli $\rho$ can be stabilized for $\Delta =\kappa^{1/2}(\rho
+\rho ^\dagger )=2$ (since ${\cal K}=0$ for a real inflaton) and,
therefore, one has $V(\phi, \rho )=\vert \partial W/\partial \phi
\vert ^2/4$. Notice the $1/4$ prefactor which comes from stabilizing
the moduli and there is no $H^2$ contribution to the mass of the
inflaton due to the partial shift symmetry preserved by the K\"ahler
potential. In this model, we obtain chaotic inflation depending
explicitly on the shift symmetry breaking superpotential. In
particular for a quadratic superpotential, we obtain that
\begin{equation}
V(\phi)=\frac{1}{2} m^2\phi^2 \, ,
\end{equation}
which is nothing but the usual chaotic inflation potential and where
one can check from Eq.~(\ref{kinetic}) that the real field $\phi $ is
correctly normalized since $K_{\phi \phi ^\dagger }=1/2$. However, it
happens that the present model is not the one favored in string theory
as ${\cal G}$ has the wrong sign. In the following section we examine
cases where the K\"ahler potential has a structure dictated by string
theory.

\subsection{Runaway potential}

\begin{figure*}
\includegraphics[width=.95\textwidth,height=.85\textwidth]{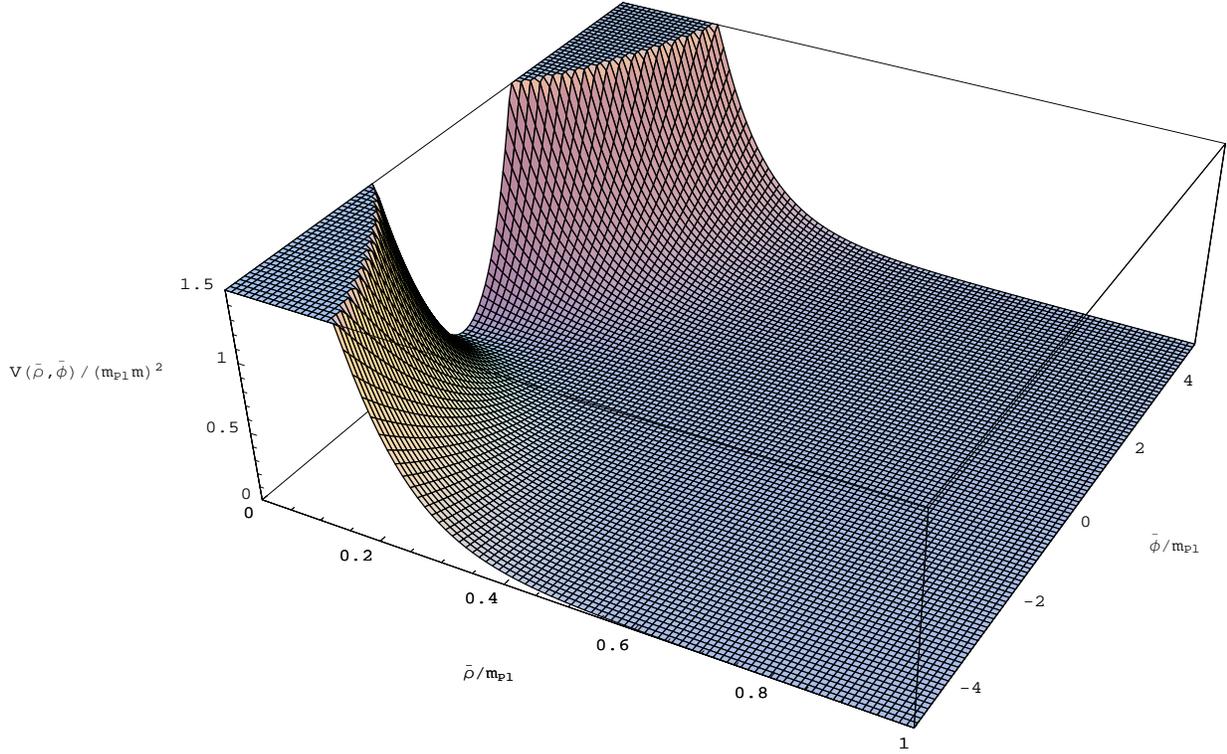}
\caption{Potential $V(\bar{\rho}, \bar{\phi} )$ given by
Eq.~(\ref{potmod}) represented in terms of the normalized field
$\bar{\phi }$ and $\bar{\rho }$. The potential possesses a minimum in
the $\bar{\phi }$ direction but, as is obvious from the figure, not in
the $\bar{\rho }$ direction. Hence, the moduli is not stabilized in
this model.}
\label{potnonstab}
\end{figure*}

In this subsection, we illustrate the stabilization problem on a
simple example. As mentioned above, we now consider a model where the
function ${\cal G}$ possesses an overall minus sign, \ie where the
K\"ahler potential has a form which can be justified in string
theory. The superpotential is chosen to be the same as in the previous
subsection. In this case, we show below that the moduli can no longer
be stabilized. Therefore, when the K\"ahler potential is chosen
according to the string-motivated considerations, it is necessary to
consider more complicated forms for the superpotential (depending
explicitly on the moduli). This will be done in the following
section. Here we choose
\begin{equation}
{\cal K}=0\, , \quad {\cal G}=-\frac{1}{2}\left(\phi -\phi
^{\dagger }\right)^2\, ,\quad W(\rho, \phi )=2\sqrt{2}m\phi^2\, ,
\end{equation}
where, again, the factor $2\sqrt{2}$ in front of the inflationary
superpotential has been chosen for future convenience. This leads to
the following scalar potential
\begin{equation}
V(\rho ,\phi)=\frac{8}{\Delta
^3}m^2\phi ^2\, .
\end{equation}
Redefine the fields in order to have properly normalized fields using
$K_{\phi \phi ^\dagger}=1$ and $K_{\rho \rho ^\dagger }=3/\Delta ^2$
as given by Eq.~(\ref{kinetic}), we find that the normalized field
$\bar{\phi }$ and $\bar{\rho }$ (assuming that $\rho =\rho ^{\dagger
}$) are given by $\bar{\phi }=\sqrt{2}\phi $ and $\kappa ^{1/2}\bar
{\rho}= \sqrt{3/2}\ln \left(\kappa ^{1/2}\rho \right)$. The potential
for the redefined fields finally reads
\begin{equation}
\label{potmod}
V(\bar{\rho },\bar{\phi }) =\frac{1}{2} m^2 \bar{\phi }^2 \exp
\left(-\sqrt{6}\kappa ^{1/2}\bar{\rho }\right)\, .
\end{equation}
This potential is represented in Fig.~\ref{potnonstab}. Notice that
$\rho$ (or $\bar{\rho }$) is not stabilized and is given by a runaway
potential. This is the usual problem for moduli fields. In the
following section, we give examples of potentials leading to inflation
and a stabilization of the moduli. At this point, a last remark is in
order. We will see that the KKLT procedure consists in adding a term
$1/\rho ^2$ and/or $1/\rho ^3$ to the potential. It is clear here that
this mechanism would not be enough to stabilize the moduli.

\section{Mutated chaotic inflation}

\subsection{Giving a mass to the inflaton}

In this section, we discuss a successful model of inflation combining
moduli stabilization and chaotic inflation. To motivate the
introduction of a mass term for the inflaton field, let us consider
the D3/D7 system in string theory~\cite{D3,Hsu,koyama}.  This system
can be modeled at low energy using three fields, the inflaton $\phi$
measuring the inter--brane distance and two charged fields $\phi^\pm$
representing the open strings between the two types of branes. The
fields interact according to the superpotential
\begin{equation}
\label{superpotpm}
W\left(\phi, \phi^+,\phi ^-\right)=\sqrt 2 g\left( \phi^+ \phi \phi^-
-\zeta ^2\phi \right)\, ,
\end{equation}
where $g$ is the $U(1)$ gauge coupling and $\zeta $ is a constant term
which is turned on when the compactification looks like a resolution
of an orbifold singularity $\setC^2/\setZ_2$ locally. The term $\zeta
^2$ is stabilized at the same time as the complex structure
moduli. This is not the case of the Fayet--Iliopoulos term which
depends on the K\"ahler moduli. To simplify we consider the case where
there is no Fayet-Iliopoulos term.

\par

Let us also give the K\"ahler function of the model. As before, for
the inflaton field $\phi $, we focus on the case where the function
${\cal K}$ vanishes and where ${\cal G}=-1/2 \left(\phi-\phi^{\dagger
}\right)^2$. On the other hand, the K\"ahler functions for the charged
fields are standard and, therefore, the total K\"ahler function is
given by
\begin{equation}
K\left(\phi, \phi^+,\phi ^-\right)=-\frac12 \left(\phi-\phi^{\dagger
}\right)^2+\phi ^+\left(\phi ^+\right)^{\dagger}+\phi ^-\left(\phi
^-\right)^{\dagger}\, .
\end{equation}

\par

As we discussed in the introduction, stabilizing moduli can be
achieved by considering non-perturbative superpotentials springing
from gaugino condensation. In this case, the superpotential contains a
part which explicitly depends on the moduli and reads
\begin{equation}
W(\rho )=W_0-A\exp \left(-\beta \kappa ^{1/2}\rho \right)\, ,
\end{equation}
where $W_0$, $A$ are free constants of dimension three and $\beta $ is
dimensionless. The superpotential $W(\rho )$ has already been studied
in Refs.~\cite{KKLT} and~\cite{Tye}. In string theory, the constant
term springs from the stabilization of both the complex structure
moduli and the dilaton. The exponential term is of non-perturbative
origin. We take the total superpotential of the system to be the sum
of the two previous expression, namely
\begin{equation}
\label{Wtotal}
W\left(\phi, \phi^+,\phi ^-,\rho \right)=W\left(\phi, \phi^+,\phi
^-\right)+W(\rho )\, .
\end{equation}
It is known from string considerations that the coupling constant $g$
and/or $\zeta $ can not depend on the moduli $\rho $~\cite{D3}.
Therefore, this superpotential is the simplest way of coupling the
moduli to inflation.

\par

Finally, the total K\"ahler potential is also the sum of the K\"ahler
potentials in the inflaton and moduli sector. It can be expressed as
\begin{equation}
K\left(\phi, \phi^+,\phi ^-,\rho \right)=K\left(\phi, \phi^+,\phi
^-\right)-\frac{3}{\kappa }\ln \left[\kappa ^{1/2}\left(\rho +\rho
  ^{\dagger} \right)\right]\, .
\end{equation}

\par

Having specified what the K\"ahler and super potentials of the model
are, one can now determine the corresponding scalar potential. It
reads
\begin{widetext}
\begin{eqnarray}
\label{pottotalpm} V &=& \frac{{\rm e}^{\kappa \left(\vert
\phi^+\vert^2 +\vert
\phi^-\vert^2\right)}}{\kappa^{3/2}\left(\rho+\rho^{\dagger}\right)^3}
\biggl\{\left\vert\sqrt{2}g\phi\phi^- +\kappa
\left(\phi^+\right)^{\dagger }W\right\vert ^2 +
\left\vert\sqrt{2}g\phi\phi^++\kappa \left(\phi^-\right)^{\dagger
}W\right\vert ^2 +2g^2\left\vert \phi^+\phi^--\zeta ^2\right\vert^2
\nonumber \\ & &-2\kappa^{3/2} \beta A \left(\rho+
\rho^\dagger\right)\Re \left[{\rm
e}^{-\beta\kappa^{1/2}\rho^\dagger}W\left(\phi,\phi^+,\phi^-\right)\right]\biggr\}
+ {\rm e}^{\kappa \left(\vert \phi^+\vert^2 +\vert
\phi^-\vert^2\right)} \tilde{\cal V}(\rho
)+\frac{g^2}{2}\left(\left\vert \phi^+\right \vert^2 - \left\vert
\phi^-\right \vert^2\right)^2\, ,
\end{eqnarray}
\end{widetext}
where $y\equiv \beta \kappa^{1/2} \rho$ and where $W$ in the previous
formula denotes the total superpotential defined by
Eq.~(\ref{Wtotal}). The function $\tilde{\cal V}(\rho )$ is defined by
\begin{equation}
\label{defV}
\tilde{\cal V}(\rho )\equiv \frac{\kappa A\beta ^2}{2y}{\rm e}^{-y}
\left[\frac{A\beta }{3}{\rm e}^{-y}-\frac{\beta }{y} \left(W_0-A{\rm
e}^{-y}\right)\right]\, .
\end{equation}
The true vacuum of the above potential is obtained for $ \phi^+=
\phi^-= \zeta $, $\phi=-\kappa W(\rho _{{\rm min},\tilde{\cal
V}})/(\sqrt{2}g)$, $\rho= \rho_{{\rm min},\tilde{\cal V}}$ being the
minimum of the function $\tilde{\cal V}$. This leads to an AdS vacuum
with unbroken supersymmetry~\cite{KKLT}.

\par

For large $\phi$ compared to $\zeta $, the potential has a flat
inflationary valley where $\phi^\pm=0$ and the slope of the potential
can be lifted by radiative corrections. When $\phi$ becomes small
compared to $\zeta $, the field $\phi^\pm$ run towards the true
minimum $\phi^+=\phi^-=\zeta $ and $\phi=-\kappa W/(\sqrt{2}g)$. This
realizes a hybrid inflation scenario in string theory.

\par

Now, let us consider the situation where $\phi$ is small enough and
the charged fields are stuck at the minimum. In this regime, the field
$\phi $ can be expanded according to
\begin{equation}
\phi \simeq -\frac{\kappa W\left(\rho \right)}{\sqrt{2}g} +\delta \chi
\, .
\end{equation}
Then the field $\delta \chi $ becomes massive with a potential
\begin{equation}
\label{potmodelpm}
V=\tilde{\cal V}(\rho) +\frac{4g^2 \zeta ^2
}{\kappa^{3/2}(\rho+\rho^{\dagger })^3} \left \vert \delta \chi \right
\vert ^2\, .
\end{equation}
We now notice that this potential can be also obtained using the much
simpler superpotential
\begin{equation}
W(\rho,\phi)= W(\rho) + g\zeta (\delta \chi) ^2\, ,
\end{equation}
in the global supersymmetry limit where the Planck mass is taken
to be very large compared to the inflation scale. This provides a
motivation to use the above superpotential in a regime where
supergravity corrections are taken into account and where one can
be far from the true minimum.  We show in the next subsection that
the potential becomes very  interesting and leads to ``mutated
chaotic inflation''.

\subsection{The model}

Following the considerations presented before, we assume that the
inflaton is a massive field like in chaotic inflation and discuss
its coupling to the moduli stabilization sector. Since our model is
a string inspired, we take ${\cal K}=0$ and ${\cal G}=-1/2\left(\phi
-\phi ^\dagger \right)^2$ for the K\"ahler potential. For the
superpotential, we use the calculation of the previous subsection
section and assume that
\begin{eqnarray}
W(\rho ,\phi) &=& W(\rho )+W_{\rm inf}(\phi )\nonumber \\
&=& W_0-A\exp \left(-\beta \kappa ^{1/2}\rho \right)+\frac{\alpha
}{2}m\phi^2\, ,
\end{eqnarray}
where the mass $m$ can be related to $\zeta $, namely $m=2g\zeta
/\alpha $. Then, we use the general formula established before, see
Eqs.~(\ref{potgeneral}) and (\ref{potgeneral2}), and we obtain
\begin{equation}
\label{ourpot}
V\left(\rho, \phi \right)=\tilde{{\cal V}}(\rho ) +\tilde{{\cal
U}}(\rho )\phi ^2\, ,
\end{equation}
where the function $\tilde{\cal V}$ has already been defined above in
Eq.~(\ref{defV}) and $\tilde{\cal U}$ given by
\begin{equation}
\tilde{{\cal U}}(\rho )=\frac{\alpha m\beta ^3}{4y^2}
\left(\frac{\alpha m}{2y}-\kappa A{\rm e}^{-y}\right)\, ,
\end{equation}
and where we remind that $y\equiv \beta \kappa ^{1/2}\rho $. Let us
notice that, if $\kappa \rightarrow 0$, then $\tilde{\cal U}\sim
1/\rho ^3$ in accordance with Eq.~(\ref{potmodelpm}). The function
$\tilde{\cal U}(\rho )$, from the point of view of the field $\phi $,
plays the role of an effective squared mass. The function $\tilde{\cal
V}(\rho )$ is not multiplied by a function of the field $\phi $ and,
therefore, can be viewed as an ``offset''. In order for the above
potential to be relevant, it is necessary for the effective squared
mass to be positive at the extremum of $\tilde{\cal U}$ where the
moduli is stabilized.

\begin{figure*}
\includegraphics[width=.95\textwidth,height=.65\textwidth]{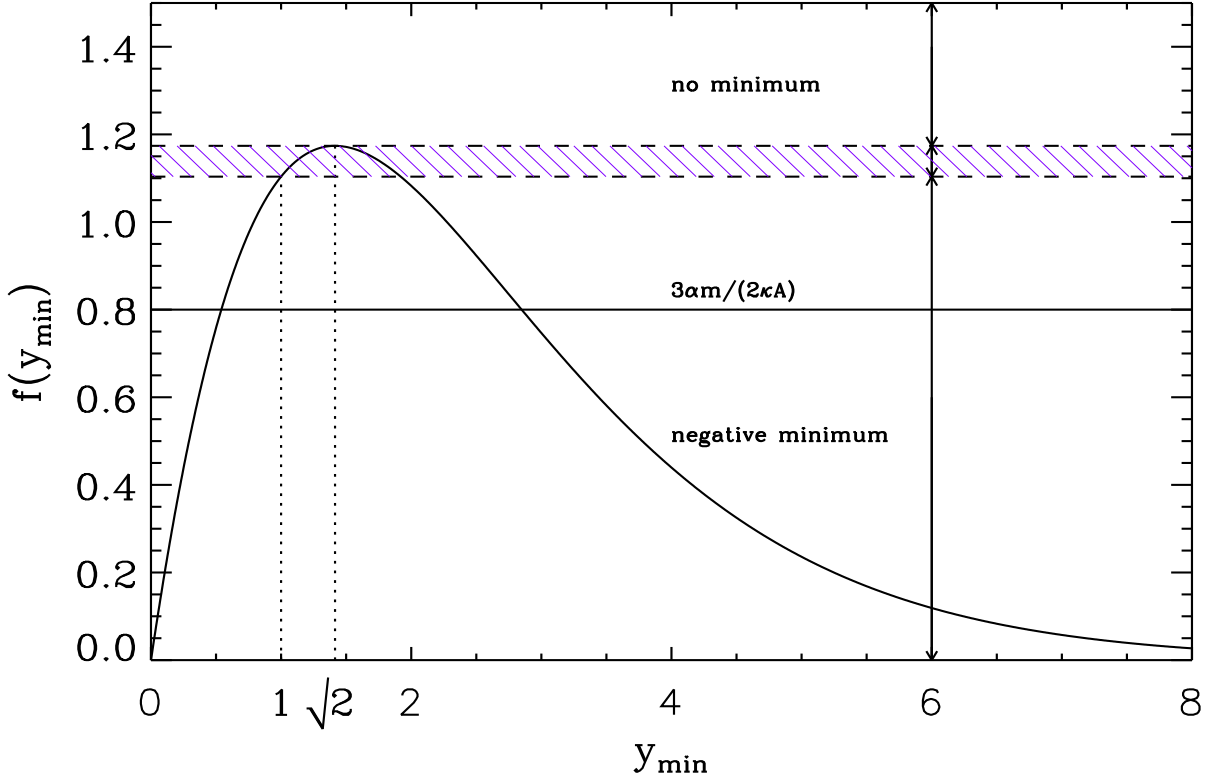}
\caption{The function $f\left(y_{{\rm min},\tilde{\cal U}}\right)$
defined in Eq.~(\ref{deff}). The maximum of this function is located
at $y_{{\rm min},\tilde{\cal U}}=\sqrt{2}$ and is equal to
$1.17387$. The intersection of $f\left(y_{{\rm min},\tilde{\cal
U}}\right)$ with the horizontal line (solid line) $3\alpha m/(2\kappa
A)$ determines the value(s) of $\rho $ corresponding to the extrema of
the function $\tilde{\cal U}$. According to the values of the
parameter $m$ and $A$, the line moves up or downwards while the
function $f\left(y_{{\rm min},\tilde{\cal U}}\right)$ remains the
same. As explained in the text, if the line is above the maximum of
$f(y_{{\rm min},\tilde{\cal U}})$ then there are no extrema. If the
line is below the maximum, there are always two extrema. The first
intersection between $f\left(y_{{\rm min},\tilde{\cal U}}\right)$ and
the horizontal line [in the increasing part of the function
$f\left(y_{{\rm min},\tilde{\cal U}}\right)$] corresponds to a minimum
while the second intersection corresponds to a maximum. However, if
the line is below the value $3/{\rm e}\simeq 1.10364$, the
corresponding value of $\tilde{\cal U}$ is negative. The effective
squared mass $\tilde{\cal U}$ is positive at the minimum if the
horizontal line is in the dashed area which, therefore, represents the
allowed region.} \label{studyu}
\end{figure*}

Let us also give the potential in terms of the normalized fields
$\bar{\rho }$ and $\bar{\phi }$. These fields are given by the
formulas in the text before Eq.~(\ref{potmod}) since the K\"ahler
potential in the present section is the same as in the subsection of
Eq.~(\ref{potmod}). Below, for convenience, we reproduce the relation
between canonical and non-canonical fields
\begin{equation}
\phi =\frac{\bar{\phi }}{\sqrt{2}}\, ,\quad y=\beta
\exp\left(\sqrt{\frac23}\kappa ^{1/2}\bar{\rho }\right)\, .
\end{equation}
Therefore, the potential reads
\begin{widetext}
\begin{eqnarray}
V\left(\bar{\rho} ,\bar{\phi }\right) &=& \frac{\kappa A\beta }{2}{\rm
e}^{-\sqrt{2/3}\kappa ^{1/2}\bar{\rho }} \exp \left(-\beta {\rm
e}^{\sqrt{2/3}\kappa ^{1/2}\bar{\rho }}\right) \biggl\{ \frac{A\beta
}{3}\exp \left(-\beta {\rm e}^{\sqrt{2/3}\kappa ^{1/2}\bar{\rho
}}\right) \nonumber \\ & & -{\rm e}^{-\sqrt{2/3}\kappa ^{1/2}\bar{\rho
}}\left[W_0 -A\exp \left(-\beta {\rm e}^{\sqrt{2/3}\kappa
^{1/2}\bar{\rho }}\right) \right]\biggr\} \nonumber \\ & &
+\frac{\alpha m\beta }{8} {\rm e}^{-2\sqrt{2/3}\kappa ^{1/2}\bar{\rho
}} \left[\frac{\alpha m}{2\beta } {\rm e}^{-\sqrt{2/3}\kappa
^{1/2}\bar{\rho }} -\kappa A \exp \left(-\beta {\rm
e}^{\sqrt{2/3}\kappa ^{1/2}\bar{\rho }}\right) \right]\bar{\phi }^2\, .
\end{eqnarray}
\end{widetext}
Notice that, since the link between $(\rho, \phi )$, on one hand, and
$(\bar{\rho }, \bar{\phi })$, on the other hand, is monotonic, the
above change of variables does not modify the properties of the
minima. In particular, in order to study how these properties depend
on the free parameters, it is sufficient to work in terms of the
non-canonically normalized fields which is simpler.

\subsection{The squared mass function $\tilde{\cal U}(\rho )$}

We now analyze whether the moduli can be stabilized to a value
corresponding to a positive potential. For this purpose, we study the
effective squared mass function $\tilde{\cal U}$ and requires that
this function has a positive minimum. We will see that the valley
where the moduli is stabilized is not exactly given by the minimum of
the function $\tilde{\cal U}$ because, for small values of $\phi $,
the offset function $\tilde{\cal V}$ also plays a role.  In fact,
strictly speaking, the minimum of $\tilde{\cal U}$ is the valley of
stability for $\phi/\mP \rightarrow +\infty$ only. However, we
emphasize that it is mandatory that the true minimum of $\tilde{\cal
U}$ be positive since this term is multiplied by $\phi ^2$. Otherwise
the inflaton field becomes tachyonic. It is easy to calculate the
derivative of the effective mass $\tilde{\cal U}$. This gives
\begin{equation}
\frac{{\rm d}\tilde{\cal U}}{{\rm d}y}=
\frac{\alpha m\beta ^3}{4y^4}\left(-\frac32\alpha m+2\kappa Ay{\rm e}^{-y}
+\kappa Ay^2{\rm e}^{-y}\right)\, .
\end{equation}
It vanishes at $y=y_{{\rm min},\tilde{\cal U}}$ where $y_{{\rm
min},\tilde{\cal U}}$ satisfies the following equation
\begin{equation}
\label{deff}
f\left(y_{{\rm min},\tilde{\cal U}}\right)=y_{{\rm min},\tilde{\cal
U}}\left(y_{{\rm min},\tilde{\cal U}}+2\right){\rm e}^{-y_{{\rm
min},\tilde{\cal U}}}=\frac{3\alpha m}{2\kappa A}\, .
\end{equation}
{}From this expression, one gets a constraint on the parameters $m$
and $A$, coming from the fact that $3\alpha m/(2\kappa A)$ must be
smaller than the maximum value of the function $y_{{\rm
min},\tilde{\cal U}}(y_{{\rm min},\tilde{\cal U}}+2){\rm e}^{-y_{{\rm
min},\tilde{\cal U}}}$, otherwise the above equation has no solution,
see Fig.~\ref{studyu}. This function vanishes at the origin, increases
and reaches a maximum at $y_{{\rm min},\tilde{\cal U}}=\sqrt{2}$ and
then exponentially decreases towards zero. Therefore, the function
$\tilde{\cal U}$ possesses an extremum if and only if $0<3\alpha
m/(2\kappa A)<\sqrt{2}\left(2+\sqrt{2}\right){\rm e}^{-\sqrt{2}}\simeq
1.17387 $. The value of $\tilde{\cal U}$ at the extremum can be easily
derived and one obtains
\begin{equation}
\label{minu}
\tilde{\cal U}\left(y=y_{{\rm min},\tilde{\cal U}}\right)= \frac{\alpha
^2m^2\beta ^3}{8y_{{\rm min},\tilde{\cal U}}^3}\frac{y_{{\rm
min},\tilde{\cal U}}-1}{y_{{\rm min},\tilde{\cal U}}+2}\, .
\end{equation}
Therefore, the extremum corresponds to a positive potential if
$y_{{\rm min},\tilde{\cal U}}>1$ but, at this level, this does not
require new constraints on $m$ and $A$.
\begin{figure*}
\includegraphics[width=8.6cm,height=7.5cm]{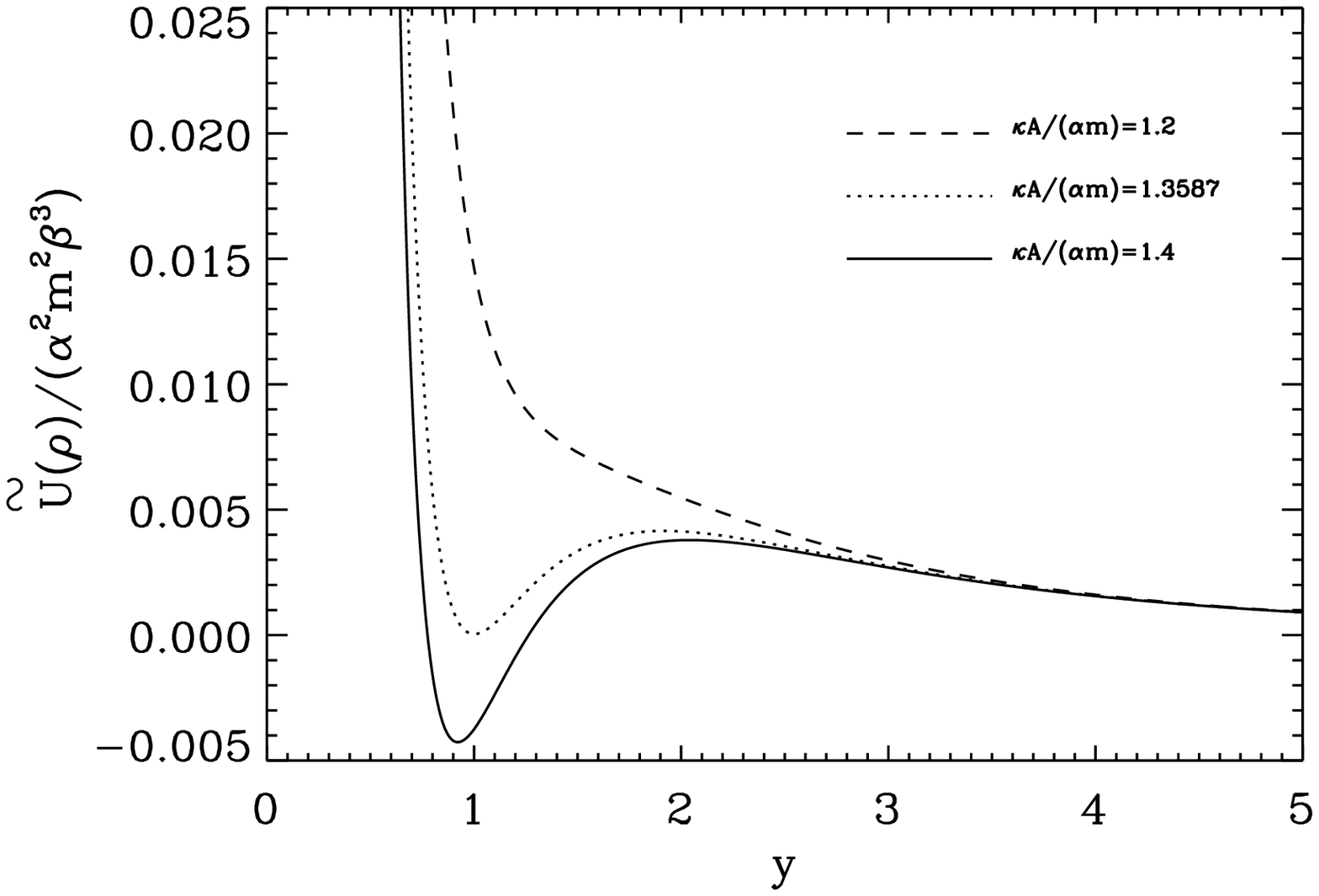}
\includegraphics[width=8.8cm,height=7.5cm]{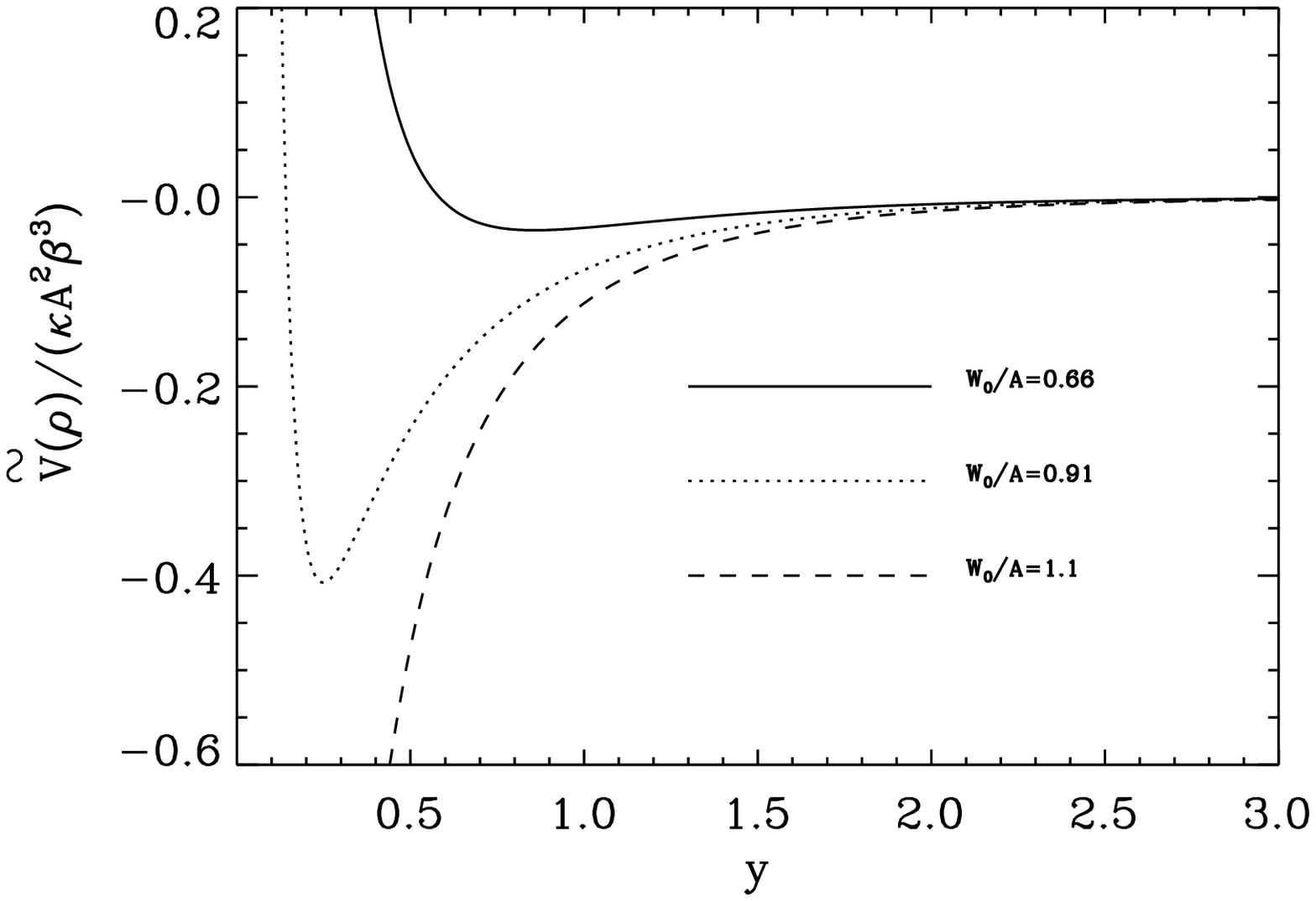}
\caption{Left panel: function $\tilde{\cal U}(\rho )$ for three
different values of the ratio $\kappa A/(\alpha m)$. The dashed line
corresponds to a situation where there is no minimum, \ie the line
$3\alpha m/(2\kappa A)$ is above the function $f\left(y_{{\rm
min},\tilde{\cal U}}\right)$, see Fig.~\ref{studyu}. The dotted line
represents the case where there is a positive minimum, \ie the line
$3\alpha m/(2\kappa A)$ is within the dashed area in
Fig.~\ref{studyu}. Finally, the solid line corresponds to a function
$\tilde{\cal U}$ with a negative minimum. Right panel: function
$\tilde{\cal V}(\rho )$ for three different values of the ratio
$W_0/A$. One can check on the figure that, if $W_0/A<1$ then there is
a negative minimum and $\lim _{y\rightarrow 0}\tilde{\cal V}=+\infty $
while, if $W_0/A> 1$ there is no minimum but $\lim _{y\rightarrow
0}\tilde{\cal V}=-\infty $.}
\label{plotuv}
\end{figure*}
Now, let us check whether this is a maximum or a minimum. For this
purpose, one calculate the second derivative of $\tilde{\cal U}$ at
the extremum. One obtains
\begin{equation}
\frac{{\rm d}^2\tilde{\cal U}}{{\rm d}y^2}\biggl\vert _{y=y_{{\rm
min},\tilde{\cal U}}}= -\frac{3\alpha ^2m^2\beta ^3}{8}\frac{y_{{\rm
min},\tilde{\cal U}}^2-2}{y_{{\rm min},\tilde{\cal U}}^5\left(y_{{\rm
min},\tilde{\cal U}}+2\right)}\, .
\end{equation}
This is positive if $y_{{\rm min},\tilde{\cal
U}}<\sqrt{2}$. Therefore, we have a positive minimum if the parameters
$m$ and $A$ are such that $1<y_{{\rm min},\tilde{\cal U}}<\sqrt{2}$
which in turn implies that one must have $3/{\rm e}\simeq
1.10364<3\alpha m/(2\kappa A)<\sqrt{2}\left(2+\sqrt{2}\right){\rm
e}^{-\sqrt{2}} \simeq 1.17387$ or
\begin{equation}
\label{consparam}
 1.27782 <\frac{\kappa A}{\alpha m}< 1.35914\, .
\end{equation}
This interval is represented in Fig.~\ref{studyu} by the dashed
region. It is quite clear from the above considerations that the ratio
$m/(\kappa A)$ has to be adjusted precisely. However, this does not
mean $m$ and/or $\kappa A$ must be tuned very accurately. As a matter
of fact, they can a priori change over a large range of values,
provided of course that their ratio satisfies the constraint derived
above. The function $\tilde{\cal U}$ for various values of $m/(\kappa
A)$ is represented in Fig.~\ref{plotuv}.

\subsection{The offset function $\tilde{\cal V}(\rho )$}

Let us now study the effective offset function $\tilde{\cal V}(\rho )$
in more details. Firstly, let us evaluate
\begin{equation}
\frac{{\rm d}\tilde{\cal V}}{{\rm d}y}=-\frac{\kappa A^2\beta
^3}{3y^3} {\rm e}^{-2y}\left[y^2+\frac72 y +3 -\frac{3W_0}{2A}{\rm
e}^y(y+2)\right]\, ,
\end{equation}
and, therefore, there is a minimum if the following equation is
satisfied
\begin{equation}
\label{condder}
y_{{\rm min},\tilde{\cal V}}^2+\frac72y_{{\rm min},\tilde{\cal
V}}+3=\frac{3W_0}{2A}{\rm e}^{y_{{\rm min},\tilde{\cal V}}}
\left(y_{{\rm min},\tilde{\cal V}}+2\right)\, .
\end{equation}
The existence of a solution to the above equation is controlled by the
ratio $W_0/A$. If $W_0/A>1$, then there is no solution because the
parabola in Eq.~(\ref{condder}) cannot intersect the function
$3W_0/(2A){\rm e}^{y_{{\rm min},\tilde{\cal V}}} \left(y_{{\rm
min},\tilde{\cal V}}+2\right)$ (notice that for $W_0/A=1$ the two
functions are equal at $y_{{\rm min},\tilde{\cal V}}=0$). Moreover, in
this situation, we have $\lim _{y\rightarrow 0}\tilde{\cal V}=-\infty
$ and, therefore, the function $\tilde{\cal V}$ cannot be positive
everywhere. On the other hand, if $W_0/A<1$, then $\lim _{y\rightarrow
0}\tilde{\cal V}=+\infty $ and, in addition, Eq.~(\ref{condder})
admits a solution, hence $\tilde{\cal V}$ possesses an extremum. Since
we also have
\begin{eqnarray}
\label{massV}
& & \frac{{\rm d}^2\tilde{\cal V}}{{\rm d}y^2}\biggl \vert _{y=y_{{\rm
min},\tilde{\cal V}}}=\frac{3\kappa W_0^2\beta^3\left(y_{{\rm
min},\tilde{\cal V}}+2\right)} {8y_{{\rm min},\tilde{\cal V}}^3\left(
y_{{\rm min},\tilde{\cal V}}^2+7y_{{\rm min},\tilde{\cal
V}}/2+3\right)^2} \nonumber \\ &\times & \left(2y_{{\rm
min},\tilde{\cal V}}^3+9 y_{{\rm min},\tilde{\cal V}}^2 +12y_{{\rm
min},\tilde{\cal V}}+4\right)>0 \, ,
\end{eqnarray}
the function $\tilde{\cal V}$ possesses an extremum which is a
minimum. However, the value of this function at this minimum is given
by
\begin{eqnarray}
\label{minV}
& & \tilde{\cal V}\left( y=y_{{\rm min},\tilde{\cal V}}
\right)=-\frac{\kappa AW_0\beta ^3}{4y_{{\rm min},\tilde{\cal
V}}}
\nonumber \\ &\times &
\frac{y_{{\rm min},\tilde{\cal V}}+2}{y_{{\rm min},\tilde{\cal
V}}^2+7y_{{\rm min},\tilde{\cal V}}/2+3} {\rm e}^{-y_{{\rm
min},\tilde{\cal V}}} <0 \, .
\end{eqnarray}
As a consequence, the function $\tilde{\cal V}$ cannot be positive
everywhere since it is negative at its extremum. However, this is not
automatically a problem, as it would have been for the function
$\tilde{\cal U}$, since the offset function can always be
``renormalized'' by adding a positive cosmological constant, namely
$-\tilde{\cal V}\left( y=y_{{\rm min},\tilde{\cal V}} \right)$. We
conclude from the above analysis that the function $\tilde{\cal V}$,
regardless of the values of the free parameters $A$ and $W_0$, is
necessarily negative somewhere. The regime of interest is given by
$W_0/A<1$ since in this case $\tilde{\cal V}$ possesses a minimum
which can be ``renormalized'' by adding a constant. The function
$\tilde{\cal V}(\rho )$ is represented in Fig.~\ref{plotuv} for
various values of the ratio $W_0/A$.

\par

Another remark is in order in this subsection. In the following, we
will study another mechanism of stabilization (\ie the KKLT mechanism)
which consists in adding a term $\propto 1/\rho ^3$ to the offset
function. In this case one will show that one can always find a
positive minimum regardless of $W_0/A$. Therefore, with this other
mechanism, the above discussion of the shape of the offset function is
modified and the condition $W_0/A<1$ can be relaxed.

\par

Let us now discuss another class of models.  In our models,
inflation is only driven by the $F$ --terms originating from
non--shift symmetric superpotentials. This is enough to lift the
inflaton flat direction and leads to an inflation potential with no
${\cal O}(H)$ corrections to the inflaton mass.  Of course we need
to introduce a superpotential of chaotic inflation with a
fine--tuning of the inflaton mass scale (see below where we apply
the COBE normalization). This is the usual flatness problem in
inflation model building. It has been argued in Ref.~\cite{mac} that
such a fine--tuning is also present in string compactification where
threshold corrections are taken into account leading to a
superpotential of the form
\begin{equation}
W(\rho,\phi)=W(\rho)\left(1+ \delta \kappa \phi^2\right)\, ,
\end{equation}
where $\delta$ has to be small to guarantee a small enough mass of the
inflaton. This requires a tuning of the complex structure moduli. For
this model the scalar potential becomes
\begin{equation}
V\left(\rho ,\phi \right)=\tilde{\cal V} (\rho)\left(1+\delta \kappa
\phi ^2\right)^2 + \frac{4\kappa ^2\delta ^2}{3\Delta^2}\left\vert
W(\rho)\right\vert^2 \phi^2\, ,
\end{equation}
where the K\"ahler potential corresponds to ${\cal G}=0$ and ${\cal
K}= -(\phi-\phi^{\dagger})^2/2$ and where $\tilde {\cal V}$ is the
same function as studied above. For large $\phi$ the potential reduces
to
\begin{equation}
V\left(\rho ,\phi \right)\simeq \tilde{\cal V}(\rho)\delta^2 \kappa^4
\phi ^4 \, .
\end{equation}
Now, when $\rho$ is at the minimum of ${\tilde {\cal V}}$, the
coupling constant of the $\phi^4$ term becomes negative implying
that this potential is not suitable to obtain inflation.  This
exemplifies how difficult it is to find a satisfactory model of
inflation in supergravity.

\subsection{Renormalizing the offset function by a constant}

\begin{figure*}
\includegraphics[width=.95\textwidth,height=.75\textwidth]{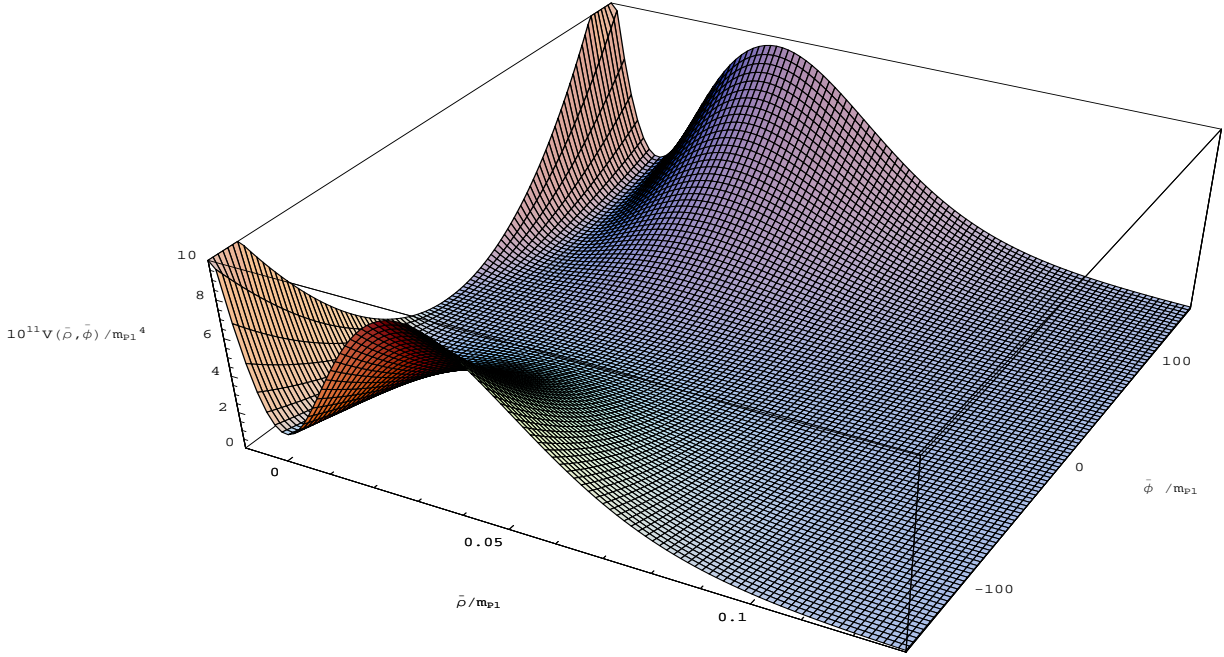}
\caption{Potential $V(\bar{\rho}, \bar{\phi })$ [what is actually
  plotted is $10^{11}V(\bar{\rho}, \bar{\phi })/\mP^4$] for the
  following choice of parameters: $\alpha =\sqrt{2}$, $\beta =1$,
  $m=10^{-6}\mP$, $\kappa A/(\alpha m)=1.35135$ and
  $W_0/A=0.41111$. This gives an absolute minimum of the potential
  located at $\bar{\phi }=0$, $\bar{\rho }_{\rm min, \tilde{\cal
  V}}\simeq 0.1181\times \mP$. These parameters satisfy the
  constraints derived in the text. The moduli is stabilized at the
  value $y=y_{\rm min, \tilde{\cal U}}\simeq 1.016$ or $\bar{\rho
  }_{\rm min, \tilde{\cal U}}\simeq 0.0040\times \mP$ (for $\bar{\phi
  }\gg \mP$) as can be checked in the figure where the inflationary
  valley is clearly seen.}
\label{potrhophi2}
\end{figure*}

Having studied the functions $\tilde{\cal V}$ and $\tilde{\cal U}$
in some details, let us now turn to the properties of the full
potential when shifted by a positive cosmological constant. It is
represented in Fig.~\ref{potrhophi2}.
The derivatives of the potential in the two directions $\bar{\rho }$
and $\bar{\phi }$ are given by
\begin{eqnarray}
\label{potder}
\frac{\partial V}{\partial \bar{\phi} } &=& \frac{{\rm d}\phi }{{\rm
    d}\bar{\phi}}\frac{\partial V}{\partial \phi }=
\frac{1}{\sqrt{2}}\times 2\tilde{\cal U}\phi\, , \\
\label{potder2}
\frac{\partial V}{\partial \bar{\rho }} &=& \frac{{\rm d}y}{{\rm
    d}\bar{\rho }}\frac{\partial V}{\partial y}=
\sqrt{\frac23}\kappa ^{1/2}y\left(\frac{{\rm d}\tilde{\cal V}}{{\rm
d}y} +\frac{{\rm d}\tilde{\cal U}}{{\rm d}y}\phi ^2\right)\,  .
\end{eqnarray}
{}From these expressions, we deduce that the potential possesses an
absolute minimum located at
\begin{equation}
\bar{\phi }=0\, , \quad
\bar{\rho }={\bar \rho}_{\rm min,\tilde{\cal V}}\, .
\end{equation}
At this minimum, the potential vanishes exactly.

\par

\begin{figure*}
\includegraphics[width=.95\textwidth,height=.65\textwidth]{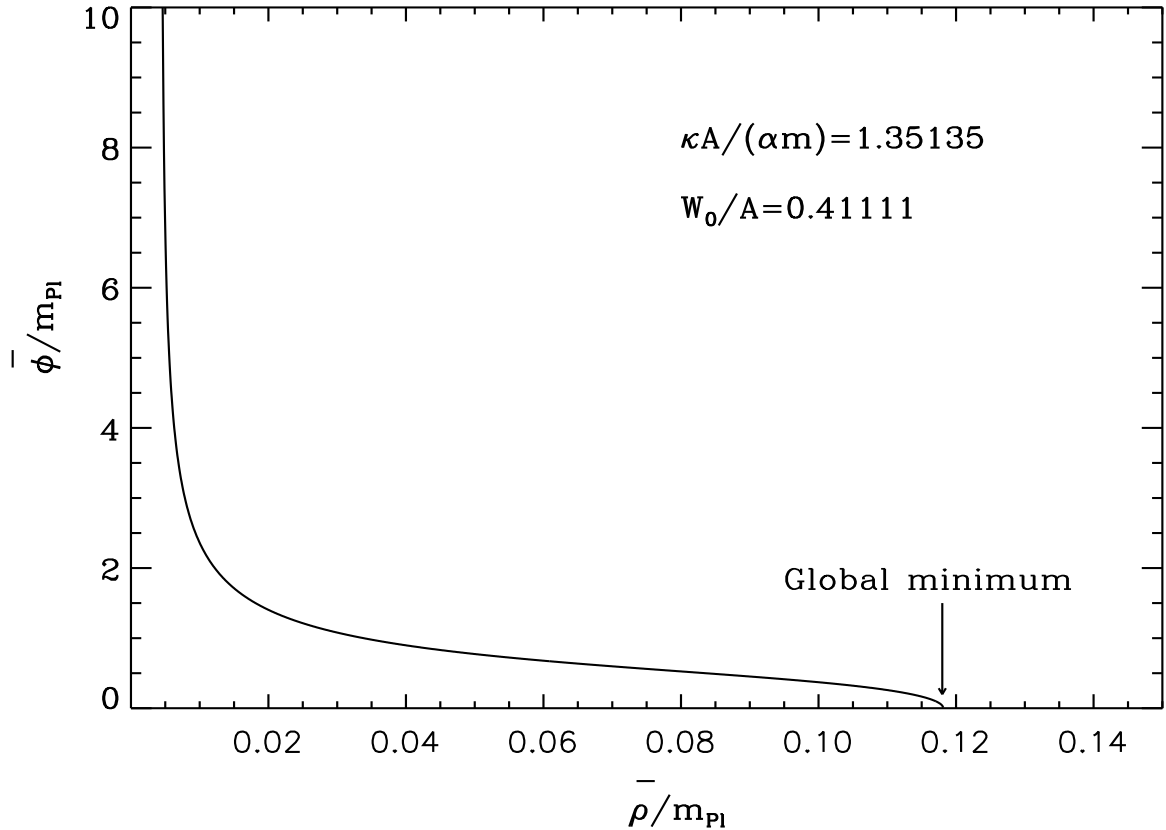}
\caption{Trajectory of the valley of stability in the plan
$(\bar{\rho} ,\bar{\phi} )$ (the valley is seen from above). For large
values of $\bar{\phi}$, $\bar{\phi }\gg \mP$, the valley coincides
with the minimum of the function $\tilde{\cal U}$, \ie is just given
by the equation $\bar{\rho} =\bar{\rho }_{\rm min, \tilde{\cal
U}}$. For smaller values of the inflaton, the trajectory bends and
quickly joins the absolute minimum of the potential, namely $y=y_{\rm
min, \tilde{\cal V}}$, $\phi=0$. The parameters used here are: $\alpha
=\sqrt{2}$, $\beta =1$, $m=10^{-6}\mP$, $\kappa A/(\alpha m)=1.35135$
and $W_0/A=0.41111$. This gives $y=y_{\rm min, \tilde{\cal U}}\simeq
1.016$ or $\bar{\rho }_{\rm min, \tilde{\cal U}}\simeq 0.0040\times
\mP$ and the global minimum of the potential is located at $\bar{\phi
}=0$, $\bar{\rho }_{\rm min, \tilde{\cal V}}\simeq 0.1181\times \mP$.}
\label{courbmin}
\end{figure*}

{}From the expressions of the derivatives of the potential,
Eqs.~(\ref{potder}) and (\ref{potder2}), one deduces that there also
exists a valley of stability. This valley is also clearly seen in
Fig.~\ref{potrhophi2} and is of course of utmost importance for us. It
is given by the following trajectory in the $(\rho, \phi )$ plane
\begin{widetext}
\begin{equation}
\label{trajecnoncanon}
\kappa \phi ^2_{\rm valley}\left(y\right)=\frac43\left(\frac{\kappa
A}{\alpha m}\right)^2 y{\rm
e}^{-2y}\left[y^2+\frac72y+3-\frac32\frac{W_0}{A}{\rm
e}^y\left(y+2\right)\right]\times \left[-\frac32+2\left(\frac{\kappa
A}{\alpha m}\right)y{\rm e}^{-y}+\left(\frac{\kappa A}{\alpha
m}\right)y^2{\rm e}^{-y} \right]^{-1}\, .
\end{equation}
In terms of canonically normalized fields, the same trajectory reads
\begin{eqnarray}
\label{inftrajec}
\kappa \bar{\phi }^2_{\rm valley}\left(\bar{\rho }\right)&=&\frac83
\left(\frac{\kappa A}{\alpha m}\right)^2 \beta
{\rm e}^{\sqrt{\frac23} \kappa ^{1/2}\bar{\rho }}
\exp\left(-2\beta {\rm e}^{\sqrt{\frac23} \kappa ^{1/2}\bar{\rho }}\right)
\biggl[\beta ^2{\rm e}^{2\sqrt{\frac23} \kappa ^{1/2}\bar{\rho }}
+\frac72\beta {\rm e}^{\sqrt{\frac23} \kappa ^{1/2}\bar{\rho }}+3
\nonumber \\
& &
-\frac32\frac{W_0}{A}
\exp\left(\beta {\rm e}^{\sqrt{\frac23} \kappa ^{1/2}\bar{\rho }}\right)
\left(\beta {\rm e}^{\sqrt{\frac23} \kappa ^{1/2}\bar{\rho }}
+2\right)\biggr]\times
\biggl[-\frac32+2\left(\frac{\kappa
A}{\alpha m}\right)\beta {\rm e}^{\sqrt{\frac23} \kappa ^{1/2}\bar{\rho }}
\exp\left(-\beta {\rm e}^{\sqrt{\frac23} \kappa ^{1/2}\bar{\rho
}}\right)
\nonumber \\
& & +\left(\frac{\kappa A}{\alpha
m}\right)\beta ^2{\rm e}^{2\sqrt{\frac23} \kappa ^{1/2}\bar{\rho }}
\exp\left(-\beta {\rm e}^{\sqrt{\frac23} \kappa ^{1/2}\bar{\rho }}\right)
\biggr]^{-1}\, .
\end{eqnarray}
\end{widetext}
This trajectory is represented in Fig.~\ref{courbmin}.
{}For large values of the inflaton, $\bar{\phi }/\mP \gg 1$, the
offset function is negligible and the potential is almost given by
$V\simeq \tilde{\cal U}\phi ^2$. As a consequence, the valley of
stability reduces to the simple equation $\bar{\rho }=\bar{\rho }_{\rm
min,\tilde{\cal U}}$ as can be directly checked in
Fig.~\ref{courbmin}: in this regime (but in this regime only), as it
is the case for hybrid inflation, the waterfall field is frozen. For
small values of $\bar{\phi }/\mP$, the offset function becomes
important, the trajectory bends and, as expected, joins the global
minimum of the potential .

\par

Let us now discuss how inflation proceeds in this model. Clearly, the
potential plotted in Fig.~\ref{potrhophi2} is reminiscent of hybrid
inflation where $\bar{\phi }$ is the inflaton and where the moduli
$\bar{\rho }$ plays the role of the waterfall field~\cite{hybrid}. In
hybrid inflation, inflation proceeds along the valley either in the
vacuum dominated regime, in which case the potential is almost a
constant, or in the inflaton dominated regime, in which case $V\sim
\phi ^2$. There is also a version, well-motivated from the particle
physics point of view, where the flat valley is lifted by quantum
corrections~\cite{LR}. In this case the potential along the valley is
computed by means of the well-known Coleman-Weinberg formula.  In our
case, we have already noticed that, for $\bar{\phi }\gg \mP$, the
potential in the valley takes the form $V\sim \tilde{\cal U}\bar{\phi
}^2$. Therefore, our case belongs to the inflaton dominated regime
case.

\par

The next question is to study how inflation ends. Typically, in hybrid
inflation, inflation stops by instability. At some point along the
inflationary valley, the effective squared mass in the direction
perpendicular to the valley becomes negative and, as a consequence,
the field can no longer be kept on tracks and quickly rolls down
towards its minimum where reheating proceeds. We will show that this
is not the case here. In our case, inflation stops while still in the
valley due to a violation of the slow-roll conditions. The crucial
ingredient here is that the valley is not a straight line but
corresponds to a non-trivial path in the configuration space of the
two fields. This aspect is reminiscent of mutated
inflation~\cite{mutated} (other interesting inflationary models where
the trajectory is not trivial can be found in
Refs.~\cite{shifted}). In fact, the analogy can even been pushed
further. Indeed, in the case of mutated inflation, the potential has
typically the following form~\cite{LR}
\begin{equation}
\label{vmutated}
V\left(\psi, \phi \right)=V_0\left(1-\frac{\psi }{M}\right)
+\frac{\lambda }{4}\phi ^2\psi ^2 +\cdots \, ,
\end{equation}
where $V_0$, $M$ and $\lambda $ are constants. In the above
expression, $\phi $ is the inflaton and the second field $\psi $ plays
the role of our moduli $\rho $. One of the main feature of mutated
inflation is that an effective potential for the inflaton can be
produced even if the original potential contains no term that depends
only on $\phi $ as can be seen explicitly in the previous formula. In
the configuration space, the trajectory reads $\psi \phi
^2=2V_0/(\lambda M)$ and after having inserted this expression into
Eq.~(\ref{vmutated}) one gets
\begin{equation}
V\left(\phi \right)=V_0\left(1-\frac{V_0}{\lambda ^2M^2\phi
  ^2}\right)\, ,
\end{equation}
which is suitable for inflation. In the same manner, our potential
given in Eq.~(\ref{ourpot}) does not contain any piece depending on
the inflaton field only. However, inserting the trajectory $\rho (\phi
)$ in $V\left(\rho ,\phi \right)$ would lead to a potential $V(\phi )$
as above. The difficulty in our case is that
Eq.~(\ref{trajecnoncanon}) gives rather $\phi (\rho )$ and that the
expression is too complicated to be invertible. But, clearly, at the
level of principles this is very similar. Therefore, the model
presented here combines aspects from chaotic and mutated inflation
hence the name ``mutated chaotic inflation'' given at the beginning of
this section.

\subsection{Phenomenological constraints}

We now discuss the constraint on the parameter characterizing the
inflaton sector, \ie the mass $m$ (since the parameters $A$ and $W_0$
have already been discussed before). In order to simplify the
discussion, we will assume that the initial conditions are such that
the fields are, at the beginning of the evolution, in the valley of
stability and more particularly in the straight line part of the
valley (we will come back to the questions of the initial conditions
later on and will discuss this assumption in some detail) where $V
\sim \tilde{\cal U}\left(\bar{\rho }_{\rm min, \tilde{\cal
U}}\right)\bar{\phi }^2/2$. Since the quantum fluctuations of the
inflaton field $\bar{\phi }$ are at the origin of the CMB anisotropy
observed today, the COBE and Wilkinson Microwave Anisotropy Probe
(WMAP) normalizations fix the coupling constant of the inflaton
potential, namely the mass function $\tilde{\cal U}\left(\bar{\rho
}_{\rm min, \tilde{\cal U}}\right)$ in the present context.
Concretely, for small $\ell $, the multipole moments are given by
\begin{equation}
C_{\ell }=\frac{2H^2}{25\epsilon \mP^2}\frac{1}{\ell (\ell +1)}\, ,
\end{equation}
where $\epsilon $ is the first slow-roll parameter to be discussed
below. What has been actually measured by the COBE and WMAP satellites
is $Q^2_{\rm rms-PS}/T^2=5C_2/(4\pi )\simeq \left(18\times
10^{-6}/2.7\right)^2\simeq 36\times 10^{-12}$. The quantity $H$ is the
Hubble parameter during inflation and is related to the potential by
the slow-roll equation $H^2\simeq \kappa V/3$ evaluated at Hubble
radius crossing. Putting everything together, we find that the
inflaton mass is given by
\begin{equation}
\left[\frac{\tilde{\cal U}\left(\bar{\rho }_{\rm min, \tilde{\cal
U}}\right)}{\mP}\right]^2\simeq 45 \pi \left(N_*+\frac12 \right)^{-2}
\frac{Q^2_{\rm rms-PS}}{T^2}\, ,
\end{equation}
where $N_*\simeq 60$ (\ie the number of e-folds between the time at
which the modes of astrophysical interest today left the Hubble radius
during inflation and the end of inflation, see Ref.~\cite{number}),
that is to say
\begin{equation}
\sqrt{\tilde{\cal U}\left(\bar{\rho }_{\rm min, \tilde{\cal
      U}}\right)} \simeq 1.3\times 10^{-6}\times \mP\, .
\end{equation}
At this point, it is important to recall that the mass function
has been defined by the expression
\begin{equation}
\tilde{\cal U}\left(\bar{\rho }_{\rm min, \tilde{\cal U}}\right)
=\frac{(\alpha m)^2\beta ^3 }{4y_{\rm min, \tilde{\cal U}}}
\left(\frac{1}{2y_{\rm min, \tilde{\cal U}}}-\frac{\kappa A}{\alpha m}
{\rm e}^{-y_{\rm min, \tilde{\cal U}}}\right)\, .
\end{equation}
In this expression, all the factors but $m$ are of order one, see the
previous discussion about the constraints on the parameters $A$ and
$W_0$. Therefore, this implies that $m\simeq {\cal
O}\left(10^{-6}\right)$ and this is the value that will be used in the
following.

\par

In order to compute the inflationary observables (as the spectral
indices for instance), it is convenient to use the slow-roll
approximation. The slow-roll approximation is controlled by two
parameters (in fact, at leading order, there are three relevant
slow-roll parameters but we will not need the third one) defined
by~\cite{slowroll}
\begin{eqnarray}
\epsilon &\equiv & -\frac{\dot{H}}{H^2}\, ,\quad
\delta =-\frac{\dot{\epsilon}}{2H\epsilon }+\epsilon \, .
\end{eqnarray}
The main advantage of these definitions is that they involve the
background Hubble parameter $H$ only. Therefore, in some sense, they
are independent from the matter content, in particular they do not
require the knowledge of the number of scalar fields present in the
underlying inflationary model. If we now assume that only one scalar
field is present, then it is easy to obtain that
\begin{eqnarray}
\label{srparam}
\epsilon &\simeq & \epsilon _{\bar{\phi }}\equiv \frac{\mP^2}{16\pi
}\left(\frac{V_{,{\bar{\phi}}}}{V}\right)^2\, ,\\ \delta &\simeq &
\delta _{\bar{\phi }}\equiv -\frac{\mP^2}{16\pi
}\left(\frac{V_{,\bar{\phi}}}{V}\right)^2 +\frac{\mP^2}{8\pi
}\frac{V_{,\bar{\phi}\bar{\phi}}}{V}\, .
\end{eqnarray}
Two remarks are in order. Firstly, the condition $\epsilon <1$ is
equivalent to $\ddot{a}>0$. Therefore, in order to have inflation,
strictly speaking $\epsilon $ needs not to be small with respect to
one, it only needs to be less than one. Secondly, the parameter
$\delta $ is not positive definite contrary to the first slow-roll
parameter.

\par

In the case where two fields are present, they are different ways of
generalizing the definition of the slow-roll
parameters~\cite{isopert}. A first method consists in following the
trajectory in configuration space.  The trajectory is given by
$\bar{\phi }=\bar{\phi }(N)$ and $\bar{\rho }= \bar{\rho }(N)$, where
$N$ is the total number of e-folds counted from the beginning of
inflation (not to be confused with $N_*$). As a consequence the vector
tangent to this trajectory, ${\bf e}_{\parallel}=(e_{\bar{\phi
}},e_{\bar{\rho }})$, can be expressed as
\begin{eqnarray}
e_{\bar{\phi }} &=& \frac{\displaystyle \frac{{\rm d}\bar{\phi }}{{\rm
      d}N}} {\sqrt{\displaystyle\left(\frac{{\rm d}\bar{\phi }}{{\rm
      d}N}\right)^2 +\left(\frac{{\rm d}\bar{\rho }}{{\rm
      d}N}\right)^2}} =\cos \theta \, , \\ e_{\bar{\rho }}&=&
\frac{\displaystyle \frac{{\rm d}\bar{\rho }}{{\rm
      d}N}} {\sqrt{\displaystyle\left(\frac{{\rm d}\bar{\phi }}{{\rm
      d}N}\right)^2 +\left(\frac{{\rm d}\bar{\rho }}{{\rm
      d}N}\right)^2}}=\sin \theta \, .
\end{eqnarray}
We can then define ``directional slow-roll parameters'' by replacing
the first and second derivatives in Eqs.~(\ref{srparam}) by
directional derivatives of the potential, namely
\begin{eqnarray}
\label{directionalsr}
\epsilon _{\parallel} &=& \frac{\mP^2}{16\pi V^2}\left(\cos \theta
V_{,{\bar{\phi}}}+\sin \theta V_{,{\bar{\rho }}}\right)^2\, ,\\
\label{directionalsr2}
\delta_{\parallel} &=& -\epsilon _{\parallel}+\frac{\mP^2}{8\pi V}(\cos
^2\theta V_{,\bar{\phi}\bar{\phi}} +2\cos \theta \sin \theta
V_{,\bar{\phi}\bar{\rho }} \nonumber \\
& & +\sin ^2 \theta V_{,\bar{\rho }\bar{\rho
}})\, .
\end{eqnarray}
However, it is clear that we no longer have the equivalence between
$\epsilon _{\parallel}<1$ and $\ddot{a}>0$. For this reason, it is
also interesting to keep the original definition of $\epsilon$ (only
in terms of the ``geometry''), \ie $\epsilon =-\dot{H}/H^2$, and
express it in terms of the derivatives of the potential. This leads to
\begin{equation}
\epsilon= \epsilon _{\bar{\phi }}+\epsilon _{\bar{\rho }} =
\frac{\mP^2}{16\pi }\left(\frac{V_{,{\bar{\phi}}}}{V}\right)^2 +
\frac{\mP^2}{16\pi }\left(\frac{V_{,{\bar{\rho }}}}{V}\right)^2 \, .
\end{equation}
It is interesting to establish the link between the two types of
slow-roll parameters. For this purpose, let us introduce the vector
${\bf e}_{\perp}\equiv (e_{\bar{\rho }},-e_{\bar{\phi }})$, which is
perpendicular to ${\bf e}_{\parallel}$. Then, one can define a
directional slow-roll parameter in the direction perpendicular to the
trajectory by means of the expression
\begin{equation}
\epsilon _{\perp}=\frac{\mP^2}{16\pi V^2}\left(\sin \theta
V_{,{\bar{\phi}}}-\cos \theta V_{,{\bar{\rho }}}\right)^2\, .
\end{equation}
{}From this expression, it is not difficult to show that
\begin{equation}
\epsilon _{\parallel}+\epsilon _{\perp} = \epsilon _{\bar{\phi
}}+\epsilon _{\bar{\rho }} = \epsilon \, .
\end{equation}
Obviously, if the inflationary valley is a straight line then one has
$\epsilon \simeq \epsilon _{\parallel} \simeq \epsilon _{\bar{\phi}}$
and this is the case for the model under consideration in this article
provided that $\bar{\phi }\gg \mP$. In this situation, the model is
equivalent to chaotic inflation and therefore the slow-roll parameters
are given by
\begin{equation}
\label{chaoticsr}
\epsilon = \frac{1}{2N_*+1}, \quad
\delta = 0\, ,
\end{equation}
where $N_*\simeq 60$ (there is some freedom in the choice of this
number, see Ref.~\cite{number}) has already been defined before. In
the situation where these parameters are small, namely $\epsilon \ll
1$ and $\delta \ll 1$, the equation of motion of the inflaton field
can be easily integrated. One finds
\begin{equation}
\label{srinflaton}
\frac{\bar{\phi }}{\mP}=\sqrt{\left(\frac{\bar{\phi }_{\rm
ini}}{\mP}\right)^2 -\frac{N}{2\pi }}\, ,
\end{equation}
where $\bar{\phi }_{\rm ini}$ is the initial value of the field. Let
us emphasize again that this is valid only if the scales of
astrophysical interest leave the Hubble radius in the straight line
part of the potential. If this happens in the curved part of the
potential the above result is no longer valid.

\par

Finally, let us introduce the squared mass in the direction
perpendicular to the inflationary trajectory. This is nothing but the
second order directional derivative of the potential along ${\bf
e}_{\perp}$ given by
\begin{equation}
\label{massperp} m_{\perp}^2\equiv (\cos ^2\theta
V_{,\bar{\phi}\bar{\phi}} -2\cos \theta \sin \theta
V_{,\bar{\phi}\bar{\rho }} +\sin ^2 \theta V_{,\bar{\rho }\bar{\rho
}})\, .
\end{equation}
This quantity allows us to distinguish whether inflation ends by
instability or not. If, as it is the case for hybrid inflation, an
instability occurs, $m_{\perp}^2$ becomes negative.

\begin{figure*}
\includegraphics[width=8.8cm,height=7.5cm]{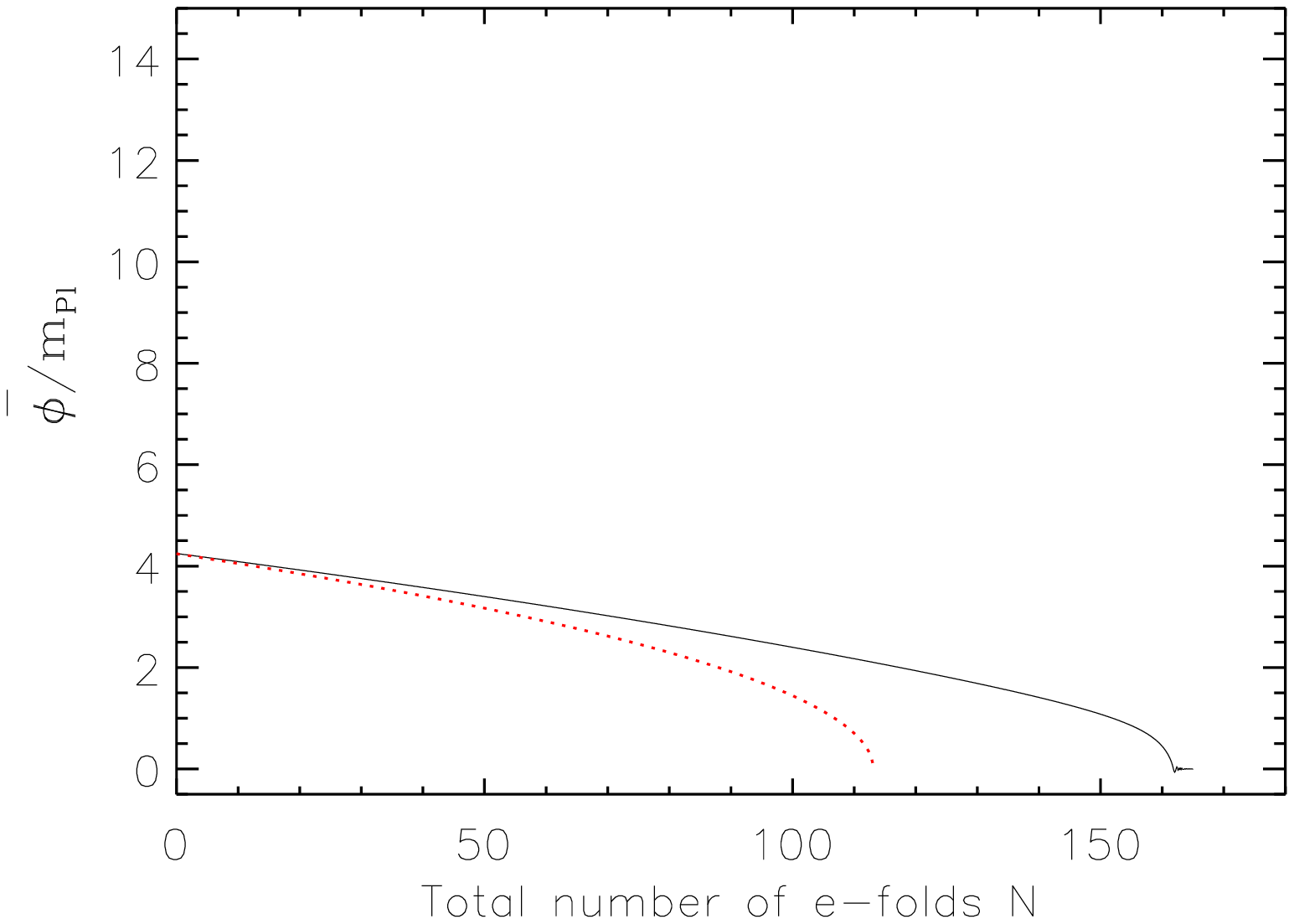}
\includegraphics[width=8.8cm,height=7.5cm]{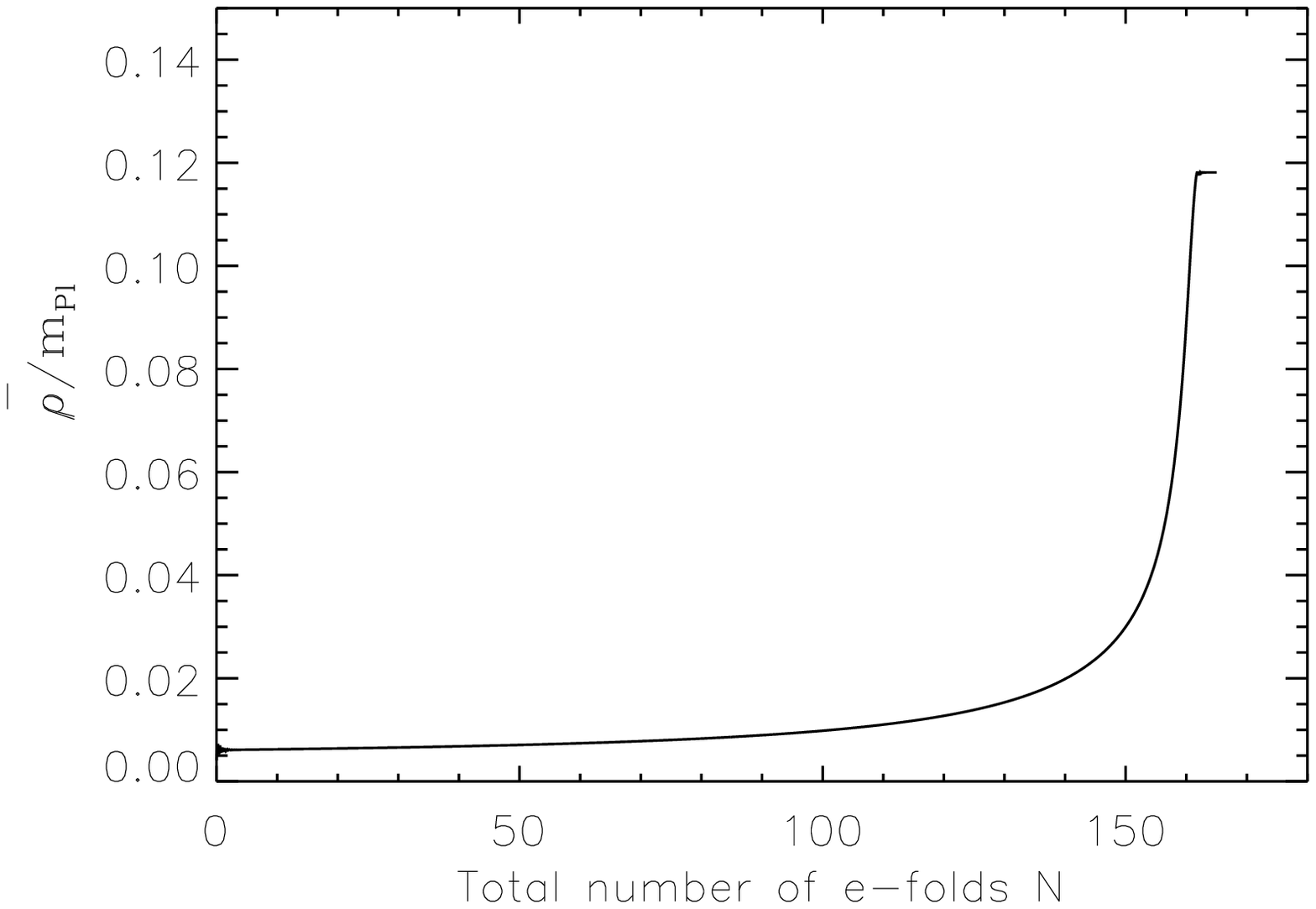}\\
\includegraphics[width=8.8cm,height=7.5cm]{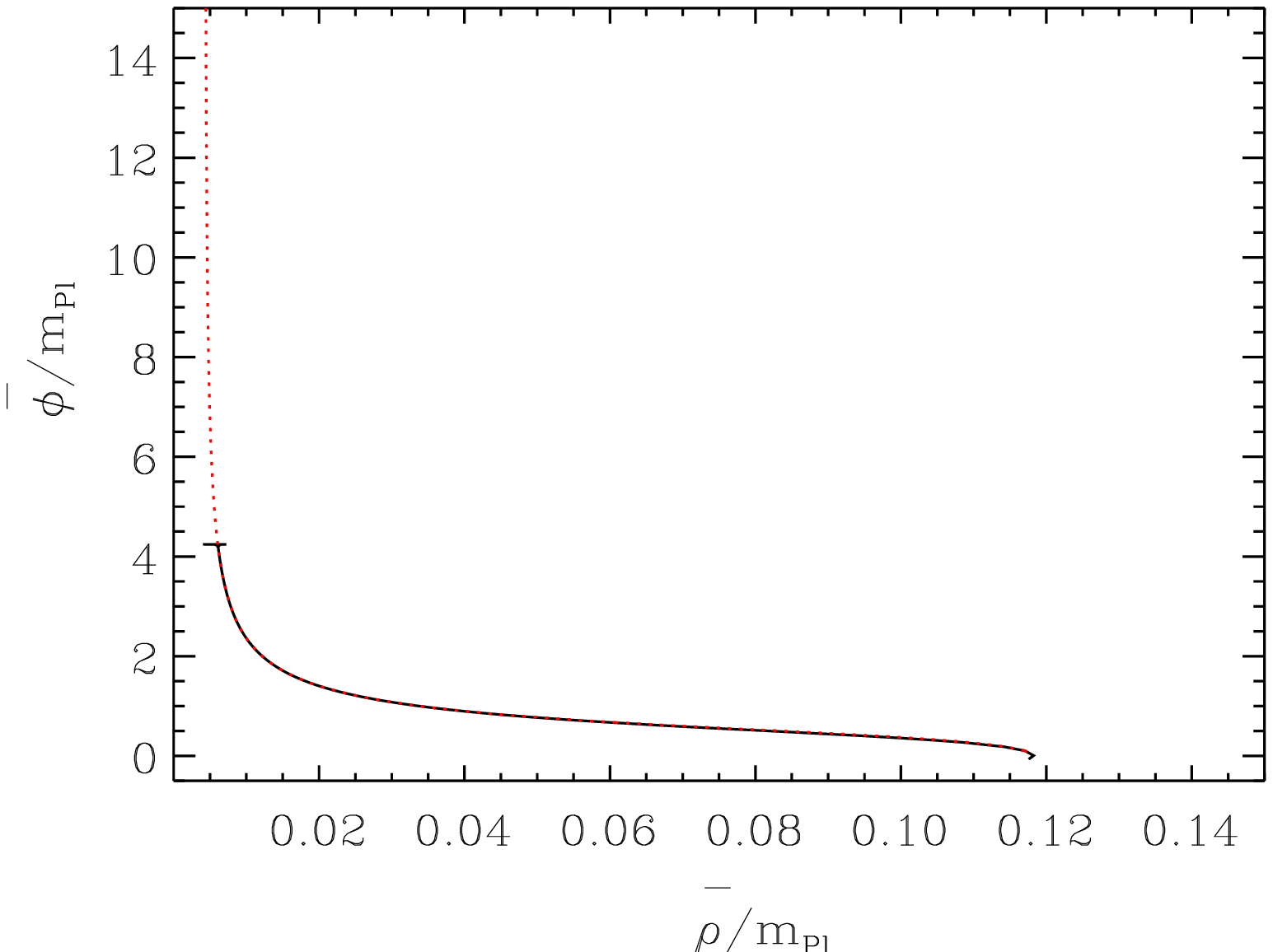}
\includegraphics[width=8.8cm,height=7.5cm]{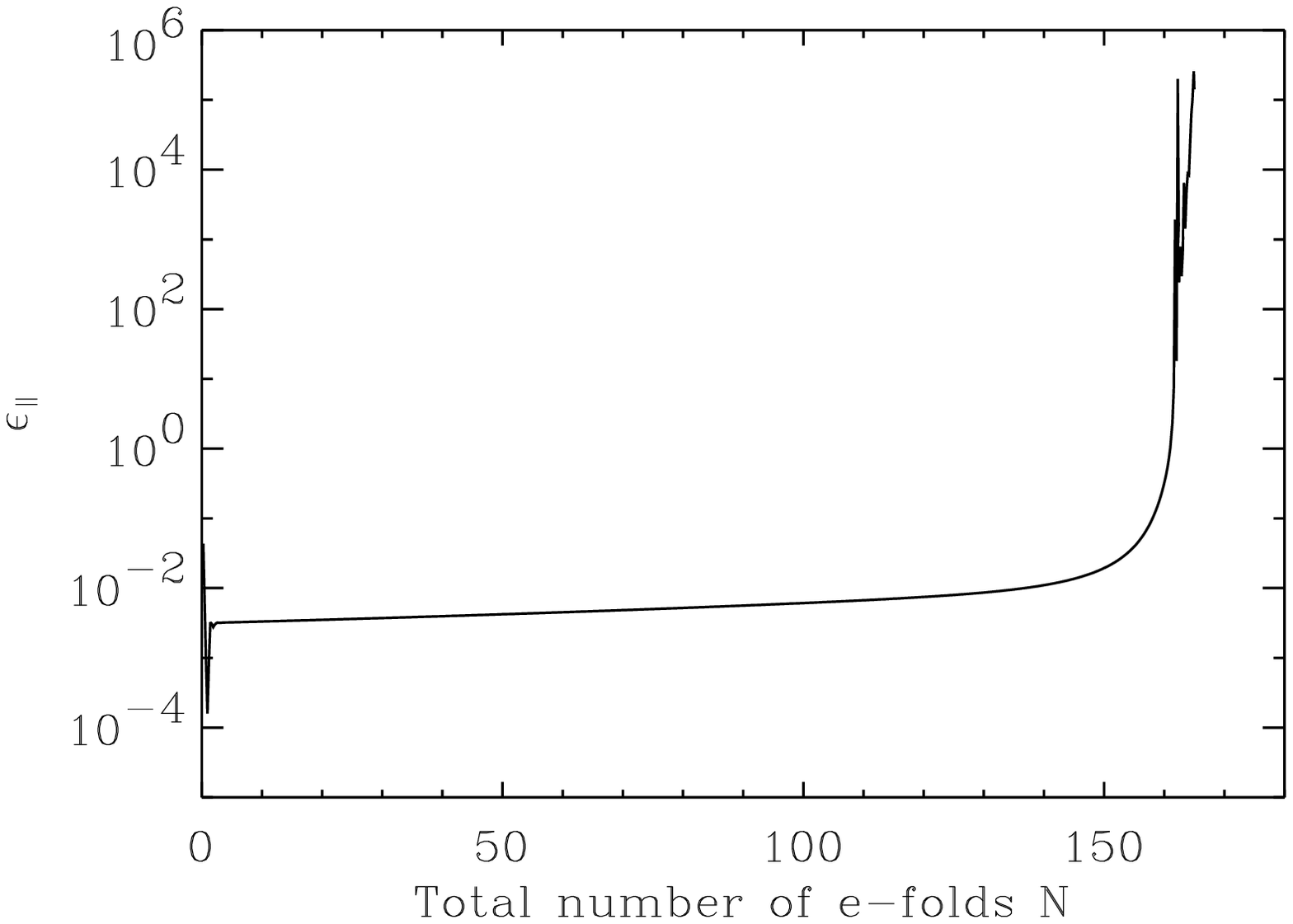}\\
\includegraphics[width=8.8cm,height=7.5cm]{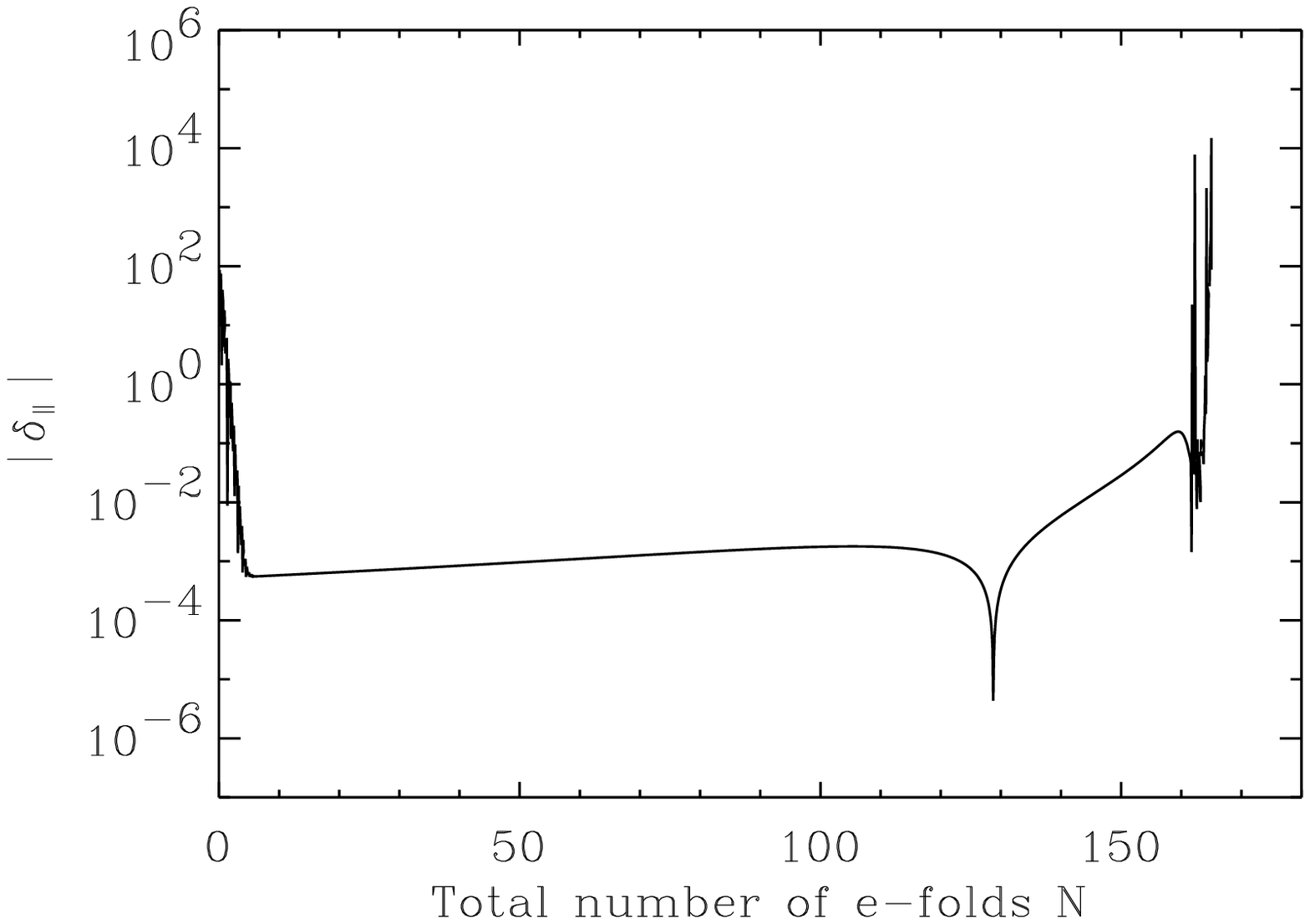}
\includegraphics[width=8.8cm,height=7.5cm]{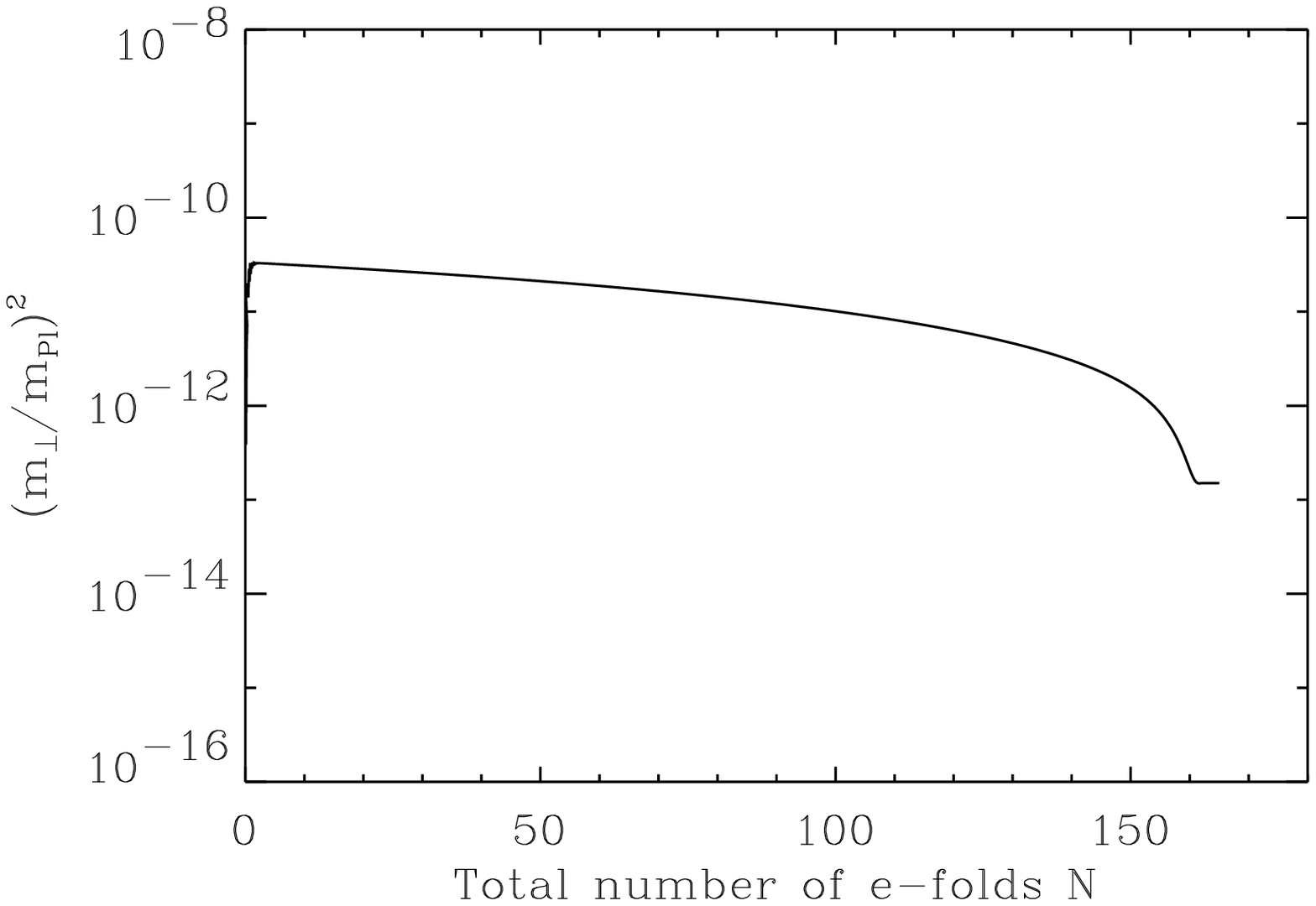}
\caption{Evolution of the fields $\bar{\phi }$ and $\bar{\rho }$
  together with the slow-roll parameters and the effective squared
  mass in the direction perpendicular to the inflationary
  trajectory. The parameters are $\alpha =\sqrt{2}$, $\beta =1$,
  $m=10^{-6}\mP$, $\kappa A/(\alpha m)=1.35135$ and $W_0/A=0.41111$.
  This gives $y_{{\rm min},\tilde{\cal U}}\simeq 1.067$ or $\bar{\rho
  }_{{\rm min},\tilde{\cal U}}\simeq 0.004 \times \mP$. The absolute
  minimum of the potential is located at $\phi =0$, $y=y_{{\rm
  min},\tilde{\cal V}}\simeq 1.622$ or $\bar{\phi }=0$, $\bar{\rho
  }=\bar{\rho }_{{\rm min},\tilde{\cal V}}\simeq 0.118 \times
  \mP$. The initial conditions are $\phi _{\rm ini}=3\times \mP$ or
  $\bar{\phi }_{\rm ini}\simeq 4.243 \times \mP$ and $y_{\rm
  ini}=y_{{\rm min},\tilde{\cal U}}\simeq 1.067$ or $\bar{\rho }_{\rm
  ini}=\bar{\rho }_{{\rm min},\tilde{\cal U}}\simeq 0.004 \times \mP$,
  \ie at the bottom of the valley exactly. The interpretation of these
  plots is discussed in the text.}
\label{case3}
\end{figure*}

\subsection{Numerical results}

To prove the above claims that inflation ends by violation of the
slow-roll conditions and not by instability, it is necessary to
determine the inflationary trajectory exactly. Clearly, the potential
is too complicated to permit an analytical integration of the exact
motion and, therefore, we will perform a numerical integration of the
two Klein-Gordon equations and of the Friedmann equation. For
convenience, as already mentioned, the total number of e-folds
$N\equiv \ln \left(a/a_{\rm ini}\right)$ will be used as time
variable. In this case, the Klein-Gordon equation reads (here for the
inflaton field; this is of course also the case for the moduli)
\begin{eqnarray}
& & \label{kgefold} \frac{{\rm d}^2}{{\rm d}N^2}\left(\frac{\bar{\phi
}}{\mP}\right) +\left(3+\frac{1}{H}\frac{{\rm d}H}{{\rm d}N}\right)
\frac{{\rm d}}{{\rm d}N}\left(\frac{\bar{\phi }}{\mP}\right)
\nonumber \\
& & +\left(\frac{\mP }{H}\right)^2 \frac{\partial
\left(V/\mP^4\right)}{\partial (\bar{\phi }/\mP)}=0\, ,
\end{eqnarray}
where $H$ is the Hubble parameter during inflation.

\par

\begin{figure*}
\includegraphics[width=8.8cm,height=7.5cm]{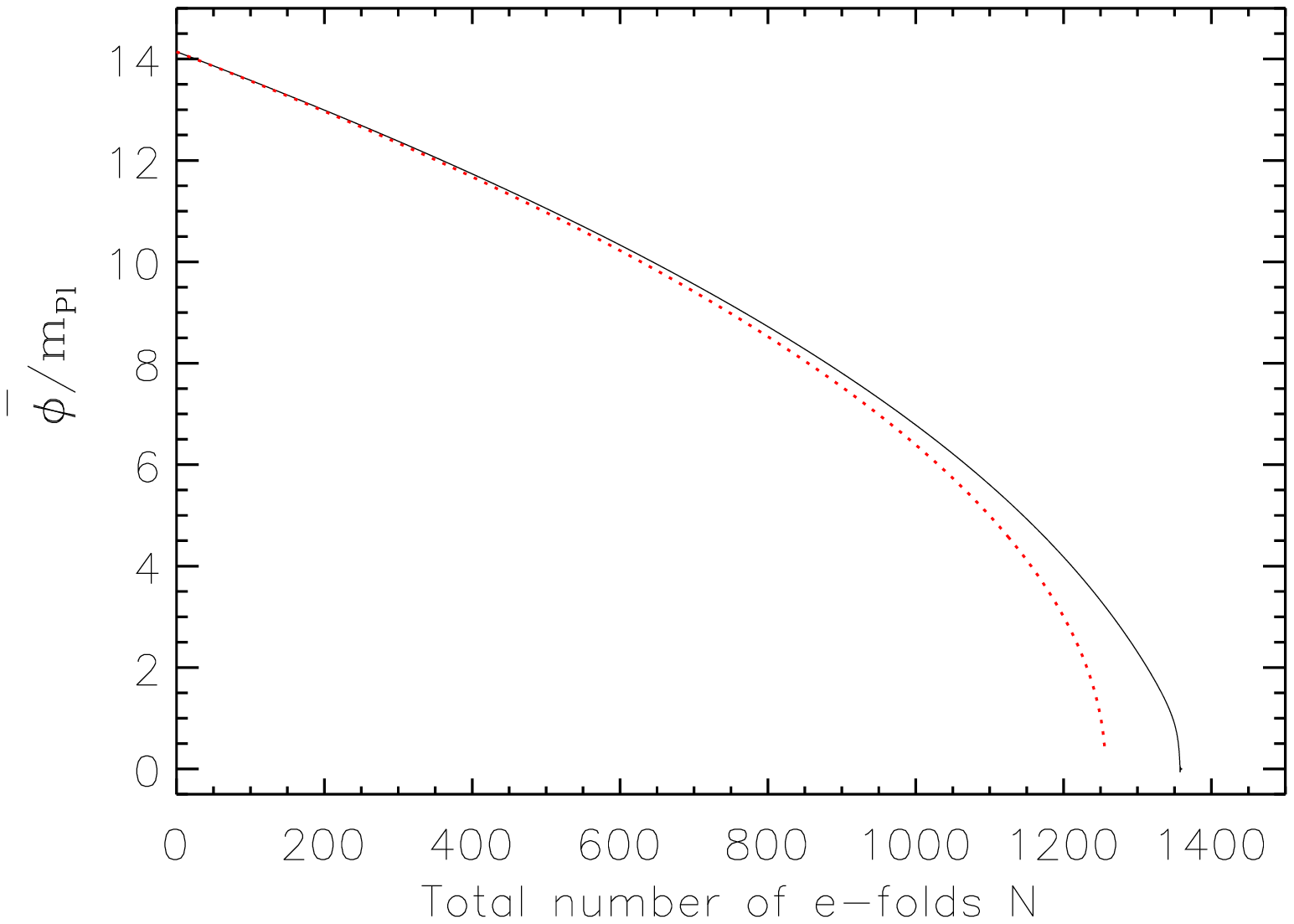}
\includegraphics[width=8.8cm,height=7.5cm]{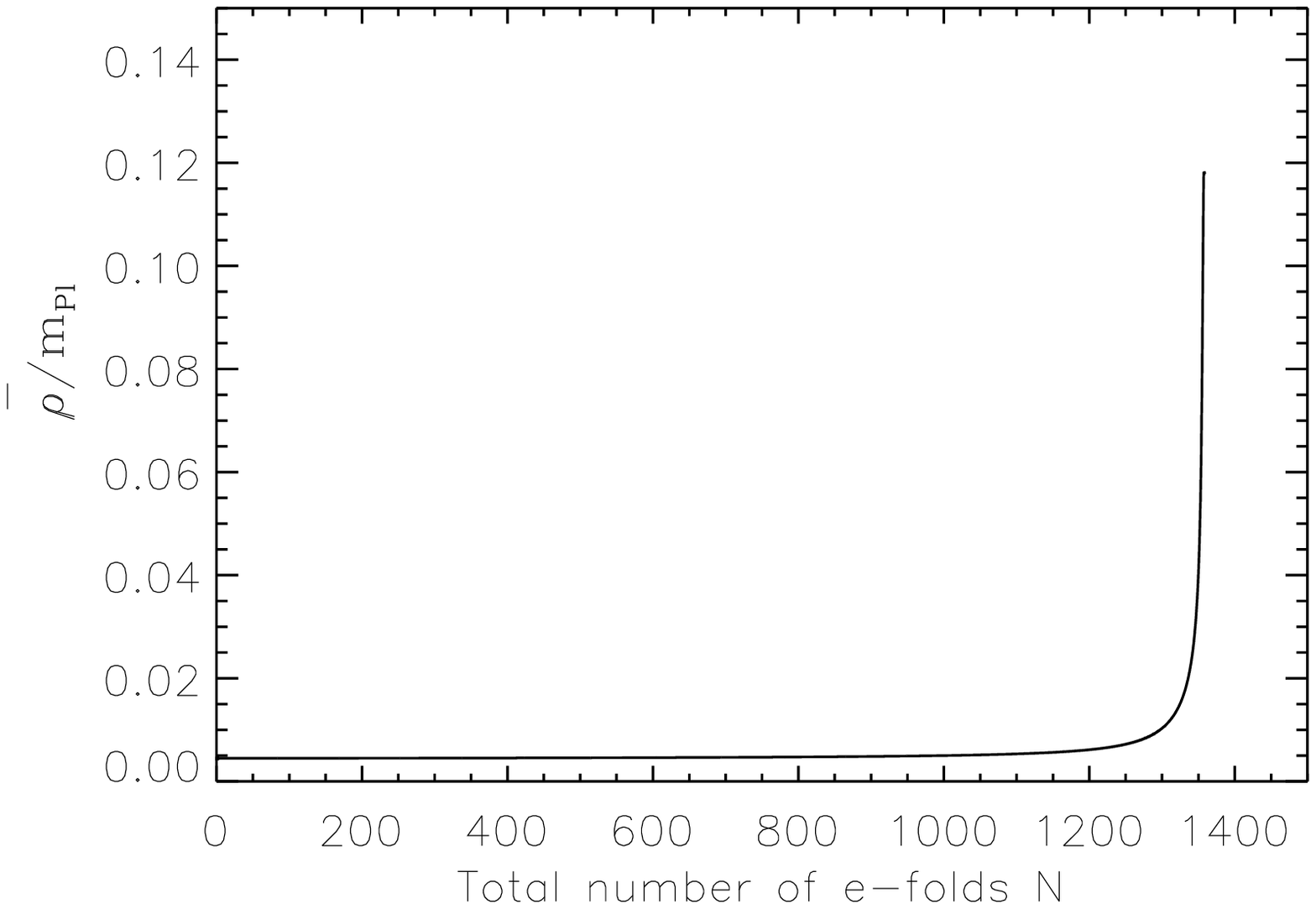}\\
\includegraphics[width=8.8cm,height=7.5cm]{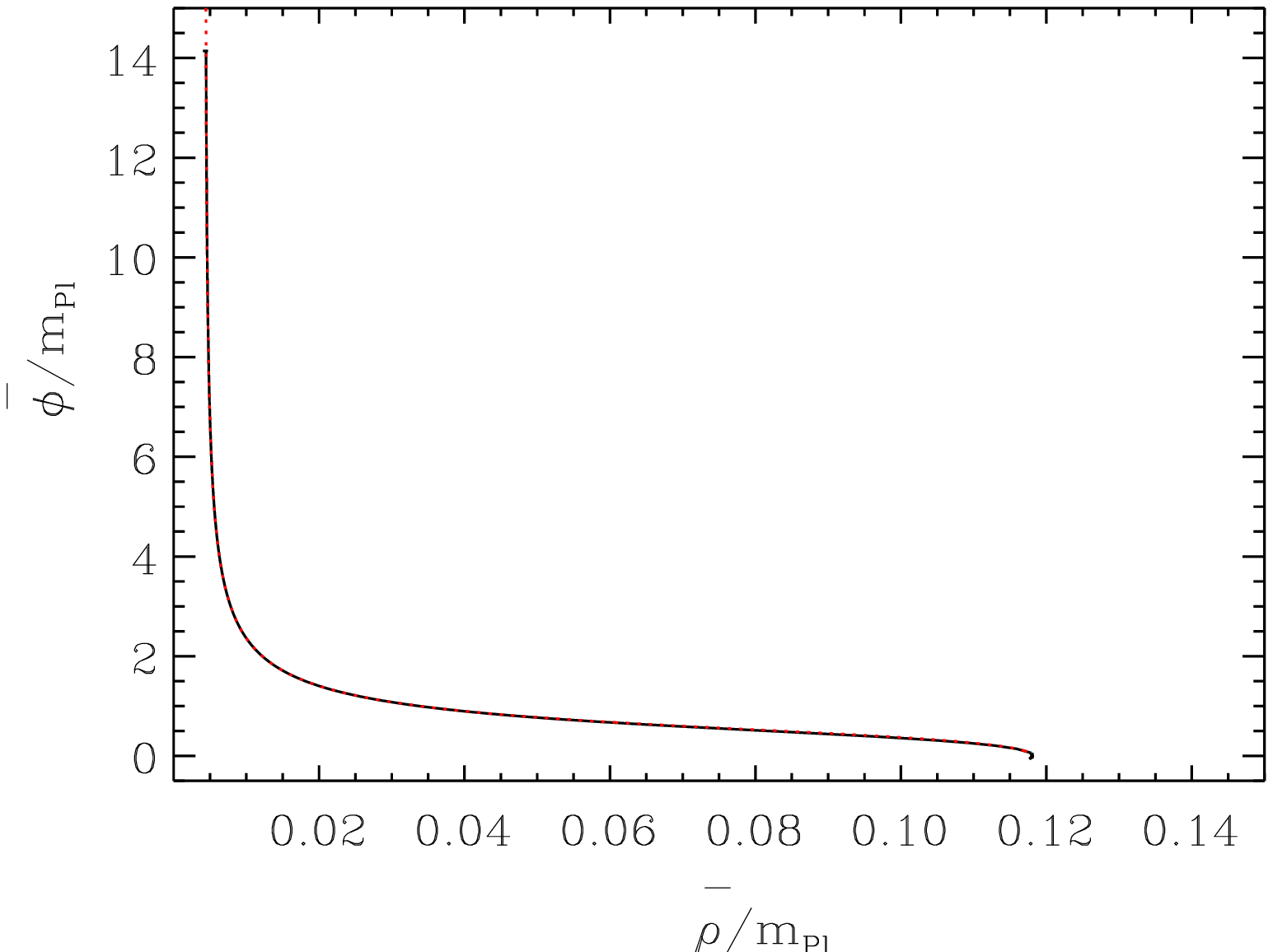}
\includegraphics[width=8.8cm,height=7.5cm]{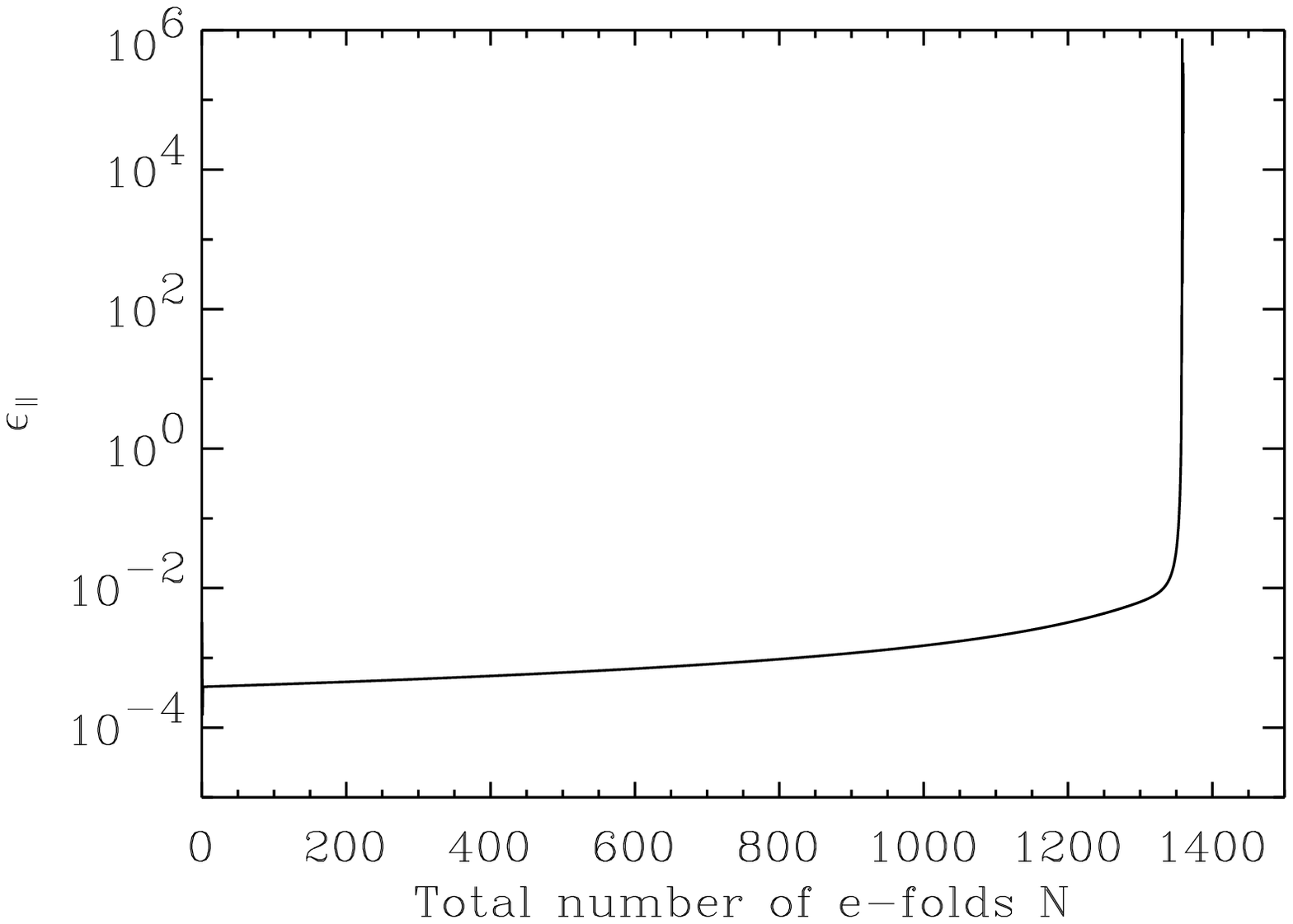}\\
\includegraphics[width=8.8cm,height=7.5cm]{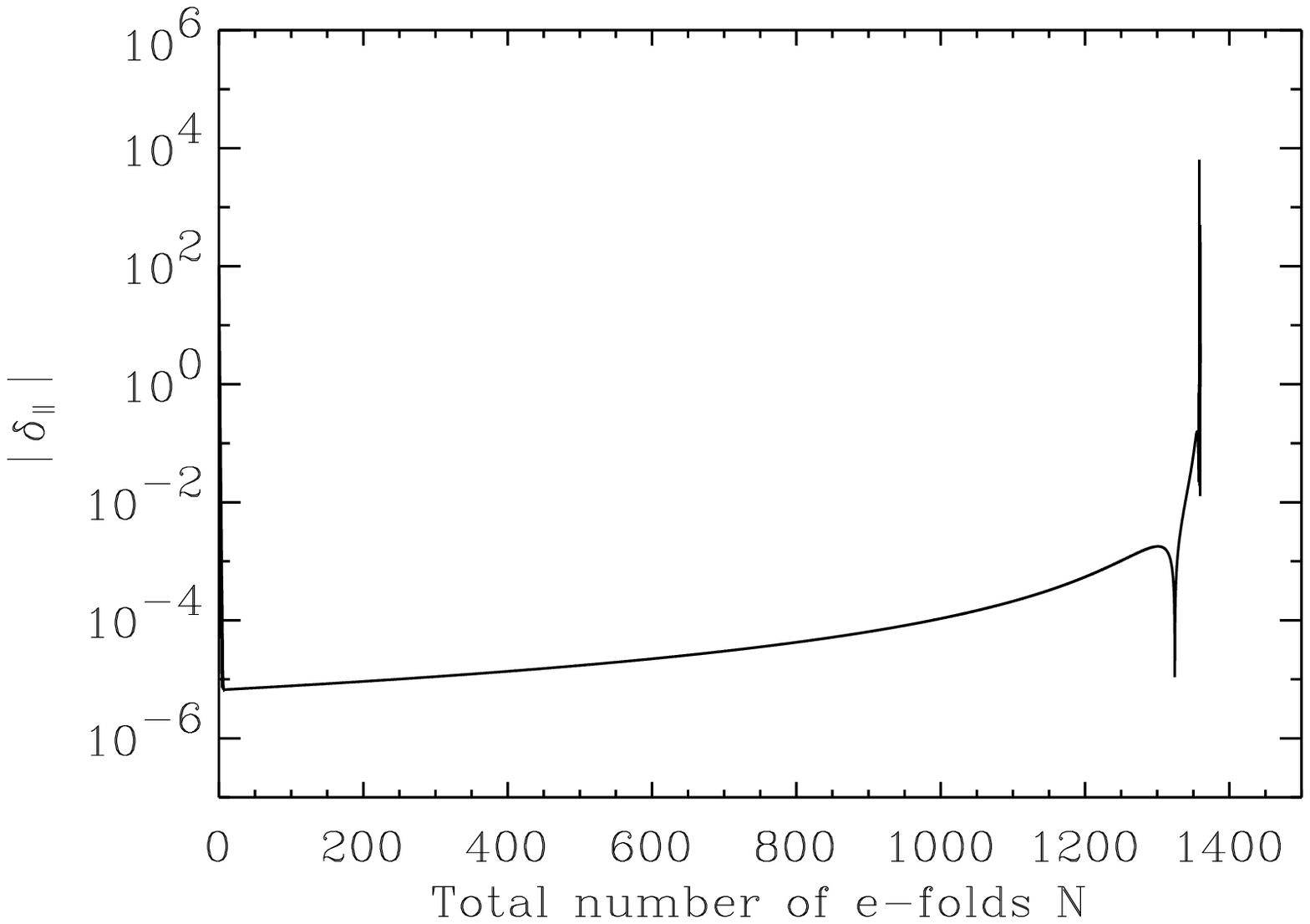}
\includegraphics[width=8.8cm,height=7.5cm]{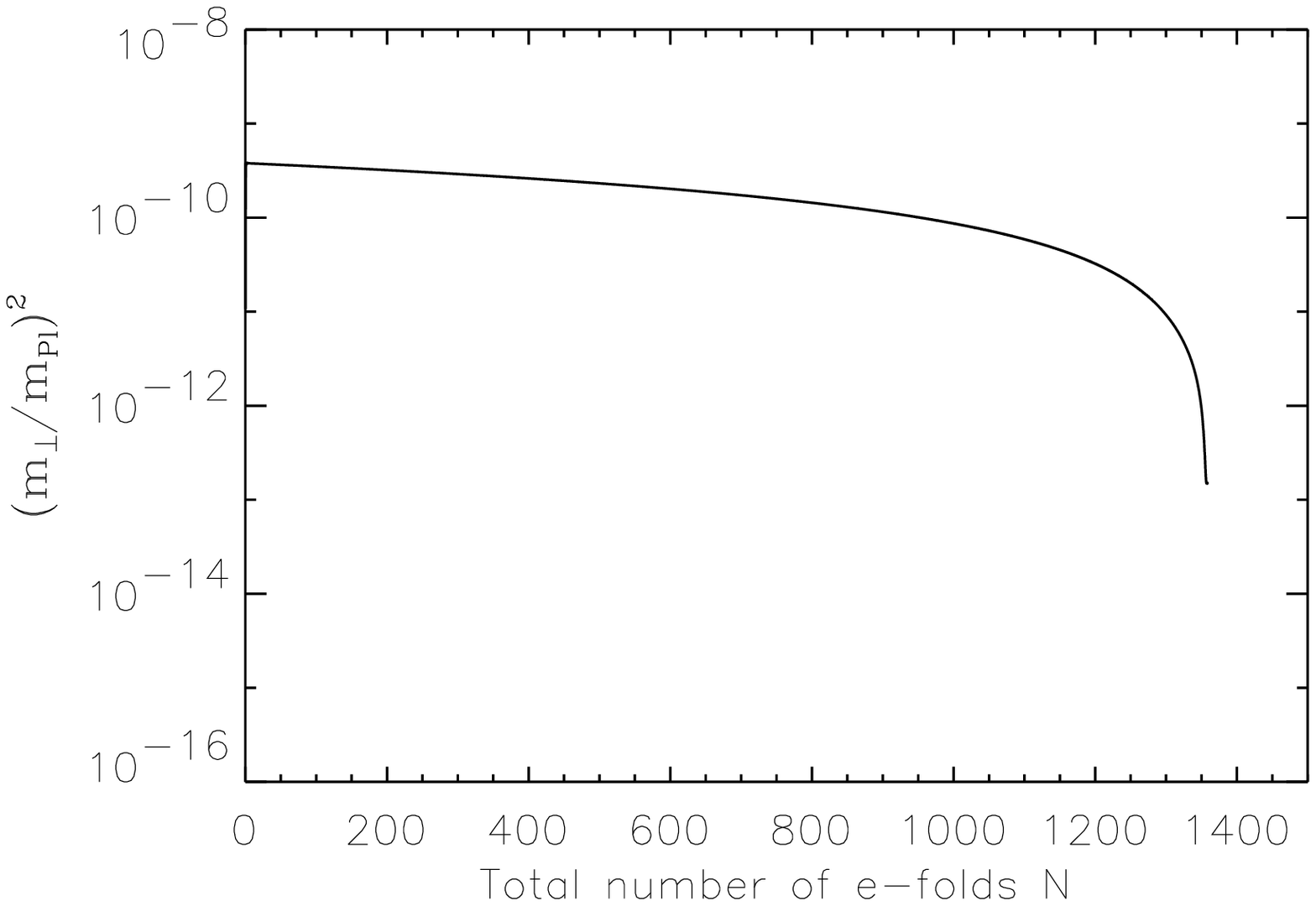}
\caption{Same as in Fig.~\ref{case3} but for different initial
  conditions. The parameters are $\alpha =\sqrt{2}$, $\beta =1$,
  $m=10^{-6}\mP$, $\kappa A/(\alpha m)=1.35135$ and $W_0/A=0.41111$.
  This gives $y_{{\rm min},\tilde{\cal U}}\simeq 1.067$ or $\bar{\rho
  }_{{\rm min},\tilde{\cal U}}\simeq 0.004 \times \mP$. The absolute
  minimum of the potential is still located at $\phi =0$, $y=y_{{\rm
  min},\tilde{\cal V}}\simeq 1.622$ or $\bar{\phi }=0$, $\bar{\rho
  }=\bar{\rho }_{{\rm min},\tilde{\cal V}}\simeq 0.118 \times
  \mP$. The initial conditions are now $\phi _{\rm ini}=10\times \mP$
  or $\bar{\phi }_{\rm ini}\simeq 14.142 \times \mP$ and $y_{\rm
  ini}=y_{{\rm min},\tilde{\cal U}}\simeq 1.067$ or $\bar{\rho }_{\rm
  ini}=\bar{\rho }_{{\rm min},\tilde{\cal U}}\simeq 0.004 \times \mP$,
  \ie still at the bottom of the valley exactly.}
\label{case10}
\end{figure*}

The result of our numerical integration is displayed in
Figs.~\ref{case3} and~\ref{case10}. In Fig.~\ref{case3}, we have
chosen the parameters such that $\alpha =\sqrt{2}$, $\beta =1$,
$m=10^{-6}\mP$ (in accordance with the COBE and WMAP normalizations,
see the discussion above), $\kappa A/(\alpha m)=1.35135$ and
$W_0/A=0.41111$. This means that the minimum of the inflationary
valley is located at $y_{{\rm min},\tilde{\cal U}}\simeq 1.067$ or, in
terms of the canonically normalized fields, at $\bar{\rho }_{{\rm
min},\tilde{\cal U}}\simeq 0.004 \times \mP$. The absolute minimum of
the potential is at $\phi =0$, $y=y_{{\rm min},\tilde{\cal V}}\simeq
1.622$ or, in terms of the canonically normalized fields, at
$\bar{\phi }=0$, $\bar{\rho }=\bar{\rho }_{{\rm min},\tilde{\cal
V}}\simeq 0.118 \times \mP$. The initial conditions have been chosen
to be $\phi _{\rm ini}=3\times \mP$ or $\bar{\phi }_{\rm ini}\simeq
4.243 \times \mP$ and $y_{\rm ini}=y_{{\rm min},\tilde{\cal U}}\simeq
1.067$ or $\bar{\rho }_{\rm ini}=\bar{\rho }_{{\rm min},\tilde{\cal
U}}\simeq 0.004 \times \mP$. This means that the evolution starts at
the bottom of the valley. The first plot (first line, on the left)
shows the evolution of the field $\bar{\phi }$ versus the total number
of e-folds (solid black line).  The red dotted line represents the
slow-roll approximation given by Eq.~(\ref{srinflaton}), valid in the
case where there is only one field. At the beginning of the evolution,
the field $\bar{\phi }$ closely follows the slow-roll equation
Eq.~(\ref{srinflaton}).  Clearly, this is because the valley is a
straight line and, therefore, everything is as if there were only one
field. Then, the valley bends and the black dotted curve separates
from the red dotted line. Interestingly enough, this is not associated
with a rapid evolution of the inflaton which is already an indication
that, although the valley is now curved, the slow-roll conditions are
probably not violated. Then, the field joins its minimum at $\bar{\phi
}=0$ and there are small oscillations around that minimum (which are
difficult to distinguish with the scales used in this particular
plot). The second figure (first line, on the right) shows the
evolution of the moduli $\bar{\rho }$ with the total number of
e-folds. At the beginning, $\bar{\rho }$ is frozen at the bottom of
the inflationary valley. When the valley bends, $\bar{\rho }$ also
joins the absolute minimum. The third plot (second line, on the left)
displays the trajectory $\bar{\phi }(\bar{\rho })$. The most striking
feature of the plot is that the trajectory exactly follows the
inflationary valley (shown as the red dotted curve). Of course, this
is maybe not so surprising given the fact that, initially, the moduli
field is at the bottom of the valley and that the initial velocities
of both fields vanish. Nevertheless, in this case,
Eq.~(\ref{inftrajec}) is an analytical expression of the non trivial
inflationary trajectory. The next plots show the directional slow-roll
parameters $\epsilon _{\parallel}$ (second line, on the right) and
$\delta _{\parallel}$ (third line, one the right). One can explicitly
check that these slow-roll parameters remain small even, and this is
crucial here, when the trajectory bends. It is only at the very end,
close to the absolute minimum of the potential, that the slow-roll
conditions are violated. Therefore, we have proven that, contrary to
the case of hybrid inflation, a variation of the waterfall field is
not associated with a violation of the slow-roll conditions. In other
words, the slow-roll conditions continue to hold even when the
trajectory is curved, except, of course, when the absolute minimum is
approached. The cusp present in the plot of the second slow-roll
parameter $\delta _{\parallel}$ (around $N\simeq 130$) is due to the
fact that $\delta _{\parallel}$ becomes negative (recall that,
contrary to $\epsilon _{\parallel}$, $\delta _{\parallel}$ is not
positive-definite). One also notices the presence of quite large
oscillations at the very beginning of the evolution and at the very
end. The oscillations at the very end are clearly the oscillations
occurring when the absolute minimum is joined and when reheating
proceeds. The oscillations at the beginning of the evolution are worth
interpreting. They do not seem to be associated with some numerical
problems since it can be checked that they remain even if the
parameter controlling the accuracy of the code is modified. Our
interpretation is the following. As discussed above, the initial value
of $\bar{\rho }$ has been chosen such that it corresponds to the
bottom of the valley in the regime $\bar{\phi }\gg \mP$, in fact
strictly speaking, in the limit $\bar{\phi }/\mP\rightarrow
+\infty$. On the other hand, the initial condition on the inflaton is
$\bar{\phi} \simeq 4.2 \times \mP$ and, for this value of the field,
the bottom of the valley is not exactly located at the value obtained
before, in the limit $\bar{\phi }/\mP\rightarrow +\infty$.  The
oscillations are nothing but a transient regime during which the
moduli field is settling at the bottom of the valley.

\par

As noticed before, the directional slow-roll parameter $\epsilon
_{\parallel }$ does not control the end of inflation. However, we have
checked that, during almost all the evolution, the difference between
the parameter $\epsilon =-\dot{H}/H^2$ and $\epsilon _{\parallel}$ is
small. As a consequence, we see that the total number of e-folds
(during which we have slow-roll inflation) is $N_{_{\rm T}}\simeq
160$. This has to be compared with the single field expression for
$N_{_{\rm T}}$,
\begin{equation}
N_{_{\rm T}}=2\pi \left(\frac{\bar{\phi }_{\rm
    ini}}{\mP}\right)^2-\frac12\, ,
\end{equation}
evaluated for the same initial conditions. In this case, this gives
$N_{_{\rm T}}\simeq 113$. This means that, for the same initial
conditions, the model under investigation in this article leads to a
larger number of total e-folds, probably because the inflationary path
is, in some sense, ``longer''. Finally, the last plot (third line, on
the right) represents $m_{\perp}^2$, see Eq.~(\ref{massperp}), versus
the total number of e-folds. This figure is important because it
proves that inflation does not end by instability, as it is the case
for hybrid inflation. This is because $m_{\perp}^2$ always remains
positive, although it is decreasing as the fields are rolling down the
valley, meaning that this valley opens up as one is approaching the
absolute minimum. Therefore, in chaotic mutated inflation, inflation
stops by violation of the slow-roll conditions in the inflationary
valley (and after this valley has bent), \ie close to the absolute
minimum of the potential.

Our next step is to study whether the previous conclusions are robust
and can be modified if we change either the initial conditions and/or
the parameters of the model. In Fig.~\ref{case10}, we have considered
another initial condition for the inflaton field, namely $\phi _{\rm
ini}=10\times \mP$ or $\bar{\phi }_{\rm ini}\simeq 14.142 \times \mP$,
the other parameters being the same as in Fig.~\ref{case3} [
i.e. $\alpha =\sqrt{2}$, $\beta =1$, $m=10^{-6}\mP$, $\kappa A/(\alpha
m)=1.35135$ and $W_0/A=0.41111$.  The initial condition of the moduli
waterfall field is $y_{\rm ini}=y_{{\rm min},\tilde{\cal U}}\simeq
1.067$ or $\bar{\rho }_{\rm ini}=\bar{\rho }_{{\rm min},\tilde{\cal
U}}\simeq 0.004 \times \mP$, \ie still at the bottom of the
valley]. As can be seen all the remarks and conclusions obtained
before remain valid for this case.  Another remark is in order at this
point. In the valley, since we have $V\sim \tilde{\cal U}\bar{\phi
}^2$, the parameter $\delta _{\parallel}$ should vanish, see
Eqs.~(\ref{directionalsr2}) and (\ref{chaoticsr}). We see in
Fig.~\ref{case3} that, on the contrary, $\epsilon _{\parallel }$ and
$\delta _{\parallel}$ are initially of the same order of magnitude,
\ie $10^{-3}$. The reason is that, initially, the two fields are not
sufficiently ``deep''in the valley. On the contrary, with the new
initial condition $\bar{\phi }_{\rm ini}\simeq 14.142\times \mP$, the
fields are really in the straight part of the valley. As a
consequence, one can check that the slow-roll parameter $\delta
_{\parallel}$ ($\simeq 10^{-5}$) is now two orders of magnitude
smaller than $\epsilon _{\parallel}$ ($\simeq 10^{-3}$), in full
agreement with the arguments presented above. One can even check that
the numerical values are consistent with the above
interpretation. Indeed, as a time-dependent function, the slow-roll
parameter $\epsilon $ is given by $\epsilon=\mP^2/(4\pi \phi^2
)$. Therefore, initially one has $\epsilon=\mP^2/(4\pi \phi _{\rm
ini}^2 )\simeq 0.39\times 10^{-3}$ which is the value seen in
Fig.~\ref{case10} for small $N$.

\begin{figure*}
\includegraphics[width=8.8cm,height=7.5cm]{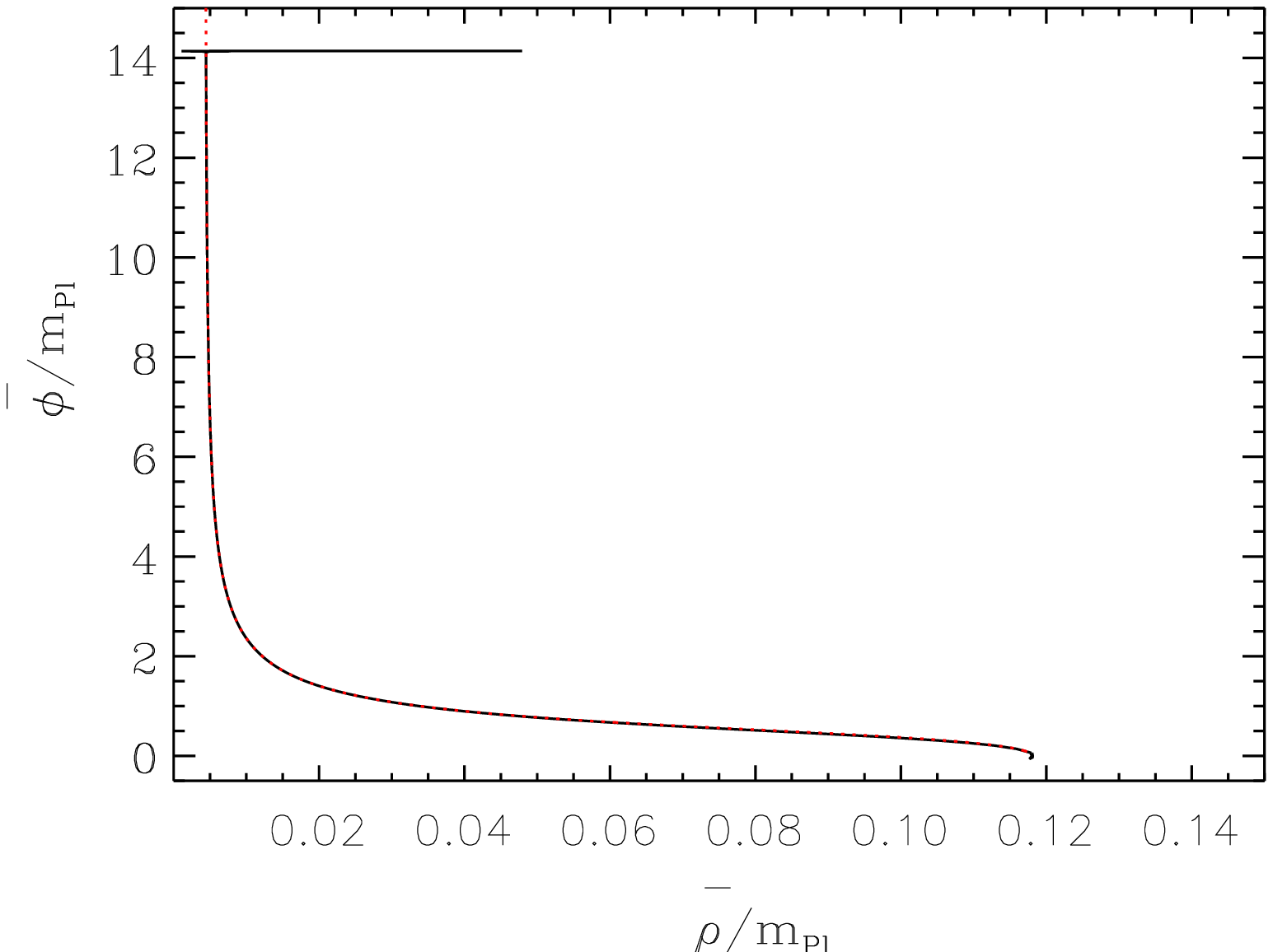}
\includegraphics[width=8.8cm,height=7.5cm]{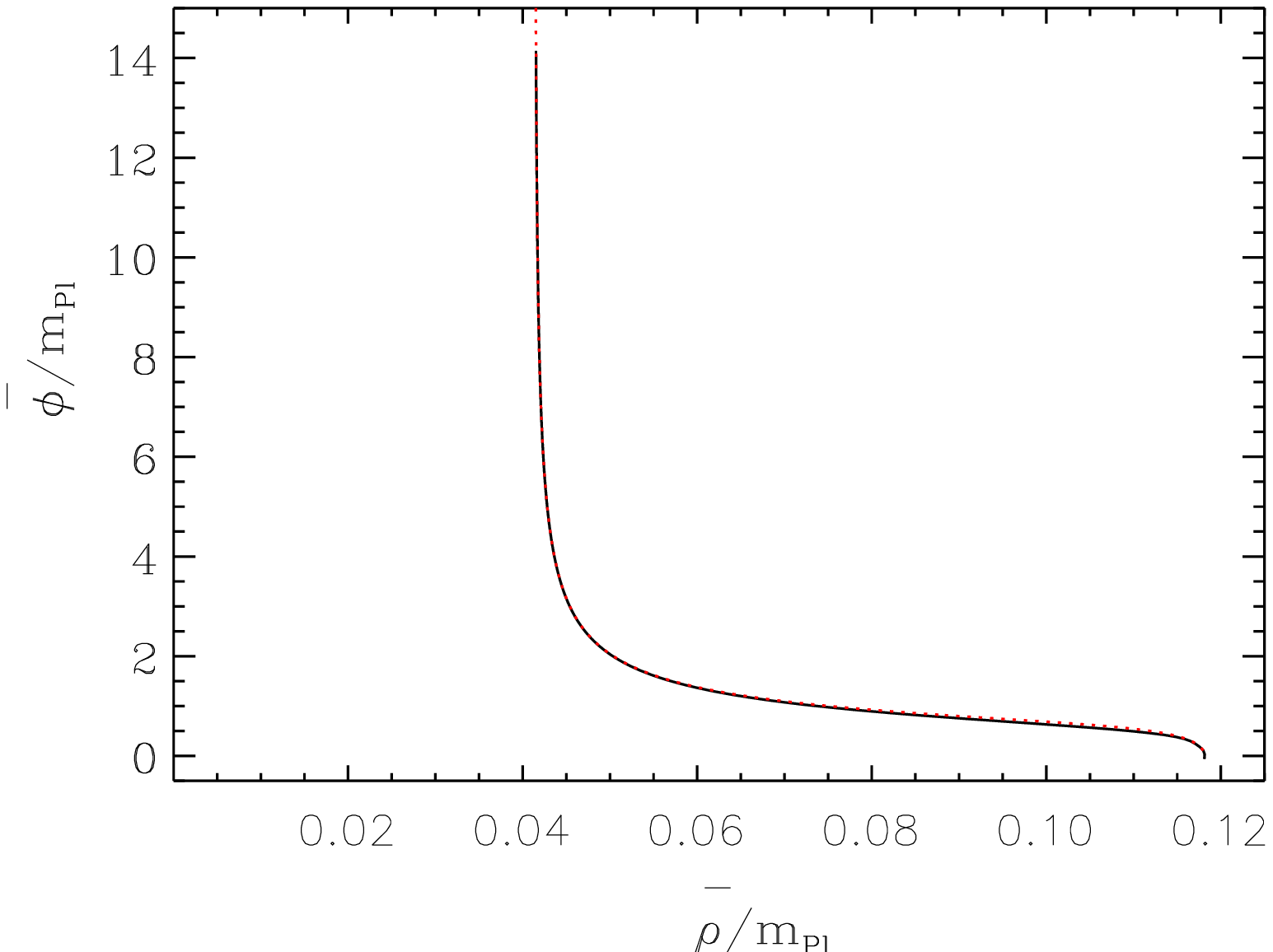}
\caption{Left panel: inflationary trajectory in the case where the
  moduli does not start from the bottom of the valley. The parameters
  are $\alpha =\sqrt{2}$, $\beta =1$, $m=10^{-6}\mP$, $\kappa
  A/(\alpha m)=1.35135$ and $W_0/A=0.41111$. The initial conditions
  are $\bar{\phi }_{\rm ini}=14.142\times \mP$ and $\bar{\rho }_{\rm
  ini}=0.0479 \times \mP$. Right panel: inflationary trajectory for
  different values of the parameters characterizing the model. The
  parameters are $\alpha =\sqrt{2}$, $\beta =1$, $m=10^{-6}\mP$,
  $\kappa A/(\alpha m)=1.30$ and $W_0/A=0.41111$, \ie only the value
  of $A$ has been modified. As a consequence, the absolute minimum of
  the potential is still located at $\phi =0$, $y=y_{{\rm
  min},\tilde{\cal V}}\simeq 1.622$ or $\bar{\phi }=0$, $\bar{\rho
  }=\bar{\rho }_{{\rm min},\tilde{\cal V}}\simeq 0.118 \times \mP$ but
  the position of the valley is changed. It is now located at
  $\bar{\rho }=\bar{\rho }_{{\rm min},\tilde{\cal U}}\simeq 0.0415
  \times \mP$. The initial conditions are $\phi _{\rm ini}=10\times
  \mP$ or $\bar{\phi }_{\rm ini}\simeq 14.142 \times \mP$ and $y_{\rm
  ini}=y_{{\rm min},\tilde{\cal U}}\simeq 1.165$ or $\bar{\rho }_{\rm
  ini}=\bar{\rho }_{{\rm min},\tilde{\cal U}}\simeq 0.0415 \times
  \mP$, \ie the field, in this case, still starts from the bottom of
  the valley exactly.}
\label{caseoff}
\end{figure*}

There exists another way of modifying the initial conditions.  Instead
of changing the initial value of the inflaton $\bar{\phi }$, we can
also study what happens if the moduli field $\bar{\rho }$ is initially
displaced from the bottom of the valley. This is especially relevant
because it is known that hybrid inflation is very sensitive to the
initial conditions and that only a very small fraction of possible
initial conditions leads to successful inflation, see
Refs.~\cite{hybridini}. In particular, it has been shown in these
articles that the waterfall field must be precisely tuned at the
bottom of the inflationary valley in order to obtain a satisfactory
subsequent evolution. In Fig.~\ref{caseoff} (left panel), we have used
the same set of parameters as in the previous figures but the initial
conditions are now $\bar{\phi }_{\rm ini}=14.142\times \mP$ and
$\bar{\rho }_{\rm ini}=0.0479 \times \mP$. The initial value of the
moduli field is approximatively one order of magnitude larger than in
Figs.~\ref{case3} and~\ref{case10}. We see that successful inflation
is still obtained. After very rapid oscillations, the moduli
stabilizes at the bottom of the valley and then the evolution proceeds
as before.  Therefore, it seems that the present model is more stable
to modifications of the initial conditions than standard hybrid
inflation. Of course, what should really be done, as in
Refs.~\cite{hybridini}, is a systematic scan of the space of initial
conditions but this is beyond the scope of the present article.

\par

Finally, we also need to study what happens if we change the
parameters of the model, \ie $A$ and $W_0$ (since $m$ is fixed by the
CMB normalization). In Fig.~\ref{caseoff} (right panel), the
trajectory in the space $(\bar{\phi },\bar{\rho })$ is displayed for
the following choice of parameters: $\alpha =\sqrt{2}$, $\beta =1$,
$m=10^{-6}\mP$, $\kappa A/(\alpha m)=1.30$ and $W_0/A=0.41111$. There
is a new value for the parameter $A$ (and in fact a new value for
$W_0$ but such that the ratio $W_0/A$ is left unchanged), of course
still compatible with the constraints derived above. The position of
the absolute minimum is unaffected because it only depends on
$W_0/A$. On the other hand, the location of the valley (\ie the
location of the minimum of the mass function) is changed, since it
depends on $\kappa A/(\alpha m)$ and is now located at $\bar{\rho
}=\bar{\rho }_{{\rm min},\tilde{\cal U}}\simeq 0.0415 \times \mP$. We
have chosen the initial conditions such that $\phi _{\rm ini}=10\times
\mP$ or $\bar{\phi }_{\rm ini}\simeq 14.142 \times \mP$ and such that
the evolution starts from the (new) bottom of the valley exactly,
namely $y_{\rm ini}=y_{{\rm min},\tilde{\cal U}}\simeq 1.165$ or
$\bar{\rho }_{\rm ini}=\bar{\rho }_{{\rm min},\tilde{\cal U}}\simeq
0.0415 \times \mP$. We notice that the above conclusions remain
unchanged: the fields follow the inflationary valley and join the
absolute minimum of the potential as it was the case in the previous
examples.

\par

Our conclusion is that the main features of the mutated chaotic
inflationary scenario seem to be robust either to modifications of the
initial conditions or to changes of the parameters of the model,
namely $A$ and/or $W_0$.

\subsection{Quantum stability}

Let us finish this section by a discussion on the quantum stability of
inflation along the valley where the inflaton rolls slowly. As can be
seen in Fig.~\ref{potrhophi2}, the valley is bordered on each side by
potential barriers. There is the infinite barrier at $\rho=0$ and a
finite height barrier located at $y_{{\rm max},\tilde{\cal U}}$
defined as the second root of Eq.~(\ref{deff}). Now this barrier
separates the inflation valley and the vacuum at infinity with
vanishing potential. There are two sources of instability of the
valley. The first one is the tunneling of the moduli field through the
barrier followed by the down roll towards $\rho=\infty$.  The second
one is stochastic evolution of the moduli field when its mass in the
valley is much less than the Hubble rate. In our case, the mass of the
moduli field is of the order of the Hubble rate implying that the
moduli field is not light and does not fluctuate like a stochastic
field with root mean square excursion $H/2\pi$.  The only possibility
for the moduli field to go through the barrier is tunneling. The
tunneling time is approximated by the Coleman-De Luccia
instanton~\cite{coleman}. In the thin-wall approximation where the
height of the barrier is large, \ie $\tilde U(y_{{\rm min},\tilde{\cal
U}})/\tilde U(y_{{\rm max},\tilde{\cal U}})\ll 1$, and the width of
the potential barrier is large in Planck units (see
Fig.~\ref{plotuv}), the tunneling time is given by
\begin{equation}
t_{\rm Decay}\simeq t_{_{\rm Pl}}{\rm e}^{24\mP ^4\pi^2/V_0}\,
,
\end{equation}
where $V_0$ is the potential in the valley and $t_{_{\rm Pl}}$ the
Planck time. Using $V_0\simeq 3\mP^2 H^2$ and $H\approx 10^{-6} \mP$,
one finds that the decay time is exponentially longer than the Planck
time. For all practical purposes the valley is quantum stable.  At the
end of the evolution along the valley, the moduli becomes sensitive to
the existence of the global minimum of the potential and rolls down
the potential towards the global minimum.

\section{Mutated chaotic inflation and KKLT stabilization}

In the previous section we have introduced a model of inflation with
moduli stabilization and mutated chaotic inflation.  Unfortunately,
the vacuum energy at the end of inflation becomes negative. We have
compensated this negative energy by introducing a constant and
positive energy of unknown origin. Here, we will combine a string
inspired stabilization mechanism (the KKLT stabilization mechanism)
with mutated chaotic inflation. This is obtained by introducing an
explicit moduli dependent potential which lifts the vacuum energy
towards positive values.

\subsection{Lifting AdS to dS}

There are two equivalent ways of lifting the potential energy for the
moduli. The first one comes from the D3/D7 system that we have already
studied at the beginning of the previous section. Instead of studying
the potential~(\ref{pottotalpm}) in the vicinity of the minimum, which
leads to Eq.~(\ref{potmodelpm}), one can focus on the regime where the
waterfall fields vanish $\phi^\pm=0$. Then, Eq.~(\ref{pottotalpm})
leads to
\begin{equation}
V=\frac{2g^2\zeta ^4}{\kappa^{3/2}\left(\rho+\rho ^{\dagger
  }\right)^3}+ \tilde{\cal V}(\rho)\, .
\end{equation}
This is a KKLT potential with a correction term $\propto 1/\rho ^3$,
with a positive minimum provided $\zeta $ is chosen appropriately.
However, we have already used the potential~(\ref{pottotalpm}) to give
a mass to the inflaton and, as discussed before, this was in another
regime. Therefore, if we want to preserve this mechanism and stabilize
the moduli by the KKLT method, the term $\propto 1/\rho ^3$ must have
another origin that we now discuss.

\par

Assume that the model possesses a $U(1)$ gauge field with a
Fayet--Iliopoulos term. The K\"ahler potential of the moduli is
modified and becomes
\begin{equation}
K=-\frac{3}{\kappa } \ln \left[\kappa^{1/2}\left(\rho+\rho ^{\dagger
}\right) + \xi V \right]\, .
\end{equation}
where $V$ is the vector superfield and $\xi $ is the Fayet--Iliopoulos
term. This could be due to an anomalous symmetry cured by the
Green-Schwarz mechanism. Moreover we assume that the gauge coupling
function reads
\begin{equation}
f(\rho )= \frac{\kappa^{1/2}\rho}{\tilde g^2}\, ,
\end{equation}
where $\tilde g$ is a constant. Assuming that no field is charged
under this $U(1)$ symmetry, the D-term associated to this gauge
symmetry is
\begin{equation}
V_{_{\rm D}}= \frac{D}{\rho^3}\, ,
\end{equation}
where $D=\tilde{g}^2\xi ^2/16$~\cite{Burgess2}. The total potential
that we obtain is the potential of Eq.~(\ref{pottotalpm}) plus
$V_{_{\rm D}}$. Then, using our usual mechanism to give a mass to the
inflaton and following the same step as before we find that the new
potential reads
\begin{equation}
\label{VKKLT}
V\left(\rho ,\phi \right)=\tilde{{\cal V}}(\rho )+\frac{D}{\rho ^3}
+\tilde{{\cal U}}(\rho )\phi ^2\, .
\end{equation}
We see that this only amounts to have a new offset function
$\hat{{\cal V}}(\rho ) \equiv \tilde{{\cal V}}(\rho )+D/\rho ^3$.
Hence, all the results obtained before on the mass function
$\tilde{\cal U}$ are still valid.

\subsection{KKLT stabilization and mutated chaotic inflation}

Let us now discuss how the parameter $D$ can be fixed. We will not
present a complete analysis of the parameter space as such an analysis
is complicated. However, we will demonstrate that the KKLT mechanism
also works in the case under consideration. The parameter $D$ must be
chosen such that, at the absolute minimum of the potential, the
potential exactly vanishes (or is equal to the value of the vacuum
energy today. Since this one is tiny and since we only study the
inflationary era, we will just assume that the minimum is zero). This
means that we have to solve simultaneously the equations
\begin{eqnarray}
& & y^2_{{\rm min},\hat{\cal V}}+\frac72 y_{{\rm min},\hat{\cal V}}+3
  = \frac{3W_0}{2A}{\rm e}^{y_{{\rm min},\hat{\cal V}}} \left(y_{{\rm
  min},\hat{\cal V}} +2\right)
\nonumber \\
& & -\frac{9D\kappa ^{1/2}}{A^2}\frac{{\rm
  e}^{2y_{{\rm min},\hat{\cal V}}}}{y_{{\rm min},\hat{\cal V}} }\, ,
  \\ & & \frac{{\rm e}^{-y_{{\rm min},\hat{\cal V}} }}{2y_{{\rm
  min},\hat{\cal V}} }\left[\frac{{\rm e}^{-y_{{\rm min},\hat{\cal V}}
  }}{3}-\frac{1}{y_{{\rm min},\hat{\cal V}}} \left(\frac{W_0}{A}-{\rm
  e}^{-y_{{\rm min},\hat{\cal V}}}\right)\right]
\nonumber \\
& & +\frac{\kappa
  ^{1/2}D}{A^2y_{{\rm min},\hat{\cal V}} ^3} =0\, .
\end{eqnarray}
The first condition is a condition on the derivative of the new offset
function and is similar to Eq.~(\ref{condder}) while the second
equation is nothing but the condition that the potential is zero at
the minimum. We see that the relevant new parameter is in fact the
dimensionless quantity $\kappa ^{1/2}D/A^2$.

\par

\begin{figure*}
\includegraphics[width=.95\textwidth,height=.65\textwidth]{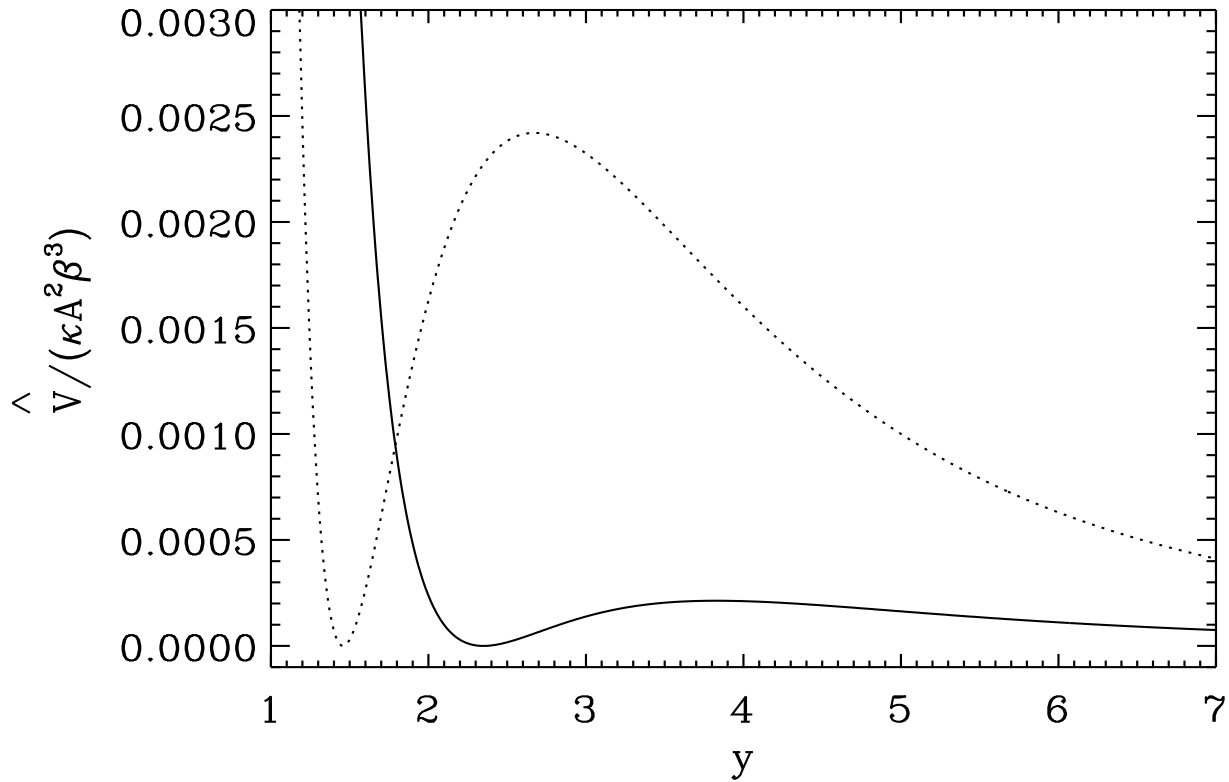}
\caption{Offset function obtained with the help of the KKLT
  stabilization mechanism. The solid line corresponds to the case
  $\alpha =\sqrt{2}$, $\beta =1$, $m=10^{-6}\mP$, $W_0/A=0.4111$ and
  $\kappa A/(\alpha m)=1.35135$ and gives $\kappa ^{1/2}D/A^2\simeq
  0.027$ and $y_{{\rm min},\hat{\cal V}}\simeq 2.349$. The dotted line
  is obtained with the following set of parameters: $\alpha
  =\sqrt{2}$, $\beta =1$, $m=10^{-6}\mP$, $W_0/A=1.2$ and $\kappa
  A/(\alpha m)=1.35135$, $\kappa ^{1/2}D/A^2\simeq 0.145$ and the
  minimum is located at $y_{{\rm min},\hat{\cal V}}\simeq 1.458$.}
\label{plotvKKL}
\end{figure*}

Let us first consider the case where the parameters are $\alpha
=\sqrt{2}$, $\beta =1$, $m=10^{-6}\mP$, $W_0/A=0.4111$ and $\kappa
A/(\alpha m)=1.35135$, \ie the case envisaged before. Then, a solution
to the two above equations can be found and reads: $\kappa
^{1/2}D/A^2\simeq 0.027$ and $y_{{\rm min},\hat{\cal V}}\simeq
2.349$. The corresponding offset function is represented in
Fig.~\ref{plotvKKL} (solid line). One can check that there is indeed a
minimum and that, at the minimum, the offset function vanishes. Let us
now consider another case, namely $\alpha =\sqrt{2}$, $\beta =1$,
$m=10^{-6}\mP$, $W_0/A=1.2$ and $\kappa A/(\alpha m)=1.35135$, \ie the
value of the ratio $W_0/A$ is now different and, most importantly,
greater than one. A solution can also be obtained and is: $\kappa
^{1/2}D/A^2\simeq 0.145$ and $y_{{\rm min},\hat{\cal V}}\simeq
1.458$. The corresponding offset function is plotted in
Fig.~\ref{plotvKKL} (dotted line) and we check again that there is a
vanishing minimum where the moduli can be stabilized. This case is in
fact more interesting than the first one. Indeed, one sees that the
ratio $W_0/A>1$ violates the bound established previously. Therefore,
the KKLT mechanism allows us to find a vanishing minimum to the
potential even if $W_0/A>1$. In other words, the constraint $W_0/A<1$
is relaxed.

\begin{figure*}
\includegraphics[width=.95\textwidth,height=.65\textwidth]{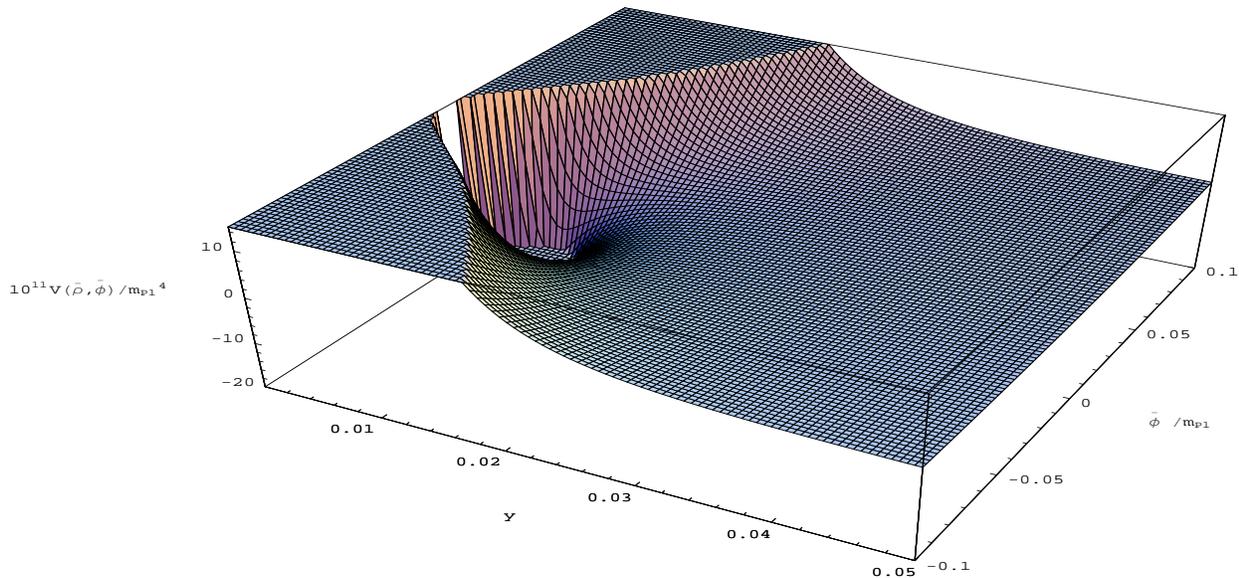}
\caption{Potential without the KKLT term $D/\rho ^3$ for the
  parameters: $\alpha =\sqrt{2}$, $\beta =1$, $m=10^{-6}\mP$,
  $W_0/A=1.2$ and $\kappa A/(\alpha m)=1.35135$. Since $W_0/A>1$ the
  offset function has no minimum around $\bar{\phi }\simeq 0$, hence
  the ``hole'' that can be seen in this region. Let us also notice
  that the potential is plotted versus $y$ rather than versus
  $\bar{\rho }$ as it was the case in Fig.~\ref{potrhophi2}.}
\label{potnonKKL}
\end{figure*}

\begin{figure*}
\includegraphics[width=.95\textwidth,height=.65\textwidth]{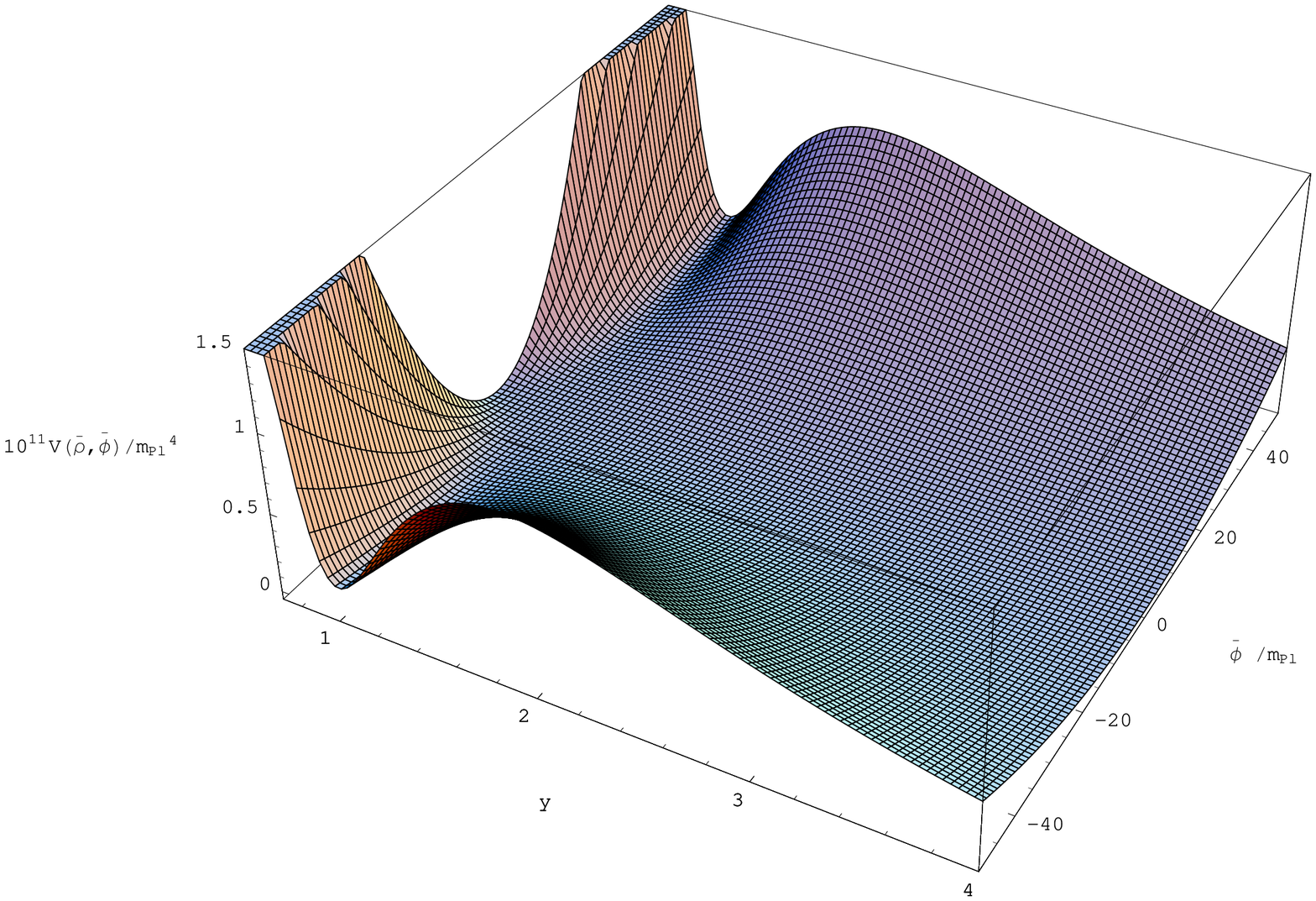}
\caption{Potential with the KKLT term $D/\rho ^3$ for the parameters:
  $\alpha =\sqrt{2}$, $\beta =1$, $m=10^{-6}\mP$, $W_0/A=1.2$, $\kappa
  A/(\alpha m)=1.35135$ and $\kappa ^{1/2}D/A^2\simeq 0.145$ (\ie same
  set of parameters than in Fig.~\ref{potnonKKL} but with a
  non--vanishing $D$). The fact that the moduli has been stabilized is
  obvious.}
\label{potKKL}
\end{figure*}

The above property is illustrated in Fig.~\ref{potnonKKL} where we
have plotted the potential $V(\bar{\rho} ,\bar{\phi} )$ for the
following set of parameters: $\alpha =\sqrt{2}$, $\beta =1$,
$m=10^{-6}\mP$, $W_0/A=1.2$, $\kappa A/(\alpha m)=1.35135$ and $\kappa
^{1/2}D/A^2=0$. As already discussed above, the moduli cannot be
stabilized in this case because $W_0/A>1$. The ``hole'' that can be
seen in this figure represents the region of instability, \ie the
region where the offset function goes to $-\infty$. In
Fig.~\ref{potKKL}, we have added the KKLT term $D/\rho ^3$. The value
of the parameters are $\alpha =\sqrt{2}$, $\beta =1$, $m=10^{-6}\mP$,
$W_0/A=1.2$, $\kappa A/(\alpha m)=1.35135$ and $\kappa
^{1/2}D/A^2\simeq 0.145$, \ie as for the dotted line in
Fig.~\ref{plotvKKL}. The ``hole'' has now disappeared and the shape of
the potential is very reminiscent to that of the potential studied in
the previous subsection, see Fig.~\ref{potrhophi2}. In particular, it
is clear that there is a new valley of stability.

\par

\begin{figure*}
\includegraphics[width=.95\textwidth,height=.65\textwidth]{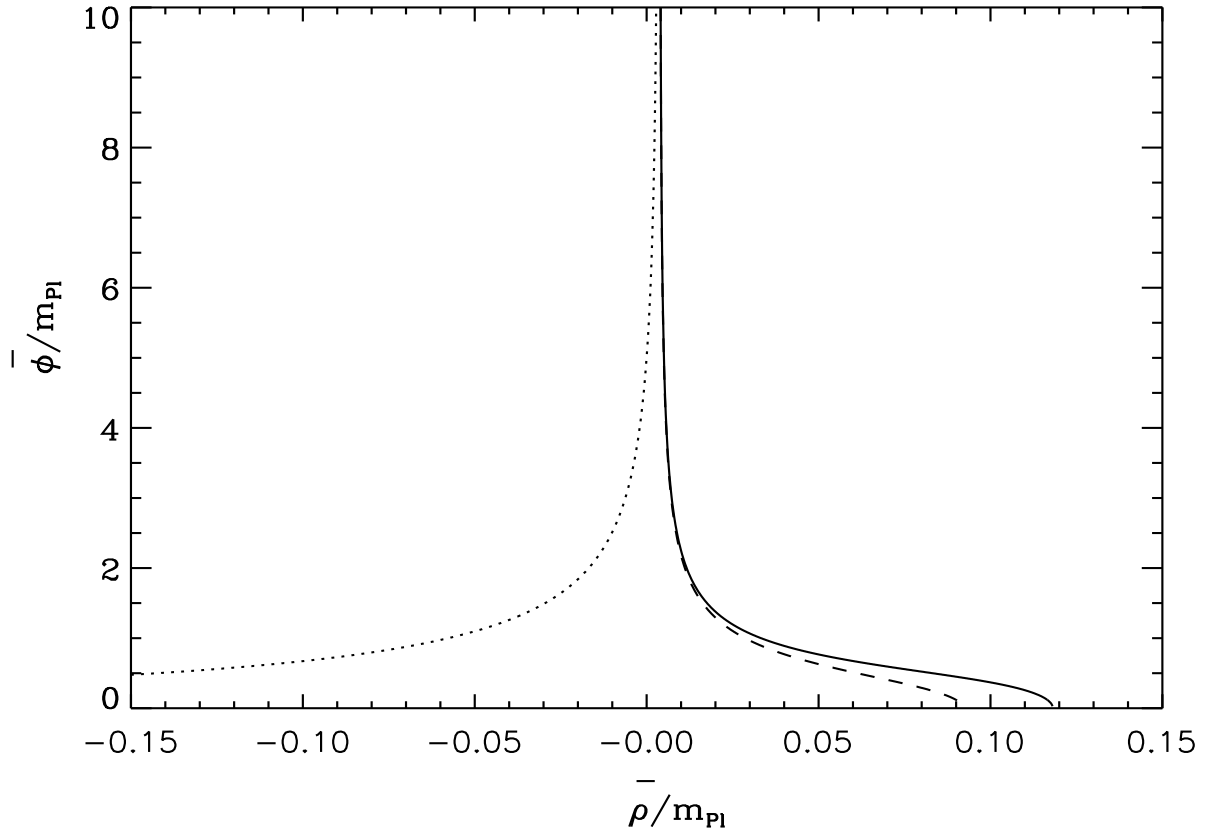}
\caption{Trajectories of the valley of stability in the plan
$(\bar{\rho} ,\bar{\phi} )$ (the valley is seen from above). The solid
line corresponds to the following parameters $\alpha =\sqrt{2}$,
$\beta =1$, $m=10^{-6}\mP$, $\kappa A/(\alpha m)=1.35135$
$W_0/A=0.41111$, $\kappa ^{1/2}D/A^2=0$. There is no KKLT term in the
potential and $W_0/A<1$. Therefore this is nothing but our
``standard'' case already displayed in Fig.~\ref{courbmin}. The dotted
line represents the valley in the case where $\alpha =\sqrt{2}$,
$\beta =1$, $m=10^{-6}\mP$, $\kappa A/(\alpha m)=1.35135$ $W_0/A=1.2$,
$\kappa ^{1/2}D/A^2=0$, \ie the KKLT term is still absent but the
value of $W_0/A$ is now changed and is such that $W_0/A>1$. In this
case the modulus is not stabilized and one sees the valley escaping to
infinity. This case corresponds to Fig.~\ref{potnonKKL}. The fact that
the valley goes to infinity is another manifestation of the ``hole''
that can be seen in Fig.~\ref{potnonKKL}. Finally, the dashed line
represents the valley in the case where $\alpha =\sqrt{2}$, $\beta
=1$, $m=10^{-6}\mP$, $\kappa A/(\alpha m)=1.35135$ $W_0/A=1.2$ and
$\kappa ^{1/2}D/A^2=0.145$. This is the same case as before except
that the KKLT is now present which allows us to have $W_0/A>1$. We see
that the valley is now very similar to the one obtained before by
simply adding a cosmological constant to the potential.}
\label{courbminKKL}
\end{figure*}

It is now interesting to study the valley in more details. Since
adding the term $D/\rho ^3$ just amounts to modifying the form of the
offset function, the analytical calculations which lead to the
equation of the valley are very similar to those which resulted in
Eq.~(\ref{inftrajec}). Straightforward manipulations yields
\begin{widetext}
\begin{eqnarray}
\kappa \phi ^2_{\rm valley}\left(y\right) &=& \frac43\left(\frac{\kappa
A}{\alpha m}\right)^2 y{\rm
e}^{-2y}\left[y^2+\frac72y+3-\frac32\frac{W_0}{A}{\rm
e}^y\left(y+2\right)+\frac{9\kappa ^{1/2}D}{A^2}\frac{{\rm e}^{2y}}{y}
\right]
\\ \nonumber
& & \times \left[-\frac32+2\left(\frac{\kappa A}{\alpha
m}\right)y{\rm e}^{-y}+\left(\frac{\kappa A}{\alpha m}\right)y^2{\rm
e}^{-y} \right]^{-1}\, .
\end{eqnarray}
As expected the only difference is the presence of the term
proportional to $D$ at the numerator of the above equation. It is also
interesting to give the trajectory expressed in terms of canonically
normalized fields, \ie the equivalent of Eq.~(\ref{inftrajec}). It can
be expressed as
\begin{eqnarray}
\label{inftrajecKKL}
\kappa \bar{\phi }^2_{\rm valley}\left(\bar{\rho }\right)&=&\frac83
\left(\frac{\kappa A}{\alpha m}\right)^2 \beta
{\rm e}^{\sqrt{\frac23} \kappa ^{1/2}\bar{\rho }}
\exp\left(-2\beta {\rm e}^{\sqrt{\frac23} \kappa ^{1/2}\bar{\rho }}\right)
\biggl[\beta ^2{\rm e}^{2\sqrt{\frac23} \kappa ^{1/2}\bar{\rho }}
+\frac72\beta {\rm e}^{\sqrt{\frac23} \kappa ^{1/2}\bar{\rho }}+3
\nonumber \\
& -& \frac32\frac{W_0}{A}
\exp\left(\beta {\rm e}^{\sqrt{\frac23} \kappa ^{1/2}\bar{\rho }}\right)
\left(\beta {\rm e}^{\sqrt{\frac23} \kappa ^{1/2}\bar{\rho }}
+2\right)+\frac{9\kappa ^{1/2}D}{A^2\beta }{\rm
  e}^{-\sqrt{\frac23} \kappa ^{1/2}\bar{\rho }} \exp\left(2\beta {\rm
  e}^{\sqrt{\frac23}
\kappa ^{1/2}\bar{\rho }}\right) \biggr]
\nonumber \\
& \times & \biggl[-\frac32+2\left(\frac{\kappa
A}{\alpha m}\right)\beta {\rm e}^{\sqrt{\frac23} \kappa ^{1/2}\bar{\rho }}
\exp\left(-\beta {\rm e}^{\sqrt{\frac23} \kappa ^{1/2}\bar{\rho
}}\right)+\left(\frac{\kappa A}{\alpha
m}\right)\beta ^2{\rm e}^{2\sqrt{\frac23} \kappa ^{1/2}\bar{\rho }}
\exp\left(-\beta {\rm e}^{\sqrt{\frac23} \kappa ^{1/2}\bar{\rho }}\right)
\biggr]^{-1}\, .
\end{eqnarray}
\end{widetext}
The valley is represented in Fig.~\ref{courbminKKL} and compared to
the valley obtained previously without the term $D/\rho ^3$. As is
clear form the figure, the two trajectories have the same features.

\subsection{Numerical Results with KKLT stabilization}

\begin{figure*}
\includegraphics[width=8.8cm,height=7.5cm]{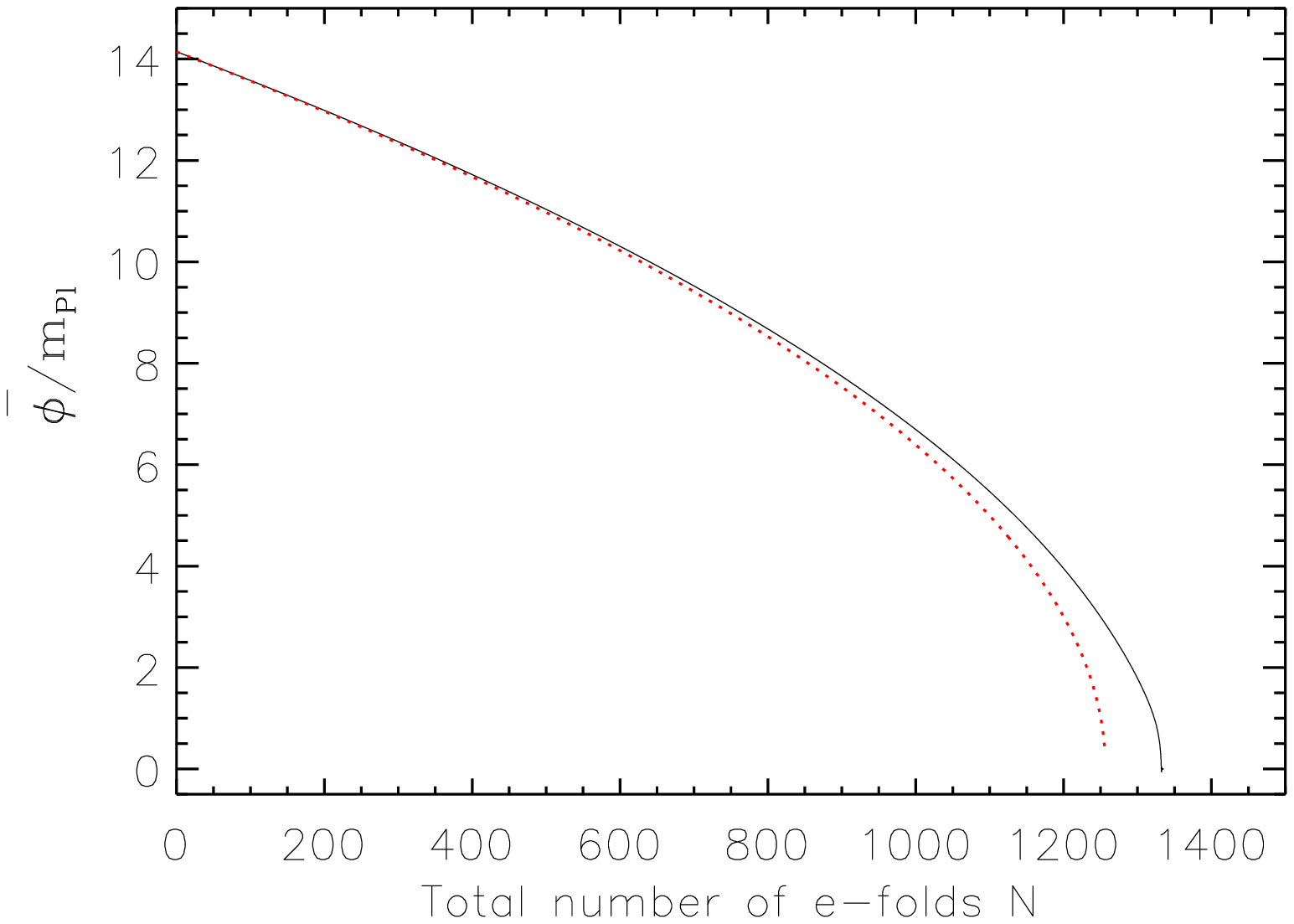}
\includegraphics[width=8.8cm,height=7.5cm]{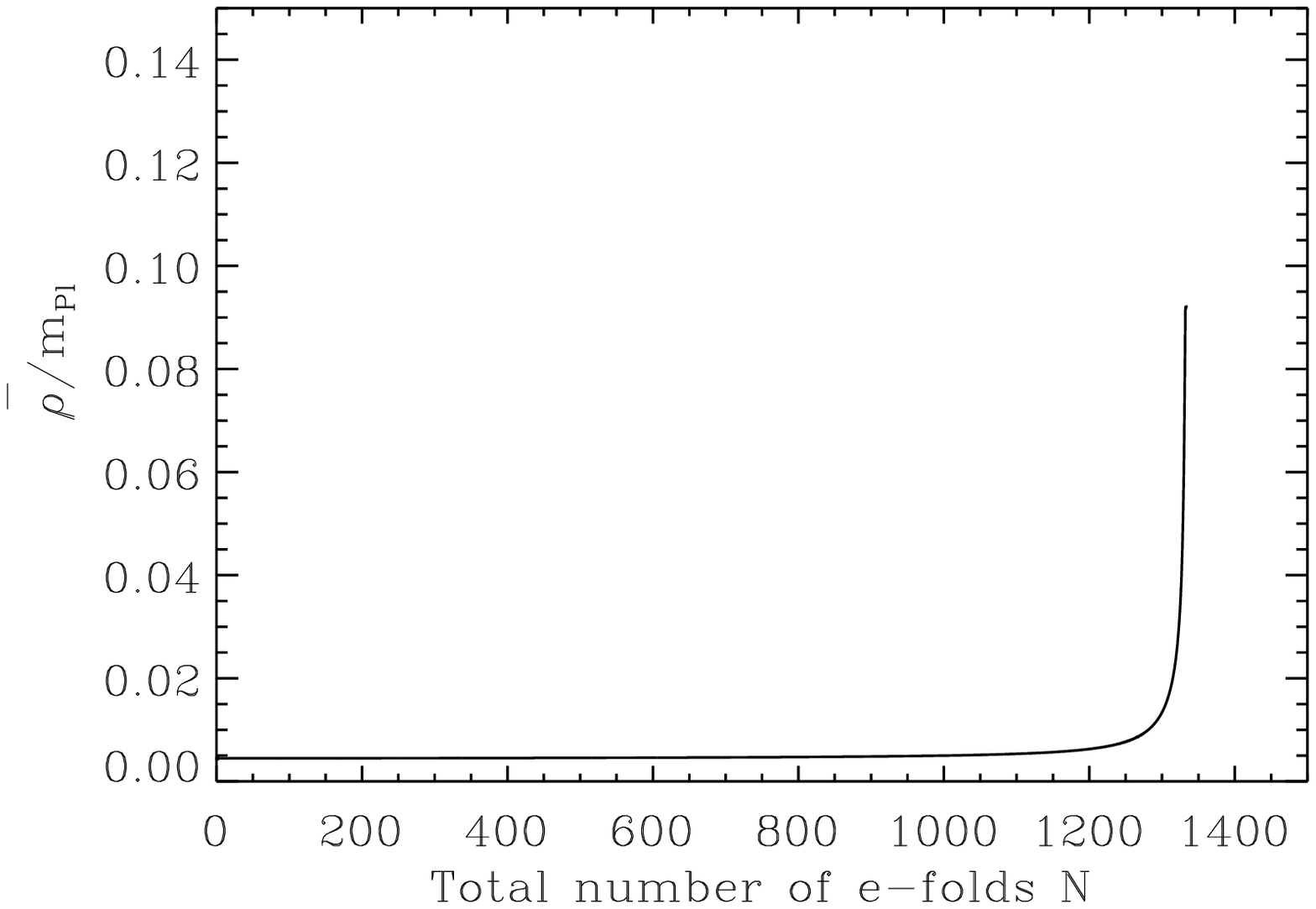}\\
\includegraphics[width=8.8cm,height=7.5cm]{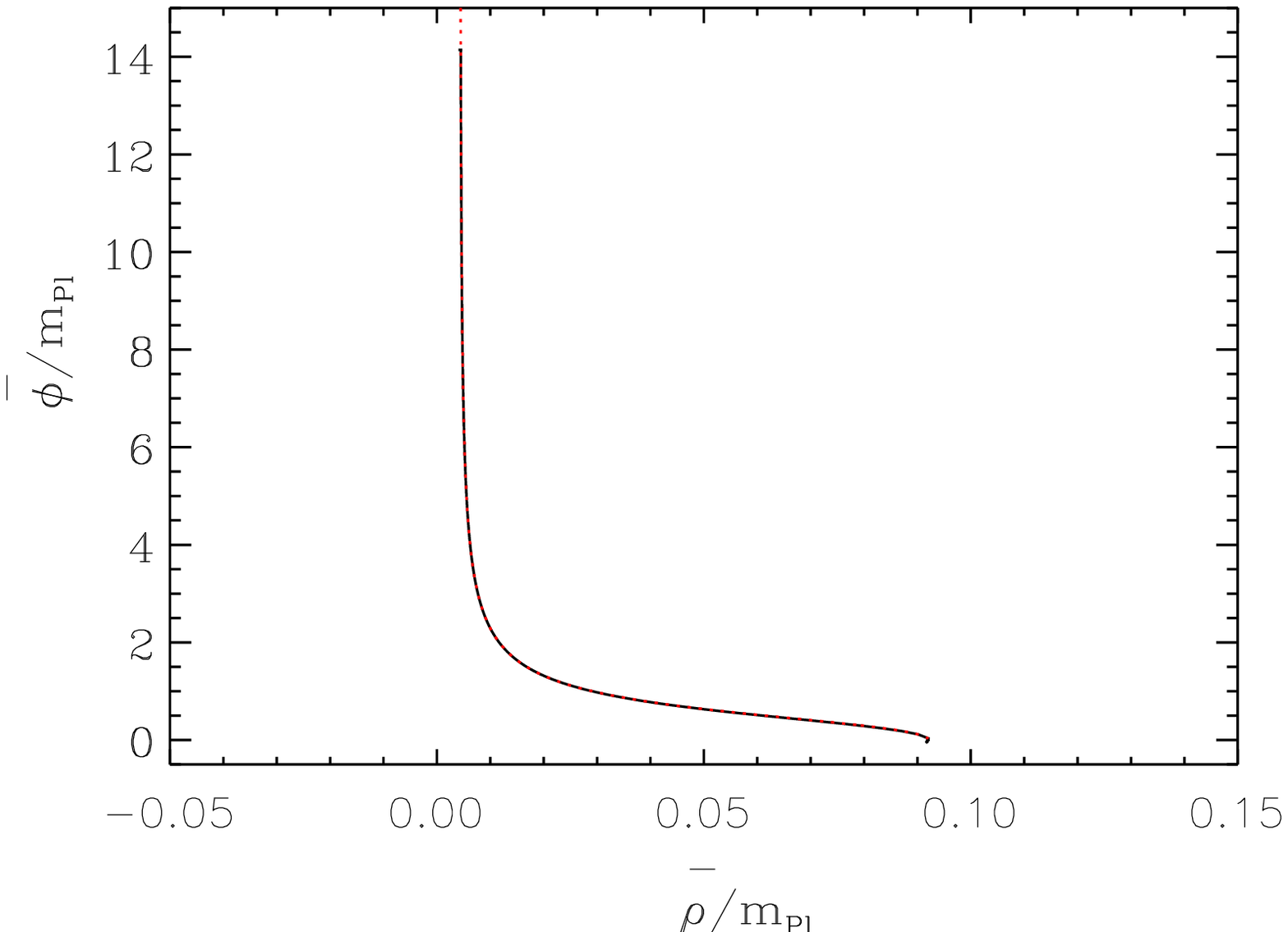}
\includegraphics[width=8.8cm,height=7.5cm]{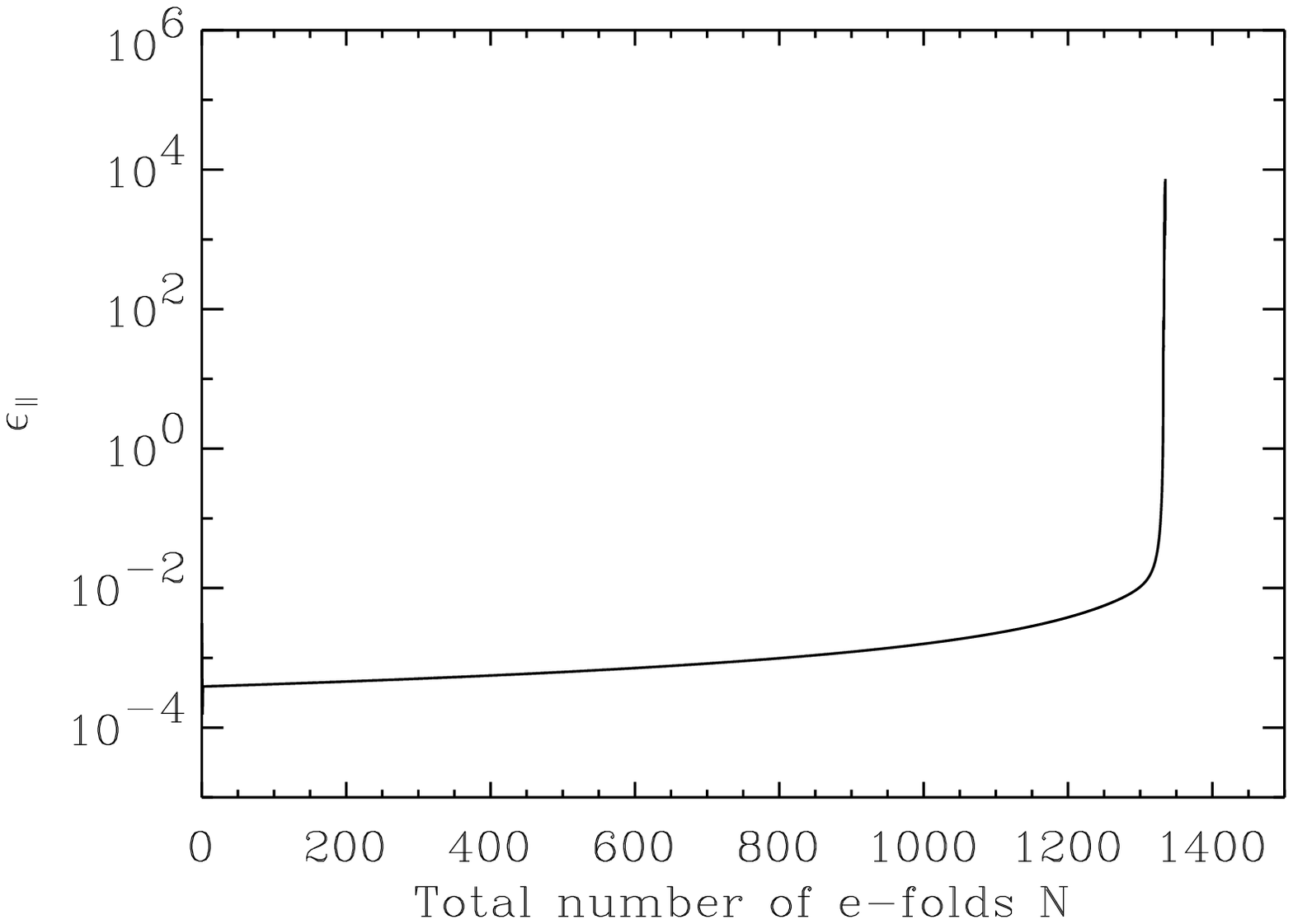}\\
\includegraphics[width=8.8cm,height=7.5cm]{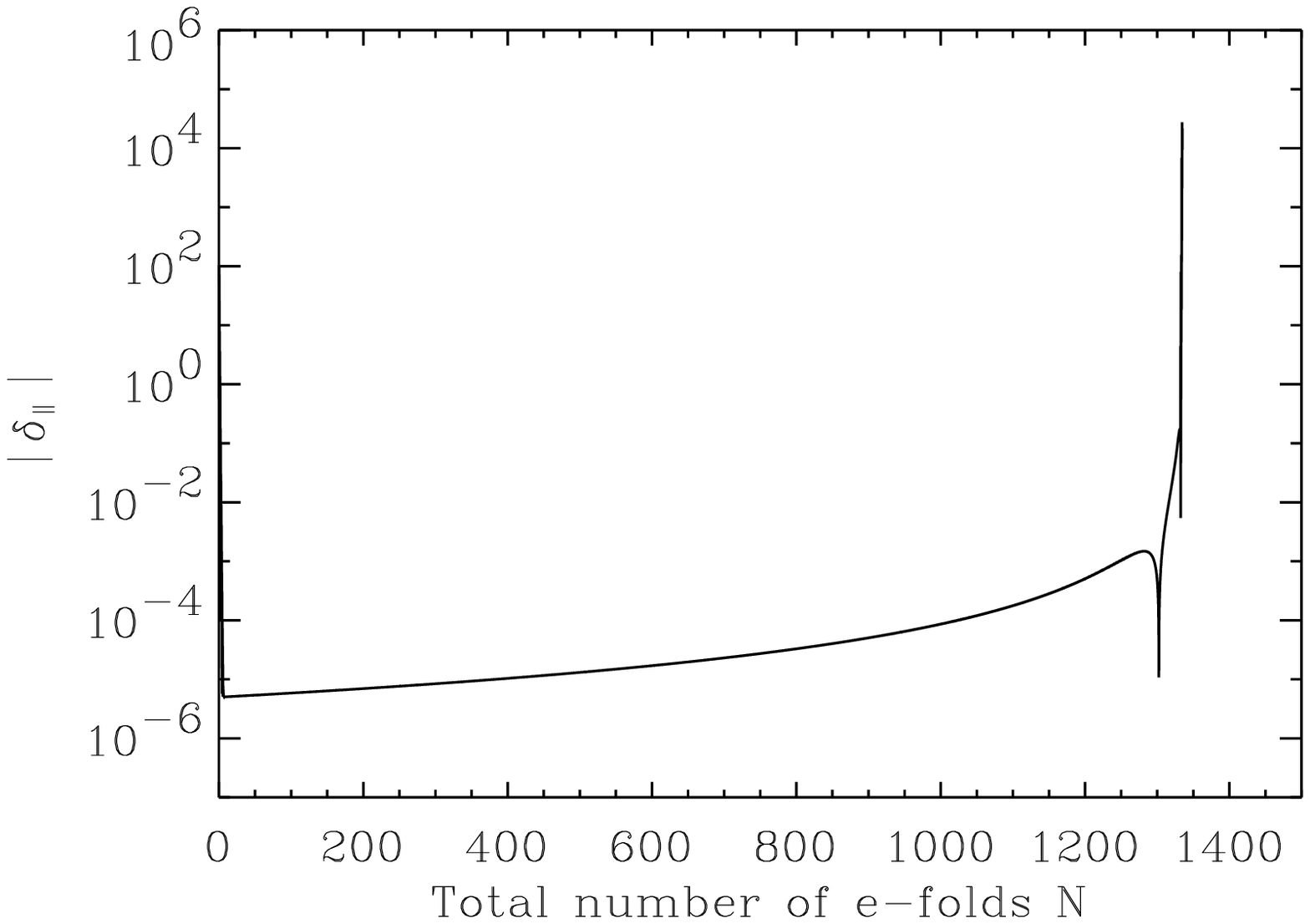}
\includegraphics[width=8.8cm,height=7.5cm]{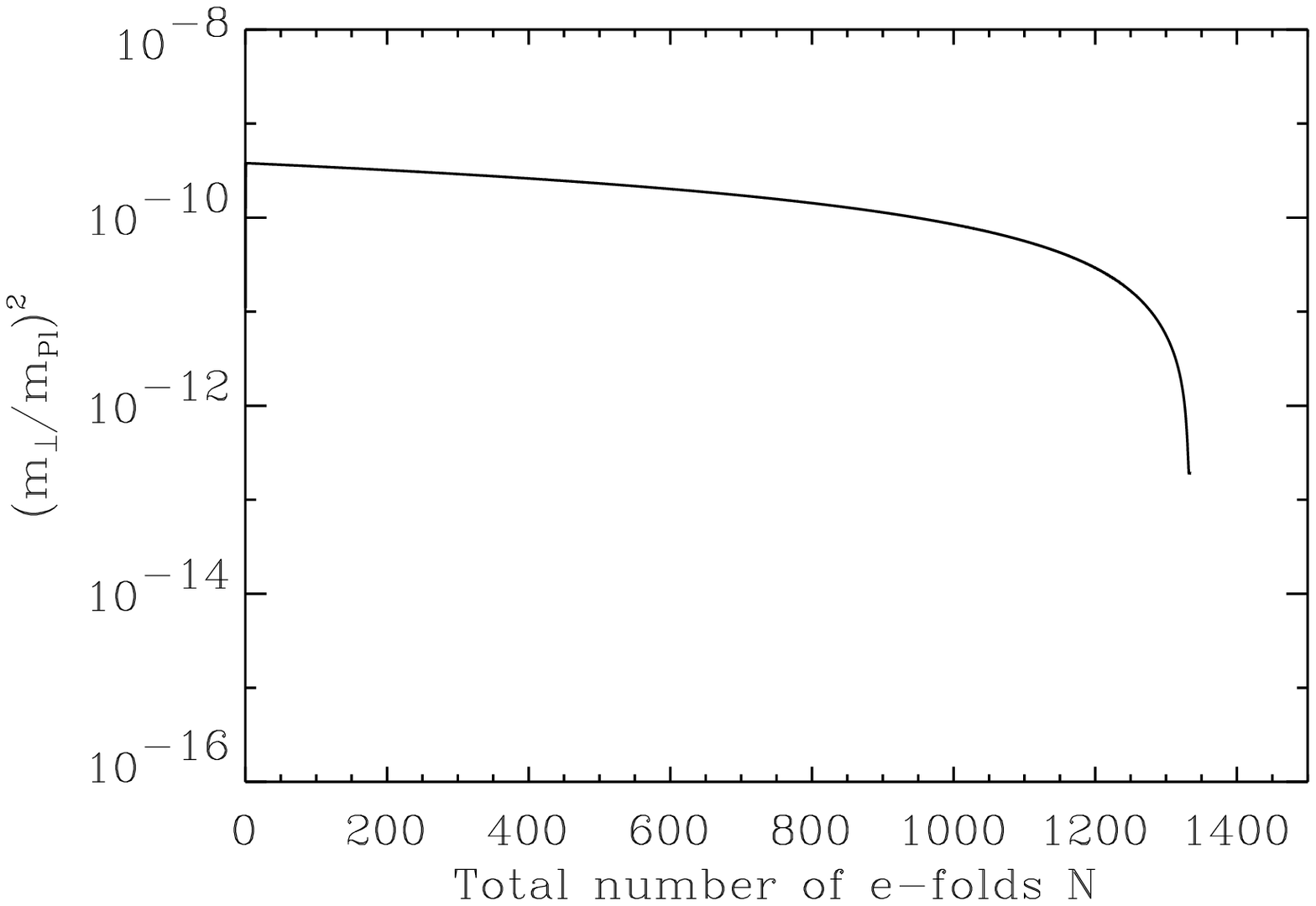}
\caption{Same as Figs.~\ref{case3} and~\ref{case10} but with the KKLT
  mechanism taking into account. The parameters are $\alpha
  =\sqrt{2}$, $\beta =1$, $m=10^{-6}\mP$, $\kappa A/(\alpha
  m)=1.35135$, $W_0/A=1.2$ and $\kappa ^{1/2}D/A^2\simeq 0.145$.  This
  gives $y_{{\rm min},\tilde{\cal U}}\simeq 1.067$ or $\bar{\rho
  }_{{\rm min},\tilde{\cal U}}\simeq 0.004 \times \mP$. The absolute
  minimum of the potential is located at $\phi =0$, $y=y_{{\rm
  min},\tilde{\cal V}}\simeq 1.457$ or $\bar{\phi }=0$, $\bar{\rho
  }=\bar{\rho }_{{\rm min},\tilde{\cal V}}\simeq 0.092\times \mP$. The
  initial conditions are $\phi _{\rm ini}=10\times \mP$ or $\bar{\phi
  }_{\rm ini}\simeq 14.142 \times \mP$ and $y_{\rm ini}=y_{{\rm
  min},\tilde{\cal U}}\simeq 1.067$ or $\bar{\rho }_{\rm
  ini}=\bar{\rho }_{{\rm min},\tilde{\cal U}}\simeq 0.004 \times \mP$,
  \ie at the bottom of the valley exactly.}
\label{case10KKLT}
\end{figure*}

Our next step is to study the potential given by Eq.~(\ref{VKKLT})
numerically. The results are displayed in Fig.~\ref{case10KKLT}. This
figure should be compared with Figs.~\ref{case3} and~\ref{case10}
where the same quantities, \ie $\bar{\phi }(N)$, $\bar{\rho }(N)$,
$\bar{\phi }(\bar{\rho })$, $\epsilon _{\parallel}$, $\delta
_{\parallel}$ and $m^2_{\perp}$, have also been displayed. The main
conclusion that can be drawn from these plots is that the inflationary
trajectory obtained in the case where the KKLT mechanism is
responsible for the stabilization of the moduli is very similar to the
trajectory obtained before (simply by adding a cosmological constant
to renormalize the true vacuum). Therefore, all the properties that
were discussed in the previous section are still valid in the present
context.

Let us now discuss in more details the inflationary scenario proposed
in this article. Previously, we have studied the properties of the
inflationary background only. However, it is clear that, if one wants
to investigate all the consequences of the model, one must compare its
predictions to the high accuracy cosmological data presently available
in cosmology, in particular to the CMB data which are likely to carry
an imprint from inflation. This requires the calculation of the
cosmological perturbations and, in the present context, this is not a
trivial task. The difficulty comes from the fact that we have more
than one scalar field and hence the standard slow-roll single field
result is not applicable here. In particular, we now have
non-adiabatic perturbations. The non-adiabatic component is expected
to be particularly important if the scales of astrophysical interest
today left the Hubble radius in the ``curved part'' of the
inflationary trajectory~\cite{wr}. This is the case, for instance, in
Fig.~\ref{case10KKLT} where $60$ e-folds before the end of inflation
corresponds to a regime where the trajectory has already bent. To be
more precise, it is quite easy to determine the spectral indices at
Hubble crossing during inflation. They are simply given by the
well-known equations~\cite{wr}
\begin{eqnarray}
n_{_{\rm S}} &=& -4\epsilon _{\parallel}+2\delta _{\parallel}\, ,
\\
n_{_{\rm T}} &=& -2\epsilon _{\parallel}\, .
\end{eqnarray}
In these equations, $n_{_{\rm S}}$ is the spectral index of the
adiabatic part of the density perturbations while $n_{_{\rm T}}$ is
the tensor spectral index. One could have also given the spectral
index of the isocurvature power spectrum but we do not need it here
and it can be found in Ref.~\cite{wr}. We see that these formulae are
a simple generalization of the one-field equations, $\epsilon
_{\bar{\phi}}$ and $\delta _{\bar{\phi}}$ just being replaced by
$\epsilon _{\parallel}$ and $\delta _{\parallel}$. We have computed
$n_{_{\rm S}}$ and $n_{_{\rm T}}$ for some cases envisaged before, see
Tab.~\ref{indices}.
\begin{table*}
\begin{tabular}{c c c c c c c c c c c}
& $\displaystyle\frac{W_0}{A}$ & $\displaystyle \frac{\kappa A}{\alpha
    m}$ & $\displaystyle \frac{\kappa ^{1/2}D}{A^2}$ & $\bar{\rho
    }_{{\rm min},\tilde{\cal U}}$ & $\bar{\rho }_{{\rm
    min},\tilde{\cal V}}$ & $\bar{\rho }_{\rm ini}$ & $\bar{\phi
    }_{\rm ini}$ & $N_{_{\rm T}}$ & $n_{_{\rm S}}$ & $n_{_{\rm T}}$ \\
    \\ \hline \hline \\ & $0.411$ & $1.3513$ & $0$ & $0.004\times \mP$
    & $0.118\times \mP$ & $0.004\times \mP$ & $14.142\times \mP$ &
    $1356$ & $0.979$ & $-0.012$ \\ \\ & $0.411$ & $1.3513$ & $0$ &
    $0.004\times \mP$ & $0.118\times \mP$ & $0.004\times \mP$ &
    $4.242\times \mP$ & $161$ & $0.979$ & $-0.012$ \\ \\ & $1.2$ &
    $1.3513$ & $0.145$ & $0.004\times \mP$ & $0.092\times \mP$ &
    $0.004\times \mP$ & $14.142\times \mP$ & $1331$ & $0.974$ &
    $-0.0139$ \\ \\ & $0.6$ & $1.3513$ & $0$ & $0.004\times \mP$ &
    $0.008\times \mP$ & $0.004\times \mP$ & $14.142\times \mP$ &
    $1258$ & $0.966$ & $-0.0167$ \\ \\ & $0.411$ & $1.330$ & $0$ &
    $0.017\times \mP$ & $0.118\times \mP$ & $0.017\times \mP$ &
    $14.142\times \mP$ & $1278$ & $0.968$ & $-0.0165$ \\ \\ \hline
\end{tabular}
\caption{Spectral indices of the adiabatic scalar and tensor power
  spectra for different initial conditions and/or parameters $W_0/A$,
  $\kappa A/(\alpha m)$ and $\kappa ^{1/2}D/A^2$. We also give the the
  minimums of the functions $\tilde{\cal V}$ and $\tilde{\cal U}$ as
  well as the initial conditions and the total number of e-folds
  predicted by the model.}
\label{indices}
\end{table*}
These numbers must be compared with those obtained in the case of
single field chaotic inflation (for a potential quadratic in the
field),
\begin{equation}
n_{_{\rm S}} \simeq 0.967\, ,\quad n_{_{\rm T}} \simeq -0.0165\, .
\end{equation}
We see they are quite similar although not identical. In fact, the
values of the spectral indices depend on the details of the
inflationary trajectory. More precisely, what is important is how the
trajectory is curved $N_*$ e-folds before the end of inflation.  For
instance, let us compare the cases corresponding to the two first
columns in Tab.~\ref{indices} to the cases corresponding to the two
last columns. One can check that, for the two first cases, the
inflationary trajectory is ``more curved'', \ie deviates from a
straight line more strongly, than for the two last cases. As a
consequence, the spectral indices obtained for the two first cases
differ more from those obtained in the standard chaotic model than the
spectral indices calculated in the two last cases do. Let us also
remark that one could have expected a smaller difference in the case
where the initial conditions are such that the fields start deep in
the valley (\ie, for instance, spectral indices closer to chaotic
inflation in the case where $\bar{\phi }_{\rm ini}=14.142 \times \mP$
than for the case $\bar{\phi }_{\rm ini}=4.242 \times \mP$). However,
what really matters is the value of the slow-roll parameters $N_*$
e-folds before the end of inflation and not at the beginning of
inflation. Therefore, the previous reasoning does not work in our
case, as confirmed by the numbers in Tab.~\ref{indices}.  Finally, one
notices that the differences observed are quite small and, although it
seems easy to interpret them as we have just done above, we have not
been able to find a simple criterion which would allow us to predict,
from the values of the free parameters $W_0$, $A$ and $D$, how far
from the fiducial model the spectral indices will be. It seems that
this really depends on the fine structure of the valley near the
Hubble scale exit.

\par

As discussed in Ref.~\cite{wr}, the point is that the previous indices
are not those that are observable. The reason is that the evolution of
the perturbations after the Hubble radius exit during inflation is non
trivial in presence of isocurvature perturbations. Technically, this
is because the standard conserved quantity (on super-Hubble scales)
$\zeta =-{\cal R}$ is sourced by the non-adiabatic pressure $\delta
p_{\rm nad}$. The curvature and entropy perturbations evolve according
to the equation
\begin{equation}
\begin{pmatrix}
{\cal R} \cr {\cal S}
\end{pmatrix}=
\begin{pmatrix}
1 & T_{{\cal R}{\cal S}} \cr
0 & T_{{\cal S}{\cal S}}
\end{pmatrix}
\begin{pmatrix}
{\cal R} \cr {\cal S}
\end{pmatrix}
_{\rm exit}\, ,
\end{equation}
where the subscript ``exit'' means the corresponding quantities
evaluated at the exit of the Hubble radius during inflation. They
correspond to the quantities given in Tab.~\ref{indices}. Then, the
next step consists in defining the correlation angle by~\cite{wr}
\begin{equation}
\cos \Delta \simeq \frac{T_{{\cal R}{\cal S}}}{\sqrt{1+T_{{\cal
    R}{\cal S}}^2}}\, ,
\end{equation}
which appears in the final expressions of the observable spectral
indices (these expressions can be found in Ref.~\cite{wr}). As shown
in Ref.~\cite{wr}, the correlation angle is the only quantity needed
in order to calculate the observable spectral indices from the
directional slow-roll parameters introduced before. Unfortunately,
this quantity is not easy to obtain. In the present context, this
would require to numerically integrate the equations governing the
evolution of the cosmological perturbations (and not only the
equations governing the evolution of the background as done
before). This is clearly beyond the scope of the present
article. However, if $\Delta $ is not too far from $\pi /2$, then the
estimates given in Tab.~\ref{indices} are sufficient to demonstrate
that the model seems to be presently compatible with the CMB data. We
hope to address the question of determining the spectral indices
exactly elsewhere. Let us finally notice that the value of the
correlation angle has to be in agreement with the CMB constraints on
the contribution originating from isocurvature perturbations obtained
from the WMAP data, see for instance Ref.~\cite{isowmap}.

\par

Another point worth discussing is the production of topological
defects at the end of inflation. As was discussed recently in
Ref.~\cite{mairiJ}, there exists quite tight constraints on the amount
of cosmic string produced at the end of hybrid inflation.  In the
present context, this problem does not exist because the models
studied here have only a single true vacuum. Hence, the production of
topological defects at the end of inflation is simply not possible.

\section{Conclusions}

We now quickly summarize our main results. Firstly, we have emphasized
the role that the shift symmetry plays in order to generate flat
enough potentials in F--term inflation supergravity. Secondly, we have
treated the issue of moduli stabilization and considered two different
possibilities, namely a simple renormalization of the potential by a
constant and the stringy motivated KKLT mechanism. Thirdly, we have
combined the two above mentioned ingredients in order to construct
inflationary models. We have shown that, quite generically, this gives
rise to models that are reminiscent of mutated inflation where the
inflationary path in the configuration space is non trivial. We have
also demonstrated that, in these models, inflation ends by violation
of the slow-roll conditions and not by instability as it is the case
in standard hybrid inflation. Finally, we have pointed out that the
calculations of cosmological perturbations may be non trivial due to
the possible presence of non-adiabatic perturbations.

\vspace{0.5cm}
\centerline{\bf Acknowledgments}
\vspace{0.2cm}

We wish to thank F.~Quevedo for several interesting comments.


\begin{thebibliography}{}

\bibitem{wmap} C.~L.~Bennet \etal, Astrophys. J. Suppl. \textbf{148},
1 (2003), {\tt astro-ph/0302207}; G.~Hinshaw \etal,
Astrophys. J. Suppl. \textbf{148}, 135 (2003), {\tt astro-ph/0302217};
L.~Verde \etal, Astrophys. J. Suppl. \textbf{148}, 195 (2003), {\tt
astro-ph/0302218}; H.~V.~ Peiris \etal
Astrophys. J. Suppl. \textbf{148}, 213 (2003), {\tt astro-ph/0302225};
A.~Kogut \etal, Astrophys. J. Suppl. \textbf{148}, 161 (2003), {\tt
astro-ph/0302213}.

\bibitem{inflation} A.~H.~Guth, Phys. Rev. D {\bf 23}, 347 (1981);
A.~D.~Linde, Phys. Lett. {\bf B108}, 389 (1982); A.~Albrecht and
P.~J.~Steinhardt, Phys. Rev. Lett. {\bf 48}, 1220 (1982); A.~Linde,
Phys. Lett. B {\bf 129}, 177 (1983).

\bibitem{pert} V.~Mukhanov and G.~Chibisov, JETP Lett. {\bf 33}, 532
(1981); Sov. Phys. JETP {\bf 56}, 258 (1982); S.~Hawking,
Phys. Lett. {\bf 115B}, 295 (1982); A.~Starobinsky, Phys. Lett. {\bf
117B}, 175 (1982); A. Guth and S.Y. Pi, Phys. Rev. Lett. {\bf 49},
1110 (1982); J.~M.~Bardeen, P.~J.~Steinhardt and M.~S.~Turner,
Phys. Rev. D {\bf 28}, 679 (1983).

\bibitem{slowroll} V.~F.~Mukhanov, H.~A.~Feldman, and
R.~H.~Brandenberger, Phys. Rep. {\bf 215}, 203 (1992); J.~E.~Lidsey
\etal, Rev. Mod. Phys. {\bf 69}, 373 (1997), {\tt astro-ph/9508078};
J.~Martin and D.~J.~Schwarz, \prd {\bf 57}, 3302 (1998), {\tt
gr-qc/9704049}; J.~Martin and D.~J.~Schwarz, \prd {\bf 62}, 103520
(2000), {\tt astro-ph/9912005}; J.~Martin, A.~Riazuelo and
D.~J.~Schwarz, Astrophys.~J.~{\bf 543}, L99 (2000), {\tt
astro-ph/0006392}; D.~J.~Schwarz, C.~A.~Terrero-Escalante, and
A.~A.~Garc\'{\i}a, Phys. Lett. B {\bf 517}, 243 (2001), {\tt
astro-ph/0106020}; S.~M.~Leach, A.~R.~Liddle, J.~Martin and
D.~J.~Schwarz, \prd {\bf 66}, 023515 (2002), {\tt astro-ph/0202094};
J.~Martin, Proceedings of the XXIV Brazilian National Meeting on
Particles and Fields, Caxambu, Brazil, (2004), {\tt astro-ph/0312492};
J.~Martin, Lecture notes of the 40th Karpacz Winter School on
Theoretical Physics, Poland, (2004), {\tt hep-th/0406011}.

\bibitem{LR} D.~Lyth and A.~Riotto, Phys. Rept. {\bf 314},
1 (1999), {\tt hep-ph/9807278}.

\bibitem{cosmological} N.~Turok, M.~Perry and P.~J.~Steinhardt, \prd
{\bf 70}, 029901 (2005), {\tt hep-th/0408083}; L.~Cornalba and
M.~S.~Costa, Fortsch. Phys. {\bf 52}, 145 (2004), {\tt
hep-th/0310099}; B.~Craps and B.~A.~Ovrut \prd {\bf 69}, 066001
(2004), {\tt hep-th/0308057}; M.~Berkooz and B.~Pioline, JCAP {\bf
0311}, 007 (2003), {\tt hep-th/0307280}; L.~Cornalba and M.~S.~Costa,
Class. Quant. Grav. {\bf 20}, 3969 (2003), {\tt hep-th/0302137};
E.~Dudas, J.~Mourad and C.~Timirgaziu, Nucl. Phys. {\bf B660}, 3
(2003), {\tt hep-th/0209176}; M.~Fabinger and J.~Mc~Greevy, JHEP {\bf
0306}, 042 (2003), {\tt hep-th/0206196}; H.~Liu, G.~Moore and
N.~Seiberg, JHEP {\bf 0210}, 031 (2002), {\tt hep-th/0206182}; H.~Liu,
G.~Moore and N.~Seiberg, JHEP {\bf 0206}, 045 (2002), {\tt
hep-th/0204168}; L.~Cornalba and M.~S.~Costa, \prd {\bf 66}, 066001
(2002), {\tt hep-th/0203031}; N.~Ohta, Int. J. Mod.  Phys. {\bf A20},
1 (2005), {\tt hep-th/0411230}; U.~H.~Danielsson, {\tt
hep-th/0409274}.

\bibitem{PBB} M.~Gasperini and G.~Veneziano, Astropart.~Phys.~{\bf 1},
317 (1993); G.~Veneziano, in {\sl The primordial Universe}, Les
Houches, session LXXI, edited by P.~Bin\'etruy {\it et al.}, (EDP
Science \& Springer, Paris, 2000); M.~Gasperini and G.~Veneziano,
Phys. Rep. {\bf 373}, 1 (2003), {\tt hep-th/0207130}.

\bibitem{ekp} J.~Khoury, B.~A.~Ovrut, P.~J.~Steinhardt and N.~Turok,
\prd {\bf 64}, 123522 (2001), {\tt hep-th/0103239}; J.~Khoury,
B.~A.~Ovrut, P.~J.~Steinhardt and N.~Turok, {\tt hep-th/0105212};
J.~Khoury, B.~A.~Ovrut, N.~Seiberg, P.~J.~Steinhardt and N.~Turok,
\prd {\bf 65}, 086007 (2002), {\tt hep-th/0108187}; J.~Khoury,
B.~A.~Ovrut, P.~J.~Steinhardt and N.~Turok, \prd {\bf 66}, 046005
(2002), {\tt hep-th/0109050}; R.~Durrer, {\tt hep-th/0112026};
R.~Kallosh, L.~Kofman and A.~Linde, \prd {\bf 64}, 123523 (2001), {\tt
hep-th/0104073}; R.~Kallosh, L.~Kofman, A.~Linde and A.~Tseytlin, \prd
{\bf 64}, 123524 (2001), {\tt hep-th/0106241}; D.~H.~Lyth,
Phys. Lett. B {\bf 524}, 1 (2002), {\tt hep-ph/0106153};
R.~H.~Brandenberger and F.~Finelli, JHEP {\bf 0111}, 056 (2001), {\tt
hep-th/0109004}; J.~Hwang, \prd {\bf 65}, 063514 (2002), {\tt
astro-ph/0109045}; D.~H.~Lyth, Phys. Lett. B {\bf 526}, 173 (2002),
{\tt hep-ph/0110007}; J.~Martin, P.~Peter, N.~Pinto-Neto and
D.~J.~Schwarz, \prd {\bf 65}, 123513 (2002), {\tt hep-th/0112128};
J.~Martin and P.~Peter, \prd {\bf 68}, 103517 (2003), {\tt
hep-th/0307077}. J.~Martin and P.~Peter, \prl {\bf 92}, 061301 (2004),
{\tt astro-ph/0312488}. J.~Martin and P.~Peter, \prd {\bf 69}, 107301
(2004), {\tt hep-th/0403173}.

\bibitem{stringinflation} J.~M.~Cline, {\tt hep-th/0501179};
  C.~P.~Burgess, Pramana {\bf 63}, 1269 (2004), {\tt hep-th/0408037};
  F.~Quevedo, Class. Quant. Grav. {\bf 19}, 5721 (2002), {\tt
  hep-th/0210292}; C.~P.~Burgess, J.~M.~Cline, H.~Stoica and
  F.~Quevedo, JHEP {\bf 0409}, 033 (2004), {\tt hep-th/0403119};
  C.~P.~Burgess, P.~Martineau, F.~Quevedo, G.~Rajesh and R.-J.~Zhang,
  JHEP {\bf 0203}, 052 (2002), {\tt hep-th/0111025}; C.~P.~Burgess,
  M.~Majumdar, D.~Nolte, F.~Quevedo, G.~Rajesh and R.-J.~Zhang, JHEP
  {\bf 0107}, 047 (2001), {\tt hep-th/0105204}; N.~Jones, H.~Stoica
  and S.~H.~H.~Tye, JHEP {\bf 0207}, 051 (2002) {\tt hep-th/0203163},
  G.~Shiu and S-H.~H.~Tye Phys. Lett. {\bf B513}, 251 (2001), {\tt
  hep-th/0105307}.


\bibitem{Dvali} G.~R.~Dvali and S.-H.~H.~Tye, Phys.~Lett.~{\bf
B450}, 72 (1999), {\tt hep-ph/9812483}.

\bibitem{Tye} H.~Firouzjahi and S.-H.~H.~Tye. Phys.~Lett.~{\bf B584},
147 (2004), {\tt hep-th/0312020}.

\bibitem{shift} J.~P.~Hsu and R.~Kallosh, JHEP {\bf 0404}, 042 (2004),
  {\tt hep-th/0402047}.

\bibitem{Hsu} J.~P.~Hsu, R.~Kallosh and S.~Prokushkin, JCAP {\bf
0312}, 009 (2003), {\tt hep-th/0311077}.

\bibitem{D3} K.~Dasgupta, J.~P.~Hsu, R.~Kallosh, A.~Linde and
M.~Zagermann, JHEP {\bf 0408}, 030 (2004), {\tt hep-th/0405247}.

\bibitem{koyama} F.~Koyama, Y.~Tachikawa and T.~Watari, \prd {\bf 69}
106001 (2004), {\tt hep-th/0311191}.

\bibitem{KKLT} S.~Kachru, R.~Kallosh, A.~Linde and S.~P.~Trivedi, \prd
{\bf 68}, 046005 (2003), {\tt hep-th/0301240}.

\bibitem{Burgess2} C.~P.~Burgess, R.~Kallosh and F.~Quevedo, JHEP {\bf
0310}, 056 (2003) {\tt hep-th/0309187}.

\bibitem{nilles} K.~Choi, A.~Falkowski, H.~P.~Nilles and
M.~Olechowski, {\tt hep-th/0503216}.

\bibitem{KKLMMT} S.~Kachru, R.~Kallosh, A.~Linde, J.~Maldacena,
L.~Mc~Allister and S.~P.~Trivedi, JCAP {\bf 0310}, 013 (2003), {\tt
hep-th/0308055}.

\bibitem{mac} L.~Mc~Allister {\tt hep-th/0502001}.

\bibitem{berg} M.~Berg, M.~Haack and B.~Kors, \prd {\bf 71}, 026005
  (2005), {\tt hep-th/0404087}; M.~Berg, M.~Haack and B.~Kors,
  contribution ot proceedings of ``PASCOS'04'', {\tt hep-th/0409282}.

\bibitem{race} J. J. Blanco--Pillado, C. P. Burgess, J. M. Cline,
C. Escoda, M. Gomez--Reino, R. Kallosh, A. Linde and F. Quevedo,
JHEP 0411 (2004) 063, {\tt hep-th/0406230}.

\bibitem{kinney} W.~H.~Kinney, {\tt gr-qc/0503017}.

\bibitem{linde} A.~Linde, {\em Particle Physics and Inflationary
Cosmology} (Harwood Academic Publishers, Chur, Switzerland, 1990).

\bibitem{hybrid} A.~Linde, \pl {\bf B259}, 38 (1991); A.~Linde, \prd
  {\bf 49}, 748 (1994), {\tt astro-ph/9307002}; E.~J.~Copeland,
  A.~R.~Liddle, D.~H.~Lyth, E.~D.~Stewart and D.~Wands, \prd {\bf 49},
  6410 (1994), {\tt astro-ph/9401011}.

\bibitem{mutated} E.~D.~Stewart, \pl {\bf B345}, 414 (1995), {\tt
astro-ph/9407040}.

\bibitem{shifted} R.~Jeannerot, S.~Khalil, G.~Lazarides and Q.~Shafi,
JHEP {\bf 0010}, 012 (2000), {\tt hep-ph/0002151}; R.~Jeannerot,
S.~Khalil and G.~Lazarides, JHEP {\bf 0207}, 069 (2002), {\tt
hep-ph/0207244}.

\bibitem{number} S.~Leach and A.~R.~Liddle, \prd {\bf 68}, 103503
  (2003), {\tt astro-ph/0305263}.

\bibitem{isopert} C.~Gordon, D.~Wands, B.~A.~Bassett and R.~Maartens,
\prd {\bf 63}, 023506 (2001), {\tt astro-ph/0009131}; S.~Groot
Nibbelink and B.~J.~van Tent, Class. Quant. Grav. {\bf 19}, 613
(2002), {\tt hep-ph/0107272}; B.~J.~W. van Tent,
Class. Quant. Grav. {\bf 21}, 349 (2004), {\tt astro-ph/0307048}.

\bibitem{hybridini} G.~Lazarides, C.~Panagiotakopoulos and
  N.~D.~Vlachos, \prd {\bf 54}, 1369 (1996), {\tt hep-ph/9606297};
  G.~Lazarides and N.~D.~Vlachos, \prd {\bf 56}, 4562 (1997), {\tt
  hep-ph/9707296}; N.~Tetradis, \prd {\bf 57}, 5997 (1998), {\tt
  astro-ph/9707214}; L.~E.~Mendes and A.~R.~Liddle, \prd {\bf 92},
  103511 (2000), {\tt astro-ph/0006020}.

\bibitem{coleman} S.~Coleman and F.~de~Lucia, \prd {\bf 21}, 3305
(1980).

\bibitem{wr} D.~Wands, N.~Bartolo, S.~Matarrese and A.~Riotto, \prd
  {\bf 66}, 043520 (2002), {\tt astro-ph/0205253}.

\bibitem{isowmap} K.~Moodley, M.~Bucher, J.~Dunkley, P.~G.~Ferreira
  and C.~Skordis, \prd {\bf 70}, 103520 (2004), {\tt
  astro-ph/0407304}.

\bibitem{mairiJ} J.~Rocher and M.~Sakellariadou, {\tt hep-ph/0405133};
J.~Rocher and M. Sakellariadou, \prl {\bf 94}, 011303 (2005), {\tt
hep-ph/0412143}.

\end{thebibliography}
\end{document}